\documentstyle[12pt,epsf,a4]{report}

\begin{document}
\def\vpu{\mathbf{p_1}}
\def\vpd{\mathbf{p_2}}
\def\vpt{\mathbf{p_3}}
\def\vpq{\mathbf{p_4}}
\def\tpu{\mathbf{\widetilde{p_1}}}
\def\tpd{\mathbf{\widetilde{p_2}}}
\def\tpt{\mathbf{\widetilde{p_3}}}
\def\tpq{\mathbf{\widetilde{p_4}}}
\def\so#1{$\underline{\smash{\hbox{#1}}}$}
\def\rap{ {1\over 2} }
\def\rapt{\scriptstyle {1\over 2} }  
\def\raps{\scriptstyle {1/2} }
\def\cg#1#2#3#4#5#6{ <{#1}~{#2}~{#3}~{#4}\mid {#5}~{#6}>}

\title{Polarization Phenomena in Hadronic and Nuclear Processes in Threshold 
Regime}
\author{ Egle Tomasi-Gustafsson and Michail P. Rekalo }
\date{\today}

\maketitle

\tableofcontents

\chapter{Introduction}

\vspace{.2 true cm}

The threshold region for processes of hadronic and nuclear
interactions is very interesting 
from a theoretical as well as an experimental point of view.

In this region one can apply different 
physical approaches, starting from classical current algebra methods for 
processes involving soft pions, through effective Lagrangian considerations or
perturbative chiral symmetry theory (ChPT).
In particular ChPT has been very successful in explaining the last exciting 
results about pion photo and electroproduction on nucleons near threshold 
\cite{Be96}.
        
The essential simplification of the spin structure of matrix elements for
threshold regime results in better understanding of the underlying mechanisms 
and allows a transparent analysis of polarization phenomena. The reason of this 
simplification is the presence, in threshold conditions, of a single independent 
physical direction, related to the initial momentum. Therefore the analysis of 
polarization effects near threshold, with evident axial symmetry, cannot be 
considered
as a limiting case of a general formalism, which applies
to  binary collisions, where the scattering plane is well defined, but a 
dedicated formalism has to be especially derived.

Such formalism is developped here for a wide class of processes
including non-binary processes as the production of pseudoscalar and 
vector mesons in nucleon-nucleon collisions.
The study of these processes allows to afford many interesting physical 
problems: hidden strangeness of nucleons and OZI-violation, $\eta N$ -and 
$\omega N$- interactions in S-state, determination of P-parity of
strange particles, identification of reaction mechanisms etc.

Special attention we devote here to the analysis of the spin structure
and polarization phenomena for nuclear processes with light nuclei which have 
important applications in fundamental astrophysics and in nuclear fusion.

\chapter{General Properties of Polarization Phenomena for Hadronic and Nuclear 
Physics at Threshold}

\section{Description of polarization properties of \\
fermions, bosons and  photons}

The most prominent feature of the threshold physics is the production of final 
particles in $S-$ state with zero three-momentum in the center of mass 
(CMS)\footnote{ Let us note that there are a few examples where the S-state 
production at threshold is forbidden, for example the process 
$\gamma+\pi\to \pi+\pi$ or $\gamma+^4\!He\to ^4\!He+{\pi^0}$. Due to angular 
momentum and 
P-parity conservation, only P-wave meson production is allowed.}. The notion of 
angular momentum, orbital and total, can be exactly defined only in CMS. Therefore  the 
following analysis will be derived in the CMS of all considered processes.

The description of the polarization properties of the different particles (with 
spin) involved in reactions at threshold is essentially simplified. The exact 
description of all the observables can be done nonrelativistically, all 
particles having zero (or relatively) small velocity. Therefore the complex 
relativistic description of spin is equivalent, here, to the nonrelativistic 
one.

This holds for spin 1/2, 1 and 
3/2, as we show in the following lines.

\underline{\bf Spin 1/2.} The relativistic description of the polarization 
properties of fermions with spin 1/2 is based on the formalism of four-component 
Dirac spinors, $u(p)$, where $p$ is the particle four-momentum. Using 
the Dirac equation in the standard form: 
$$(\hat p-M) u(p)=0,~~\hat p=p_{\mu}\gamma_{\mu},$$
where $M$ is the fermion mass and $\gamma_{\mu}$ are the Dirac matrixes,
one can find the following representation for the Dirac spinors $u(p)$ in 
terms 
of the two-component spinor $\phi$:
$$
u(p)=\sqrt{E+M}
\left (
\begin{array}{c} 
         \phi\\
         \displaystyle\frac{\vec\sigma\cdot\vec p}{E+M}\phi
\end{array}
\right )
$$
with the relativistic normalization: 
$u^{\dagger}u=2E,~\mbox{and}~ \phi^\dagger\phi=1$; $p=(E,\vec p)$, 
so $E$ is the energy of the particle, $\vec p$ is the 
three-momentum $E^2-\vec p^2=M^2$ (for a free particle).

At threshold, $\vec p=0$, $u(p)\to \phi$, i.e. the two-component spinor 
$\phi$ 
fully describes a spin 1/2 particle.

The density matrix, corresponding to the two-spinor $\phi$ has the following 
form:
$$\phi_i\phi_j^\dagger=\displaystyle\frac{1}{2}
\left({ \cal I}+\vec\sigma\cdot\vec P\right)_{ij},~i,j=1,2, $$
where $\vec P$ is the vector (more exactly the axial vector or pseudovector) of 
the fermion polarization, $\sigma_a=\sigma_x,\sigma_y,\sigma_z$ are the standard 
Pauli matrixes:
$$
\sigma_x=
\left (
\begin{array}{cc}
0 & 1 \\ 
1 & 0
\end{array}
\right ),~~
\sigma_y=
\left (
\begin{array}{cc}
0 & -i \\ 
i & 0
\end{array}
\right ),~~
\sigma_z=
\left (
\begin{array}{cc}
1 & 0 \\ 
0 & -1
\end{array}
\right ),~~
$$
with the following useful properties:
$$\sigma_a\sigma_b=\delta_{ab}+i\epsilon_{abc}\sigma_c,~~ 
\sigma_a^\dagger=\sigma_a, ~~(a,b,c=x,y,z)$$
and $\epsilon_{abc}$ is the absolute antisymmetric unit tensor, so that 
$\epsilon_{xyz}=1$.

Note that the $\vec P$-vector is odd under time-inversion (T-transformation) and 
even under space-inversion (P-transformation).

\underline{\bf Spin 1}. The relativistic description of particles with spin 1 
(i.e. the vector particles) involves the four -polarization vector, 
$U_{\alpha},~\alpha=x,y,z$ and 0, with the additional relation:
$$U\cdot p=0.$$

At threshold this relation can be written:
$$U_0M=0, ~\mbox{i.e.} ~U_0=0,$$
$$U_{\alpha}\to U_a,~~a=x,y,z.$$
Here the complete physical information is contained in the three-vector 
$\vec U$.

The density matrix for any vector particle, $\rho_{ab}$ can be defined as:
$$\rho_{ab}=U_a^*U_b.$$
This expression holds for stable particles, like the deuteron, as well as for 
unstable particles like vector mesons ($\rho$, $\omega$ or $\phi$). However in 
case of deuteron, with positive parity, the corresponding three-vector, 
$\vec U$, has to be considered as an axial vector, whereas for vector mesons, 
(which have negative parity), $\vec U$ is an usual (polar) vector.

The density matrix $\rho_{ab}$ can be parametrized in different and equivalent 
ways.
For stable particles, it is expressed in terms of the vector
($S_a$) and the tensor ($Q_{ab}$) polarizations as follows:
\begin{equation}
\rho_{ab}=\displaystyle\frac{1}{3}\left(\delta_{ab}-\displaystyle\frac{i}{2}
\epsilon_{abc}S_c-Q_{ab}\right ), ~Q_{ab}=Q_{ba},~Q_{aa}=0,
\label{eq:eqrho}
\end{equation}
therefore $\rho_{aa}=1$.

Different experimental methods exist, to measure the components of the vector 
and tensor deuteron polarization \cite{ETG99}, according to the energy of the 
scattered deuteron.

The expression (\ref{eq:eqrho}) can be applied, in principle, to unstable vector 
particles, also, but the vector and tensor polarizations, for these particles, 
are directly 
measured through the analysis of the angular distribution of their decay 
products. One can show that this angular distribution is related to the 
different elements of the density matrix, $\rho_{ab}$:
\begin{equation}
S_a=i\epsilon_{abc} \rho_{bc},
\end{equation}
\begin{equation}
Q_{ab}=-\displaystyle\frac{1}{2}\left [\rho_{ab}+\rho_{ba} -2\delta_{ab}\right 
].
\label{eq:eq3}
\end{equation}

These two descriptions (one in terms of $\rho_{ab}$ and the other in terms of 
$S_a $ and $Q_{ab}$) are equivalent, but for unstable particles one derives 
directly the elements of the density matrix.

As an example let us consider a binary process for vector 
particle production, $1+2\to 3+V$, where $1$, $2$ and $3$ are particles  
(nucleons etc..)  and $V$ a vector meson: $V=\rho,$ $\omega$, $\phi$, $J/\psi$..
The density matrix can be parametrized in the following general form which is 
valid (in case of unpolarized particles) for any reaction and reaction 
mechanism (for P-invariant interactions):
\begin{eqnarray*}
\rho_{ab}&=&\rho_1\hat m_a\hat m_b+\rho_2\hat n_a \hat n_b +\rho_3\hat k_a \hat 
k_b\\
&&+\rho_4 (\hat m_a\hat k_b+\hat m_b\hat k_a) +i \rho_5(\hat m_a\hat k_b-\hat 
m_b\hat k_a),
\\
\hat{\vec n}&=&\vec k\times\vec q/|\vec k\times\vec q|,~~
\hat{\vec k}= \vec k/|\vec k|,~~
\hat{\vec m}=\hat{\vec n}\times\hat{\vec k},
\end{eqnarray*}
where $\vec k$ and $\vec q$ are the three-momenta of particles $1$ and $3$  in 
the CMS of the considered reaction, $\rho_i=\rho_i(s,t)$, $i=1-5$, are 
the real structure functions, depending on the two Mandelstam variables $s$ and 
$t$. These structure functions determine the angular distribution of the decay 
products of the vector meson. As an example for the decays $\rho\to\pi\pi$ and 
$\phi\to K\overline{K}$ one finds:
\begin{eqnarray}
W(\theta,\phi)&=\displaystyle\frac{1-\cos^2\theta}{2} &-
\rho_3 \left( \displaystyle\frac{1-3\cos^2\theta }{2} \right)
\nonumber \\
&&+\displaystyle\frac{1}{2}\left(\rho_1-\rho_2\right )\sin^2\theta\cos 
2\phi +\rho_4\sin 2\theta\cos\phi,\label{eq:eqW}
\end{eqnarray}
where $\theta$ and $\phi$ the polar and azimuthal angles of the psudoscalar 
meson $P_1$ in the decay $V\to P_1+P_2$ (in the rest frame of the decaying 
V-meson).
We use here the normalization condition:
\begin{equation}
\rho_{aa}=\rho_{11}+\rho_{22}+\rho_{33}=1,~~ 
\mbox{i.e}~\rho_{1}+\rho_{2}+\rho_{3}=1.
\end{equation}
Eq. (\ref{eq:eqW})  shows that the measurement of $\theta$- and 
$\phi$-dependences of the decay 
products in $V\to P_1+P_2$ allows to determine, in the general case, the SFs 
$\rho_1$, $\rho_2$, $\rho_3$,and  $\rho_4$, which characterize the symmetric 
part of the density matrix for the vector mesons.

However the SF $\rho_5$, which is related to the antisymmetric part of the 
density matrix, can not be determined in this way. This is an important point 
for the analysis of the polarization properties of vector mesons through their 
decays. The main decays of the vector mesons, such as:
\begin{equation}
V\to P+P,~  P+\gamma,~ \ell^++\ell^-,~PPP~, V_1+P,~...\label{eq:eq6}
\end{equation}
are driven by strong and electromagnetic interactions with conservation of $P-$ 
parity \footnote{This differs, for example, from the decay $\Lambda\to p\pi^-$ 
which is driven by the weak interaction with strong violation of 
$P-$invariance. The $\Lambda-$hyperon (spin 1/2 particle) is a self-analyzing 
particle: its vector polarization can be determined from the decay angular 
distribution.}. As a result, an analysis of the spin structure of the matrix 
elements of the processes 
(\ref{eq:eq6}) shows that 
the presence of a single interaction constant for each of such decays, can not 
induce T-odd correlations, 
which play an essential role in the measurement of the vector polarization of 
$V-$mesons. Therefore, none of the decays (\ref{eq:eq6}) can give access to the 
antisymmetric part of the corresponding density matrix.

\underline{\bf Spin 3/2} The relativistic description of particles with spin 
3/2, 
(for example the $\Delta$ isobar), in terms of a four-component Dirac spinor 
with vector index $U_\mu(p)$ can be reduced, in near threshold 
conditions, into a two-component spinor, with three vector indexes:
$$U_\mu(p)\to \chi_a,~a=x,y,z$$
with the important constrain: $\vec\sigma\cdot\vec\chi=0$. 
Summing over the polarization states gives for the density matrix:
$$\rho_{ab}(3/2)=\overline{\chi_a^\dagger\chi_b=}=\displaystyle\frac{2}{3}\left 
(\delta_{ab}-\displaystyle\frac{i}{2}\epsilon_{abc}\sigma_c \right ).$$

\underline{\bf Photon} Real photons are vector particles, with zero mass, 
therefore their polarization properties are described in a particular way. The  
polarization can be characterized by the three-vector $\vec e$, which satisfies 
the Lorentz condition: $\vec e \cdot \vec k=0$, where 
$\hat{\vec k}$ is the unit vector along the photon three-momentum. The sum 
over the photon polarizations is given by:
$$\sum_{i=1,2}e_a^{(i)*}e_b^{(i)}=\delta_{ab}-\hat {k_a}\hat {k_b},$$
where the upper index $i$ numerates the two possible independent polarization 
vectors of the photon.

The proton states with definite value of the total angular momentum $j$
are classified in two groups: electric (E$j$) and magnetic (M$j$) types - with 
different values of $P-$parity: $(-1)^j$ for the electric type and $(-1)^{j+1}$ 
for the magnetic type.

In threshold conditions, for photoproduction processes, $\gamma +a\to b+c $, or 
radiative capture, $a+b\to c+\gamma$, the photon can be characterized by the 
smallest value of $j$. In this case we can easily write the corresponding 
combinations of polarization $\vec e$ and unit vector $\hat{\vec k}$ for the 
photon states with low multipolarity:
\begin{eqnarray*}
&\vec e&\to E1 ~\mbox{(electric~ dipole)},\\
&\vec e\times\hat {\vec k}&\to M1 ~\mbox{(magnetic~ dipole)},\\
&e_a\hat k_b+e_b\hat k_a \equiv E_{ab}&\to E2 ~\mbox{(electric~ quadrupole)},\\
&(\vec e\times\vec k)_a\hat k_b+(\vec e\times\vec k)_b\hat k_a
\equiv M_{ab}&\to M2 ~\mbox{(magnetic~ quadrupole)}.\\
\end{eqnarray*}

These are the basic formulas for the construction of the matrix elements for  
electromagnetic processes ( in near threshold conditions).

\section{Parametrization of the spin structure of the matrix elements}

Using the formalism presented in the previous section, we can write in a direct 
way the matrix element for any threshold process, in a general form, using only 
the symmetry properties of the strong and the electromagnetic interactions. This
problem can be exactly solved without any dependence of reaction mechanism.

Let us recall the most important symmetry principles of fundamental interactions which allow us to establish the spin structure of the matrix element:
\begin{itemize}
\item Isotropy of space (conservation of total angular momentum).
\item Invariance of relative inversion of the space coordinates 
(P-transformation and P-invariance).
\item The gauge invariance (in case of photoproduction processes, or radiative 
capture reactions).
\item The Pauli principle for identical fermions.
\item Isotopic invariance of the strong interaction and  generalized Pauli 
principle (for non-identical fermions like $n$ and $p$).
\item Invariance under charge conjugation (C-invariance)
\end{itemize}

The first step in the determination of the spin structure of the matrix element for threshold processes, is the analysis of the 
possible multipole (for processes involving photons) or partial 
transitions, which are allowed for the considered process by the above mentioned
symmetry properties. 

From our previous discussion it follows that, at threshold, the most general 
form 
of the matrix element can contain only one three-momentum $\vec k$ (in necessary 
combinations with the polarization vectors of $V-$mesons, photons, two-component 
spinors and $\sigma$-matrixes), which is the momentum of the initial state, in 
case of nonzero threshold energy, or the momentum in the final state for the 
capture or the annihilation processes at rest.

The degree of this three-vector $\hat{\vec k}$ is directly related to the value 
of the orbital angular momentum of colliding particles or the multiplicity of 
the photon : the zero degree in $\hat{\vec k}$ describes the interaction 
of the initial particles in $S$-state, the first degree in $P$-state, the second 
degree in $D$-state etc..

We illustrate this procedure on few typical examples: 
$p+d\to ^3\!He+\pi^0$ and $\pi+N \to N+V$ (strong interaction) and $\gamma+N\to 
N+V$ (vector meson photoproduction on the nucleon in the threshold region).

For $p+d\to ^3\!He+\pi^0$ the spin and parity of the particles are 
$1/2^+ +1^+\to 1/2^+ +0^-$. Therefore at 
threshold ($\pi^0$ is produced in S-state) only one value of total 
angular momentum and P-parity is allowed, ${\cal J}^P=1/2^-$ for the final 
state. Due to the conservation of the total angular momentum and P-parity 
(P-invariance of the strong interaction) the spin and parity of the initial 
state has also to be ${\cal J}^P=1/2^-$. Therefore the orbital angular momentum 
of the colliding $p+d-$system must be equal to 1: $\ell_i=1$, and the following 
partial transitions are allowed:
$$S_i=1/2,~~\ell_i=1\to {\cal J}^P=1/2^-;~
S_i=3/2,~~\ell_i=1\to {\cal J}^P=1/2^-,$$
where $S_i$ is the total spin of the $p+d$-system.

The resulting threshold matrix element can be written as:
\begin{equation}
{\cal M}(d+p \to ^3\!He+\pi^0)=\chi_2^{\dagger}
\left (i f_1\hat{\vec k}\cdot \vec D + 
f_2\vec\sigma\cdot\hat{\vec k}\times\vec D\right )\chi_1,
\label{eq:eq7}
\end{equation}
where $\chi_1$ and $\chi_2$ are the two-component spinors of the initial proton 
and the produced nucleus $^3\!He$, $\vec D$ is the polarization vector of the 
deuteron, $f_1$ and $f_2$ are the partial amplitudes of the considered process, 
which are complex functions of the excitation energy.

Their linear combinations give origine to the partial amplitudes with a definite 
value of the initial total spin $S_i$. Let's build the two possible initial 
states with  $S_i=1/2$ and $S_i=3/2$:
\begin{equation}
\vec{\psi}_{3/2}=(-2i\vec D+\vec\sigma\times\vec D)\chi_1,~~
\vec{\psi}_{1/2}=(i\vec D+\vec\sigma\times\vec D)\chi_1,
\label{eq:eq8}
\end{equation}
with the following properties:

- $\vec\sigma\cdot\vec{\psi}_{3/2}=0$: necessary condition for spin 3/2;

- $\vec\psi_{1/2}^{\dagger}\cdot\vec{\psi}_{3/2}=0$: orthogonality of states 
with different values of $S_i$.

Comparing Eqs. (\ref{eq:eq7}) and (\ref{eq:eq8}) one can find:
$$f_1=f_{1/2}-2f_{3/2},$$
$$f_2=f_{1/2}+2f_{3/2},$$
where $f_{1/2}$ and $f_{3/2}$ are the partial amplitudes corresponding to 
$S_i=1/2$ and $S_i=3/2$.

The parametrization (\ref{eq:eq7}) of 
the spin structure of the threshold matrix element 
for the process $p+d\to ^3\!He+\pi^0$ holds for any reaction mechanism and for 
any theoretical model which can be used to describe the amplitudes.

The process of threshold vector production in $\pi N$-collisions, 
$\pi+N\to N+V$, has the same combination of spins of interacting particles, but 
different P-parities: $0^-+1/2^+\to 1/2^+ + 1^-,$ 
which induce a different matrix element.
The S-state production implies: 
${\cal J}^P=1/2^-~\mbox{ ~and ~}{\cal J}^P=3/2^-$, with the two 
following partial transitions:
$$S_i=1/2,~~\ell_i=0\to {\cal J}^P=1/2^-;~~
S_i=1/2,~~\ell_i=2\to {\cal J}^P=3/2^-,$$
i.e. due to P-parity conservation, two different values of initial orbital 
momentum, $\ell_i=0$ and $\ell_i=2$ contribute. So the resulting matrix element 
can be written as follows:
$${\cal M}_{th}(\pi N\to NV)=\chi_2^{\dagger}\left (\vec\sigma\cdot\vec U^* g_1
+\vec\sigma\cdot\hat{\vec k}\hat{\vec k}\cdot\vec U^*g_2\right )\chi_1,$$
where $\chi_1$ and $\chi_2$ are the two-component spinors of the initial and 
final nucleons, $\vec U$ is the three-vector polarization of the $V$-meson, 
$g_1$ and $g_2$  are the partial amplitudes for the considered process.

As a state with orbital momentum $\ell=2$ can be described by a traceless and 
symmetrical tensor: $\ell_{ab}=\hat k_a\hat k_b -1/3\delta_{ab},$ 
the partial amplitudes corresponding to $\ell=0$ and $\ell=2$, $g^{(0)}$ and 
$g^{(2)}$, are defined as follows:
$$g^{(0)}=g_1+\displaystyle\frac{1}{3}g_2,~~g^{(2)}=g_2.$$
This procedure can be applied to the spin structure of any process of strong 
interaction.

Let us consider, now, as an example, an electromagnetic process : the threshold photoproduction of vector mesons on the nucleons, $\gamma+N\to N+V$.
There are two possible final states, corresponding to ${\cal J}^P=1/2^-$ and 
${\cal J}^P=3/2^-$. The conservation of angular momentum and P-parity allows the
following multipole transitions:
\begin{eqnarray}
E1&\to &{\cal J}^P=1/2^-, \nonumber\\
&\to &{\cal J}^P=3/2^-,\label{eq:eq9}\\
M2&\to &{\cal J}^P=3/2^-, \nonumber
\end{eqnarray}
i.e. the threshold matrix element has to contain three different combinations 
of polarization vectors $\vec e$ and $\vec U$:
$${\cal M}=\chi_2^\dagger\left [i\vec e\cdot\vec U^*f_1+\vec\sigma\cdot\vec 
e\times\vec U^*f_2+
\left (\vec\sigma\cdot\hat{\vec k}\vec U^*\cdot\vec e\times 
\hat{\vec k}+\vec\sigma\cdot\vec e\times \hat{\vec k}\vec U^*\cdot\hat{\vec k}
\right ) f_3 \right ]\chi_1,$$
where the complex conjugation of $\vec U$ means that we are describing the 
production of the $V-$meson. The amplitudes $f_1$ and $f_2$, being in zero 
degree in $\hat{\vec k}$ describe the absorption of electric dipole $\gamma$, 
and correspond respectively to the ${\cal J}^P=1/2^-$ and to the 
${\cal J}^P=3/2^-$ transitions. The amplitude $f_3$, characterizing a spin 
structure which is quadratic in $\hat{\vec k}$, describes the $M2$ absorption.

With the help of formulas (\ref{eq:eq8}), one can find the following relations 
between the 
amplitudes $f_i$ and the multipole amplitudes $e_1$, $e_3$ and $m_3$, 
corresponding to the transitions (\ref{eq:eq9}): 
$m_3=f_3,~3e_1=2f_2-f_1,~3e_3=f_1+f_2.$
The two sets of amplitudes $f_1-f_3$ from one side and the multipole 
amplitudes $e_1,~e_3$ and $m_3$ from 
another side, give equivalent descriptions of the spin structure of the 
threshold matrix element. But from the physical point of view, the multipole 
amplitudes description seems preferable: the T-invariance of hadron 
electrodynamics can be expressed in a convenient way namely in terms of these 
amplitudes, in the form of the rigorous theorem of Christ and Lee \cite{Ch66}.
Following this theorem the relative phase of the amplitudes $e_3$ and $m_3$, 
corresponding to different multipolarities and to the same value of 
${\cal J}^P=3/2^-$, must be equal to 0 (or $\pi$). 
We can the write:
$$e_1=|e_1|e^{i\delta_1},~~e_3=|e_3|e^{i\delta_3},~~m_3=|m_3|e^{i\delta_3},$$
where $\delta_1$ and $\delta_3$ are the phases for ${\cal J}^P=1/2^-$ and
${\cal J}^P=3/2^-$.

Therefore all threshold observables for any process $\gamma+N\to N+V$ are 
characterized by three moduli of multipole amplitudes and by one relative phase, 
$\delta_3-\delta_1$, only. The complete experiment, for the full reconstruction 
of 
the 
spin structure of the matrix element, has to contain three different 
polarization measurements, in addition to the differential cross section (with 
unpolarized particles).

\section{Polarization observables}

The main feature of polarization phenomena in the near threshold region is an 
essential simplification due to the presence of a single physical direction: 
the three-momentum of the colliding particles. A similar situation occurs for the capture or annihilation processes, in case of two-body reaction, like 
$\overline{p}+p\to P+P,~P+V,~V+V-$annihilation ($P$ ($V$) is a pseudoscalar 
(vector) meson), or $K^-+p\to\Lambda+\pi^-$ capture, for example,  
where the direction of the three-momenta of the final particles is the unique 
physical vector.

The consequence of such axial symmetry of threshold kinematics is that the 
standard formalism for the analysis of polarization phenomena \cite{Ohlsen}, 
which is currently used for binary processes, in case of general kinematics, 
has to be fully revised.

The main ingredients of the threshold polarization analysis are the following symmetry properties:
\begin{itemize}
\item the axial symmetry of kinematics (i.e. the presence of a single 
three-momentum results in the absence of a scattering plane);
\item the P-invariance of the strong and electromagnetic interactions of 
hadrons;
\item definite transformation properties of vector and tensor polarizations 
with respect to T- and P-transformations.
\end{itemize}
Therefore, at threshold, there are rigorous general properties of polarization observables, which 
can be formulated as follows:
\begin{itemize}
\item  All T-odd one-spin polarization observables ( such as the vector 
analyzing powers for polarized beam or polarized target and the vector 
polarization of the final particles) are identically zero for any process and 
any reaction mechanism.
\item The tensor analyzing power ${\cal T}$ (for reactions with polarized 
deuteron beam or polarized deuteron target) is nonzero, and is related to the 
cross section $\sigma$ by:
$$\sigma=\sigma_0(1+{\cal T}Q_{ab}\hat k_a\hat k_b),$$
where $\sigma_0$ is the cross section for non-polarized particles.

\item The tensor polarization of deuterons, produced in the collisions of 
unpolarized particles, can be parametrized as follows:
$$Q_{ab}=(\hat k_a\hat k_b -\displaystyle\frac{1}{3}\delta_{ab})Q,$$
i.e. the tensor $Q_{ab}$ is characterized by a single real quantity, $Q$, which 
is function of excitation energy, only.
\item The density matrix $\rho_{ab}$ for the vector mesons produced in the 
collisions of unpolarized particles has the following form:
$$\rho_{ab}=\hat k_a\hat k_b+\rho(\delta_{ab}-3\hat k_a\hat k_b),~~\rho_{aa}=1$$
where $\rho$ is a real dynamical parameter, characterizing the angular 
dependence of the decay products. For $V\to P+P$ one can find:
$$W(\theta)\simeq 1+a\cos^2\theta,~~a=-3+\displaystyle\frac{1}{\rho},$$
where $\theta$ is the angle between $\hat{\vec k}$ and 
the three-momentum of the pseudoscalar meson in the rest frame of the $V-$meson.
\item The dependence of the cross section on the vector polarizations 
$\vec P_1$ and $\vec P_2$ of the colliding particles can be  written as:
\begin{equation}
\sigma(P_1,P_2)=\sigma_0\left(1+{\cal A}_1\vec P_1\cdot\vec P_2+{\cal 
A}_2\hat{\vec k}\cdot \vec P_1\hat{\vec k}\cdot \vec P_2\right
),\label{eq:poli}
\end{equation}
where the real quantities ${\cal A}_1$ and ${\cal A}_2$ characterize the spin 
correlation coefficients:
$$C_{zz}={\cal A}_1+{\cal A}_2 ,~~C_{xx}=C_{yy}={\cal A}_1,$$
if the $z-$axis is along the $\hat{\vec k}$ direction.
\item The values $\sigma_0$,  $\sigma_0{\cal A}_1$, and $\sigma_0{\cal A}_2$, in 
the case of NN-collisions are related to the cross sections of the 
NN-interaction in the singlet state ($\sigma_s$) and in the triplet state - with 
two 
different possible projections of total spin: $\sigma_{t,0}~(\lambda=0)$ and 
$\sigma_{t,1}~(\lambda=\pm 1)$.

In order to give relations between these two sets of polarization observables, 
we introduce the following projective operators:
$$\Pi_s=\displaystyle\frac{1-\vec P_1\cdot\vec P_2}{4},$$
$$\Pi_{t,1}=\displaystyle\frac{1+\hat{\vec k}\cdot \vec P_1\hat{\vec k}\cdot 
\vec P_2}{2},$$
$$\Pi_{t,0}=\displaystyle\frac{1+\vec P_1\cdot \vec P_2-2\hat{\vec k}\cdot 
\vec P_1\hat{\vec k}\cdot \vec P_2}{4}.$$
As a result, the cross section $\sigma(\vec P_1,\vec P_2)$ can be expressed in 
terms of $\sigma_s$, $\sigma_{t,0}$ and $\sigma_{t,1}$ as:
\begin{eqnarray*}
\sigma(P_1,P_2)&=&\sigma_s\displaystyle\frac{1-\vec P_1\cdot\vec P_2}{4}
+ \sigma_{t,1}\displaystyle\frac{1+\hat{\vec k}\cdot \vec P_1\hat{\vec k}\cdot 
\vec P_2}{2} + \nonumber\\
&&\sigma_{t,0}\displaystyle\frac{1+\vec P_1\cdot \vec P_2-2 \hat{\vec k}
\cdot \vec P_1\hat{\vec k}\cdot \vec P_2}{4};
\end{eqnarray*}
with the relations: $\sigma_0{\cal A}_1=(-\sigma_s+\sigma_{t,0})/4;~~ 
\sigma_0{\cal A}_2=(-2\sigma_{t,0}+2\sigma_{t,1})/4$, and 
$\sigma_0=(\sigma_s+\sigma_{t,0}+2\sigma_{t,1})/4$.
\item The collisions of polarized deuteron with polarized nucleon is 
characterized by the following formula:
$$\sigma(\vec d+\vec p)=\sigma_0\left ( 1+Q_{ab}\hat k_a\hat k_b 
{\cal A}+{\cal A}_1\vec S\cdot \vec P +{\cal A}_2 \hat{\vec k}\cdot\vec S 
\hat{\vec k}\cdot\vec P +{\cal A}_3  \hat{\vec k}\times \vec P\cdot\vec Q \right
),$$
where $Q_a=Q_{ab}\hat k_b$ and $\vec S$ is the vector deuteron polarization.
Note that the ${\cal A}_3$-contribution is the simplest possible T-odd 
polarization observable for $\vec d+\vec p$-threshold collisions.
\item The dependence of the $V-$meson density matrix on the vector polarization 
of the beam (or target) can be parametrized as:
$$\rho_{ab}(p)=\rho_{ab}^{(0)}+\rho_{ab}^{(1)},$$
$$\rho_{ab}(1)=i\epsilon_{abc} P_c\rho_1+i\epsilon_{abc}\hat k_c\vec 
P\cdot\hat{\vec k}\rho_2+\left [ \hat k_a(\hat{\vec k}\times\vec P)_b+\hat 
k_b(\hat{\vec k}\times\vec P)_a\right ]\rho_3.$$
The real coefficients $\rho_1$ and $\rho_2$, depending on the reaction 
mechanism, 
characterize T-even effects and the coefficient $\rho_3$ T-odd effects. Note 
also that only the T-odd contribution to $\rho_{ab}^{(1)}$ 
(i.e. the SF $\rho_3$) can be measured 
through the decays (\ref{eq:eq6}): the decay $V\to P+P$ has the following 
angular dependence: $W(\theta,\phi)\simeq P_x\sin2\theta\sin \phi,$ where 
$\theta$ and $\phi$ are the polar and azimuthal angles for the decay 
products, relative to the plane defined by the vectors $\hat{\vec k}$ and $\vec 
P$, and $\vec P$ is in the $xz-$plane).

\item The dependence of the $V-$meson density matrix on the tensor polarization 
(beam or target) can be parametrized in the following form:
$$\rho_{ab}(Q)=q_1Q_{ab}+q_2\hat k_a\hat k_b \vec Q\cdot\hat{\vec k}
+q_3\delta_{ab}\vec Q\cdot \hat{\vec k}+
q_4(Q_a\hat k_b+Q_b\hat k_a)+iq_5(Q_a\hat k_b-Q_b\hat k_a),$$
where $q_1-q_5$ are real coefficients.
\item The dependence of the vector polarization of the final particles on the 
vector and tensor polarizations of the beam (or target) can be described by the 
following formula:
$$\vec P_f=t_1\vec P+t_2\hat{\vec k}(\hat{\vec k}\cdot \vec P)+t_3\hat{\vec 
k}\times\vec Q,$$
where $t_3$ is the T-odd correlation of the initial quadrupole polarization and 
the final vector polarization.

\item The dependence of the tensor polarization $Q_{ab}^{(f)}$ of the emitted 
deuteron on the vector and tensor polarizations of the initial deuteron can be 
parametrized in the following way:
\begin{eqnarray*}
Q_{ab}^{(s)}&=& i\epsilon_{abc}S_c c_1 +i\epsilon_{abc}\hat k_c\hat{\vec 
k}\cdot\vec S c_2 +\nonumber\\
&& +\left [ \hat k_a(\hat{\vec k} \times S)_b +\hat k_b(\hat{\vec k}
\times S)_a\right ]c_3 +\delta_{ab}\vec Q\cdot \hat{\vec k} c_4+\nonumber\\
&& \hat k_a\hat k_b \vec Q\cdot\hat{\vec k} c_5+(Q_a\hat k_b+Q_b\hat k_a)c_6 +
(Q_a\hat k_b-Q_b\hat k_a)c_7
\end{eqnarray*}
where the real coefficients $c_i$ determine the corresponding coefficients of 
polarization transfer from the initial to the final deuteron.
\item The polarization of the initial photon in any threshold process 
$\gamma+a\to b+c$ can not induce any observable effect. The collisions of 
linearly polarized photons with any vector polarized target is characterized by 
the same cross section as collisions of unpolarized particles. Only the collisions of 
circularly polarized photons with vector polarized target can induce 
a non-trivial asymmetry: $\sigma(\vec\gamma\vec a)=
\sigma_0(1+\lambda P_z {\cal A})$, 
where $\lambda=\pm 1$ is the photon helicity, and $P_z$ is the component of the 
target polarization along the photon three-momentum, and ${\cal A}$ is the
corresponding asymmetry.

The linear photon polarization manifests itself only in collisions with a 
tensorially polarized target: 
$\sigma(\vec\gamma \vec d ) =\sigma_0(1+Q_{ab}e_ae_b^*{\cal A}_q).$
\end{itemize}

Finally we can mention that it is possible to describe, in the same formalism, all other more complicated polarization observables, which are present in threshold conditions. The expressions, however, are always simpler in comparison with the case of general kinematics.

Having a definite parametrization of the spin structure of the matrix element of 
any concrete process, it is possible to find the expressions for all these 
polarization observables, in terms of the corresponding partial (or multipole) 
amplitudes.

\chapter{Application to Hadronic Interaction}

\section{The $\eta$-meson production in NN-Collisions}

\subsection{Polarization phenomena for the S-state $\eta$-production in the 
reactions $N+N\rightarrow N+N+\eta.$}

We discuss here the polarization effects in processes of $\eta-$production in 
$NN-$collisions near threshold:
\begin{eqnarray}
p~+~p&\rightarrow & p~+~p~+~\eta, \nonumber\\
n~+~p&\rightarrow & n~+~p~+~\eta.
\end{eqnarray}
There are few experimental data about $\vec p p$-collisions \cite{Tati00}.
Data exist on the  differential and total cross sections with unpolarized
particles in the initial and final states, in particular on the energy 
dependence of the total cross section for $p+p\rightarrow p+p+\eta$ 
\cite{Be93,Ch94b,Ca96}. 

The cross section of 
$\eta-$production in $np-$collisions is
much larger than in the case of $pp-$production \cite{Ch94}, namely
\begin{equation}
R_{\eta}=\frac{\sigma(n+p \rightarrow n+p+\eta)}
{\sigma(n+p\rightarrow p+p+\eta)}=10\pm 2 ~(5\pm 
1)~\mbox{at}~E_{kin}=1.3~(1.5)~\mbox{GeV}
\label{ratio}
\end{equation}
The standard assumptions \cite{Ge90,La91a,Ve91} about the mechanism of $\eta-$
production in 
$NN-$interactions are based on different models of one-boson exchanges 
($\pi,~\rho,~\omega,$ or $\eta$), including the effects of strong final state 
interaction. In particular, near threshold, the excitation of the 
$S_{11}(1535)-$resonance
is dominant. The values of $R_{\eta}$ in (\ref{ratio}) 
can be explained as the result of 
special interference effects of different contributions to the amplitudes of 
the corresponding processes: $\pi-$ and $\rho-$contributions must interfere 
constructively in the case of $np-$collisions and destructively for 
$pp-$collisions.

The precise definition of the
threshold energy region for the process $N+N\rightarrow N+N+\eta$ is 
$\ell_1=\ell_2=0$, where $\ell_1$ is the orbital angular 
momentum of the relative motion of the two 
produced nucleons and
$\ell_2$ is the orbital momentum of the $\eta-$meson relative to the CMS  of these two nucleons. As the isotopic structure of the 
amplitudes for $p+p \rightarrow p+p+\eta$ and  $n+p \rightarrow n+p+\eta$ is 
different, we analyze separately these two processes.
\vspace*{0.5 true cm}

$\underline{p+p \rightarrow p+p+\eta}$
\vspace*{0.2 true cm}

Taking into account the Pauli principle for the $pp-$system in the initial and 
final states, the conservation of the total angular momentum and the 
conservation of the $P-$parity, only one partial transition is allowed at threshold:
$$
L=1,~ S_i(pp)=1 ~~\rightarrow~~{\cal J}^{P}
=0^-\rightarrow~S_f(pp)=1,~~\ell_1=\ell_2=0 , 
$$
where $S_{i,f}(pp)$ is the total spin of both protons in the initial and final 
states and $L$ is the orbital momentum of the colliding protons.
The matrix element corresponding to this transition can be written in the
following form (in the CMS of the considered reaction):
\begin{equation}
{\cal M}(pp\rightarrow pp\eta)=f_1(\tilde {\chi}_2~\sigma_y ~\vec{\sigma}
\cdot\vec{k}\chi_1)~(\chi^{\dagger}_4 \sigma_y\ \tilde {\chi}^{\dagger}_3 
)\label{eq:eta7},
\end{equation}
\noindent where $\chi_1$ and $\chi_2$ ( $\chi_3$ and $\chi_4$) are the
two-component spinors of the two incoming (outgoing) protons; $\vec k$ is
the unit vector along the 3-momentum of the initial proton; $f_1$ is the S-wave 
partial amplitude corresponding to the total isotopic spin of the channel equal to 1.
In the general case the amplitude $f_1$ is a complex function depending on three 
kinematical variables, namely $\sqrt{s}$, $E_p$ and $E_{\eta}$, where $E_p$ 
($E_{\eta}$) is the energy of the produced nucleon ($\eta-$meson). A dynamical 
model is needed to describe this function, but any polarization observable can 
be calculated without any model, using only the expression (\ref{eq:eta7}) for the matrix 
element. It is important to stress that all polarization observables do not 
depend on these kinematical variables and have an universal character. In 
particular all polarization observables have the same value for $\eta$, $\eta'$ 
and for any possible radial excitation of the $\eta-$meson.

From Eq. (\ref{eq:eta7}) it appears that all one-spin polarization observables in the 
near-threshold region  must be zero. as well as the coefficients of polarization 
transfer. On the other 
hand the collision of polarized protons (with polarizations  $\vec P_1$ and 
$\vec P_2$) can produce nonzero asymmetries:
\begin{equation}
\displaystyle \frac {d\sigma}{d\omega}(\vec P_1, \vec P_2)=\left ({\frac 
{d\sigma}{d\omega}}\right )_0~(1+\vec P_1\cdot\vec P_2-2 ~\hat{\vec k}\cdot\vec
P_1~\hat{\vec k}\cdot\vec P_2)\label{eta8},
\end{equation}
 where ${\displaystyle \left({\frac {d\sigma}{d\omega}}\right )_0}$ is the 
differential cross section with unpolarized particles, $d\omega$ is the phase 
space volume of the produced particles, i.e.: 
$C_{xx}=C_{yy}=1$, $C_{zz}=-1$, where
the $z$-axis is along $\vec k$. 

\vspace*{0.5 true cm}
$\underline{n+p \rightarrow n+p+\eta}$ 
\vspace*{0.2 true cm}

The total isotopic spin of the $n+p-$system can take two values, $I=0$ and 
$I=1$. The derivation for the case of $I=1$ is similar to $p+p\rightarrow 
p+p+\eta$. From the isotopic invariance of the strong interaction it follows 
that: $f_1^{np}=\frac{1}{2}f_1^{pp}=\frac{1}{2}f_1.$ In the 
case of $I=0$, the generalized Pauli principle requires  the 
$np-$system in the final S-state to be in a triplet spin state. The total 
angular momentum $ {\cal J}$ and parity $ P$ in this channel must be equal to  
${\cal J}^{P}~=~1^-$. Therefore an additional transition is allowed:
$$
L=1,~ S_i(np)=1 ~~\rightarrow~~{\cal J}^{P}=
1^-\rightarrow~S_f(np)=1,~~\ell_1=\ell_2=0 , 
$$
with the following matrix element:
$$
{\cal M} (n+p\rightarrow n+p+\eta)=\frac{1}{2}f_0(\tilde {\chi}_2~\sigma_y 
\chi_1)~(\chi^{\dagger}_4 ~\vec\sigma\cdot \vec k~\sigma_y\ \tilde 
{\chi}^{\dagger}_3 ),
$$
where  $f_0$ is the  amplitude of the singlet interaction of the colliding 
particles.
This amplitude is responsible for the difference in  the polarization effects in 
the two reactions $p+p \rightarrow p+p+\eta$ and $n+p \rightarrow n+p+\eta$.

The parameters ${\cal A}_1$ and ${\cal A}_2$ 
(for polarized nucleon collisions) depend only on $|f_0|^2$ and 
$|f_1|^2$:
$$
{\cal A}_1 =\frac {-|f_0|^2+|f_1|^2}{|f_0|^2+|f_1|^2};~~
{\cal A}_2 =\frac {-2|f_1|^2}{|f_0|^2+|f_1|^2}; 
$$
i.e.
$$-1\le {\cal A}_1\le 1,~~~~ -2\le {\cal A}_2\le 0.$$

It is important to note that the amplitudes  $f_0$ and $f_1$ do not interfere in 
the unpolarized differential cross section:
$$
\displaystyle \left (\frac {d\sigma}{d\omega}\right )_0\simeq 
|f_0|^2+|f_1|^2,
$$
i.e.
$$ R_{\eta}=\frac{\sigma (n+p \rightarrow p+p+\eta)}
{\sigma (p+p\rightarrow p+p+\eta)}=\frac{1}{4}+\frac{1}{4}\frac 
{|f_0|^2}{|f_1|^2}\geq \frac{1}{4}.$$

Therefore both asymmetries ${\cal A}_1$ and ${\cal A}_2$ can be related to the 
ratio $R$ of the cross sections for the production processes with unpolarized 
particles:
$$
{\cal A}_1=-1+\frac{1} {2R_{\eta}},~~~{\cal A}_2=-\frac{1} 
{2R_{\eta}}\label{li1}.
$$

These relations are independent of any models for the description of the 
$N+N\rightarrow N+N+\eta$ processes and they are valid on the level of the 
isotopic invariance of the strong interaction. 
Using the experimental values of $ R_{\eta}$ \cite{Ch94}, one can predict the 
following numerical values the asymmetries ${\cal A}_1$ and ${\cal A}_2$
(at two proton kinetic energies):
\begin{eqnarray*}
{\cal A}_1=-0.95\pm 0.01,& {\cal A}_2=-0.05\pm 0.01,&E_k=1.3~\mbox{GeV}, \\
{\cal A}_1=-0.90\pm 0.02,& {\cal A}_2=-0.10\pm 0.02,&E_k=1.5~\mbox{GeV}. 
\end{eqnarray*}
From the values of $R_{\eta}$, the singlet amplitude $f_0$ is much larger than 
the triplet amplitude $f_1$, and 
shows an evident decrease away from threshold:
$$\frac{|f_0|^2}{|f_1|^2}=4R_{\eta}-1=39\pm 8 (19\pm 8)~\mbox{at}~ E_k=1.3(1.5)~\mbox{GeV}.$$

Data from CELSIUS \cite{Ca96} confirm this behavior over an 
energy range extending from 25 up to 115 MeV above threshold.

We can try to interprete this large ratio in terms of the presence of
$s\overline{s}$-quarks in nucleons: the singlet $pp$-state (with the
$s\overline{s}$-component in a singlet state also) is the most suitable for
the production of the pseudoscalar $\eta$-meson by analogy with the $\phi$-meson 
production from the triplet state of $NN$ (or $\overline{N}N$)-collisions.

The decrease of the ratio ${|f_0|^2}/{|f_1|^2}$ when the energy of the colliding 
particles increases can be explained as a "dilution" effect \cite{El95}. The 
opening of different channels when the energy increases, in particular the 
triplet states for $np$-collisions which does not favor the $\eta$-production 
(or the singlet state in $pp$-collisions which favors the $\eta$-production) 
leads to a decreasing of the ratio $R_{\eta}$ in agreement with the experiment. 
We do not have at the moment any model to predict quantitatively these effects.

As  final remark on polarization effects in the reaction 
${n+p \rightarrow n+p+\eta}$, we note that the relative phase $\delta$ of the 
complex amplitudes $f_0$ and $f_1$ can be deduced from the coefficients of 
polarization transfer from the initial to the final nucleon: 
$K_z^{z'}\simeq {\cal R}e~f_0f_1^*=|f_0||f_1|~cos\delta.$
The ${\cal I}m~f_0f_1^*$ combination appears only in the T-odd polarization 
observables for $N+N\rightarrow N+N+\eta$, which, in the near threshold region 
are at least triple correlations, such as $\vec S_1\times\vec S_2\cdot\vec S_3$, 
or $\vec S_1\times \vec S_2\cdot\hat{\vec k}~\vec S_3\cdot\hat{\vec k}$.

\subsection{$P-$wave contributions to  $N+N\rightarrow N+N+\eta$.}

There are two possible combinations of angular momenta, in case of 
final $P-$wave production:
$$
\begin{array}{rcl}
 a) & \ell_1=1,& \ell_2=0,\\
 b) & \ell_1=0,& \ell_2=1,
\end{array}
$$
whose relative contribution is  a function of the energy of the produced 
particles.

In the reaction $p+p\rightarrow p+p+\eta$, for $ \ell_1=1,~ \ell_2=0 $, the 
produced $pp$-system must be in a triplet state, therefore we obtain: 
$~{\cal J}^{P}=0^+,~1^+$ and $2^+.$ The conservation of ${\cal J}$, $P$ and the Pauli principle allow two transitions:
$$
\begin{array}{rcl}
 S_i(pp)=0,~L=0\rightarrow &{\cal J}^{ P}=0^+,\\
 S_i(pp)=0,~L=2\rightarrow &{\cal J}^{ P}=2^+.
\end{array}
$$
The corresponding matrix elements can be written as:
\begin{equation}
\begin{array}{ll}
&p_0~(\tilde {\chi}_2~\sigma_y \chi_1)(\chi^{\dagger}_4 \vec\sigma\cdot \vec 
m~\sigma_y \tilde{\chi}_3^\dagger), \\
&p_2~(\tilde {\chi}_2~\sigma_y \chi_1)
~\chi^{\dagger}_4 (\sigma_im_j+\sigma_jm_i-\frac{2}{3}\delta_{ij}\vec 
\sigma\cdot\vec m)(k_ik_j-\frac{1}{3}\delta_{ij})
\sigma_y\tilde {\chi}_3^\dagger,
\label{eq:eta5}
\end{array}
\end{equation}
where $\vec m$ is the unit vector along the 3-momentum of a produced nucleon in 
the $CMS$, $p_0$ and $p_2$  are the two $P-$wave amplitudes, which describe the 
singlet $np-$interactions in $S~(L=0)-$ and $D~(L=2)$-states.

These new amplitudes produce an anisotropy in the angular distribution of the 
final protons with respect to the angle $\psi$, where 
$\cos \psi=\vec k \cdot \vec m$:
\begin{eqnarray*}
 |p_0|^2\rightarrow &~isotropic ~angular ~dependence, \\
 |p_2|^2\rightarrow &(1+3cos^2\psi)~ angular~ dependence,\\
 {\cal R}e~p_0p_2^*\rightarrow &(1-3cos^2\psi) ~angular~ dependence;
\end{eqnarray*}
Therefore the presence of a $cos^2\psi$-term in $(d\sigma/d\omega)_0$ shows
that the amplitude $p_2$ is different from zero.  Nethertheless, for
$p+p\rightarrow p+p+\eta$, the knowledge of the polarization observables is
necessary for the full reconstruction of the spin structure of the amplitude
in the $S+P$-waves approximation. From Eq. (\ref{eq:eta7}) and (\ref{eq:eta5}) we can do the
following remarks:
\begin{itemize}
\item As the amplitudes of the $S(P)$-wave production correspond to singlet
(triplet) $\rightarrow$ triplet (singlet) transition in the $pp$-system, no
polarization correlation coefficient in the reaction $\vec p+\vec
p\rightarrow p+p+\eta$ contains $S+P$-interference contributions.
\item The analyzing powers in the reaction $ p+\vec p\rightarrow p+p+\eta$ and 
the polarizations of any final proton produced in collisions of unpolarized 
protons must be equal to zero, for any values of the amplitudes $f_1$, 
$p_0$, and $p_2$.
\item To study the $S+P$-interference, it is necessary to measure the 
polarization transfer coefficients.
\end{itemize}
We stress again that these remarks are model-independent as they are based on 
the most general symmetry properties of the strong interaction.

The $P-$ wave in $ p+ p\rightarrow p+p+\eta$ corresponds to the singlet $pp-$ 
interaction of the colliding particles. It would then be responsible for the 
steep increasing of the cross section observed for this reaction 
near threshold, if we assume a singlet state preference for 
$\eta$-production.

The other possibility for a  $P-$wave contribution in $ p+ p\rightarrow 
p+p+\eta$, i. e. $\ell_1=0,~\ell_2=1$ can be analyzed in a similar way. In this 
case the final $pp-$system must be in a singlet state, therefore ${\cal 
J}^P=1^+$. The P-parity conservation allows only even values of the orbital 
momentum $L$ for the colliding protons, so from the Pauli principle the initial 
$pp-$system must be spin singlet. But for these quantum numbers the state ${\cal 
J}^P=1^+$ can not be obtained: $P-$wave $\eta$-production is then forbidden in 
this process.

The situation is different for the process $ n+ p\rightarrow n+p+\eta$, where
the $P-$wave $\eta$-production is possible, for $I=0$. Then the produced 
$np-$system with $\ell_1=0$ has to be in a triplet state with 
$~{\cal J}^{P}=0^+,~1^+ $ and $2^+.$ Therefore the initial $np-$system must be 
also in a triplet state, with $L=0$ and $L=2$, i.e. the following transitions 
are allowed:
\begin{equation}
\begin{array}{rcl}
 S_i(np)=1,~L=0\rightarrow &{\cal J}^{ P}=1^+,\nonumber \\
 S_i(np)=1,~L=2\rightarrow &{\cal J}^{ P}=1^+,\label{eq:eta6} \\
 S_i(np)=1,~L=2\rightarrow &{\cal J}^{ P}=2^+.\nonumber 
\end{array}
\end{equation}
From our analysis (and in agreement with the experimental data) it is shown that the enhancement of the cross section for the  $\eta$-meson production is 
directly linked to the presence of a {\bf singlet} state in the initial state of the {\it NN} system. When the initial {\it NN} state is selected in a triplet 
state such enhancement is absent.

The main results obtained above can be summarized as follows:
\begin{itemize}

\item At threshold the spin structure of the amplitudes for the processes
$p+p \rightarrow p+p+\eta$ and $n+p\rightarrow  p+p+\eta$
is different: only one amplitude ({\it triplet}) is present in the first 
reaction, while in the second reaction two complex amplitudes contribute:
$f_0~ (singlet)$ and $f_1~(triplet)$. These amplitudes do not interfere in the 
differential cross section if the particles in the initial and final state are 
unpolarized.

\item Using the experimental values of the ratio:
$ R_{\eta}=\displaystyle\frac{\sigma(n+p \rightarrow n+p+\eta)}
{\sigma(p+p\rightarrow p+p+\eta)},$
we found in a model independent way the ratio of singlet and triplet amplitudes 
for the processes $NN\rightarrow NN\eta$, in the threshold region.

\item From the ratio $R_{\eta}$ it is possible to predict the values of 
the spin correlation coefficients ${\cal A}$ for the process $\vec n+\vec 
p\rightarrow  p+p+\eta$ where both the nucleons are polarized:
${\cal A}_1 =-0.95\pm 0.01$ and ${\cal A}_2 =-0.05\pm 0.01$ (at E=1.3 GeV).
\item The abnormally large value for the ratio 
$\displaystyle \frac{|f_0|^2}{|f_1|^2}$
(near the threshold of $NN \rightarrow NN\eta$) is a clear indication that 
$\eta$ production is increased in the presence of a singlet state of the $NN$ 
system. This can be relied to the presence of a polarized 
$s\overline{s}$-component inside a polarized nucleon. 

We would like to stress that the high level of symmetry of the $NN$ state in the 
threshold $\eta$-production induces well defined polarization properties of the 
colliding nucleons, similarly to the case of $\overline{p} p\rightarrow \phi 
\pi$ annihilation.
\item The decrease of the ratio $R_{\eta}$ when the energy of 
the interacting particles is increasing may be connected to some 
'dilution' of the {\it pure} singlet states in $np$-collisions 
and with the appearance of such states in $pp$-collisions, due 
to the $P-wave$ production of the $\eta$-meson. This behavior could explain the 
observed steep rising 
of the cross section for the $p+ p\rightarrow p+ p +\eta$ reaction 
near threshold. We predict that the increasing of the cross section 
for the $n +p\rightarrow n+ p +\eta$ reaction has to be slower.
\end{itemize}

\section{Production of vector mesons  in NN-Collisions}

According to the naive quark model, the nucleon (and anti-nucleon) 
wave function contain only $u$- and $d$-quarks (antiquarks) contributions. On the other hand the 
$\phi$-meson is almost a pure $s\overline{s}-$state. Therefore $\phi$-meson 
production through the disconnected diagram of Fig. 1a is forbidden, 
contrary to the production of $\omega$-meson whose wave function has essentially no strange component.
This is the basis of the so-called OZI rule \cite{Ok63,Zw64,Ii66}. 

Slight  violation of the OZI rule, as measured by the ratio $R=\phi X/\omega X$
for production of $\omega$- and $\phi$-mesons, have been observed in various 
reactions ($R\simeq (10\div 20)\cdot 10^{-3}$), but they may be explained partly 
by the fact that the mixing angle between the $\omega$- and $\phi$-meson is not 
exactly equal to the ideal one \cite{Li76}, partly due to rescattering effects 
or multi-step processes \cite{Do89,Lo94,Bu94}; see also discussion in 
\cite{El89}.
More recently, much larger violations of the OZI rule have been reported in 
vector meson production through  $\overline {p} p$-annihilation at rest 
\cite{As91,Cr95,Cr93,Ob94,Ob95}, 
allowing to formulate the interesting hypothesis of the presence of a large 
$\overline {s} s-$component in the nucleon wave function at relatively small 
momentum transfers
\cite{Bu94,Br88} (see diagram of Fig. 1b).
In particular the abnormal yields of $\phi$-meson,($R\simeq (100\div 250) \cdot 
10^{-3}$) which was observed in the
annihilation channels: $\overline {p} + p\rightarrow \phi+\pi^0$,  $\overline 
{p} + p\rightarrow \phi+\gamma$ in liquid targets \cite{Cr95} and $\overline 
{p} + p\rightarrow \phi+\pi^{+}+\pi^-$ \cite{Ob95}, are related to the S-wave 
channel, with no large deviation from the naive OZI prediction in the P-wave 
annihilation channel.
In this case parity conservation and charge conjugation selection rules allow 
only a spin triplet  $p\overline{p}-$initial state leading to the suggestion  
\cite{El95,Al95} that polarization measurements in nucleon and anti-nucleon 
induced reactions, with polarized colliding particles, could lead to decisive 
information on the polarization of the possible $s\overline{s}-$component in the 
nucleon, which has tentatively been observed in deep inelastic lepton scattering \cite{El88}.

So it appears very interesting to study polarization effects in
different processes of $\phi$-meson production, in order to get further
evidence of the possible link between the violation of the OZI rule and the
polarization states of the interacting particles.

The simplest of such processes is the collision of a polarized proton beam
with a polarized proton target: 
$\vec p~+~\vec p \rightarrow  p~+~ p~+~\phi .$

If the spin-triplet mechanism is dominating at threshold, then the $\phi$-meson 
yield should be
much larger when the spins of the colliding protons are parallel than when they
are antiparallel.

A rather large violation of the OZI rule has also been observed in the
$\omega$- and $\phi$- production induced by the reaction $d+p\rightarrow
^3He+\phi(\omega)$. The ratio R of the cross sections for these processes
 has been found to be \cite{SP495}:
$$
R=\frac{\sigma(d+p \rightarrow ^3He+\phi) }
{\sigma(d+p \rightarrow ^3He+\omega )} = (11.6 \pm 2.9)\% .
$$
Therefore it has been proposed to study $\phi$ and $\omega$ production in
the same reaction whith  polarized  beam and  target \cite{Sa94} :
$\vec d~+~\vec p \rightarrow ^3He~+ X$, in order to enhance the OZI violation 
effects since it is supposed that here again the ratio R should be much larger 
for parallel spin states than for anti-parallel ones.

At the reaction threshold the number of partial waves is greatly reduced, resulting in a 
simplification of the spin structure of the amplitudes, therefore making the 
theoretical analysis more transparent. In most cases a general analysis of
polarization effects can be carried on, based only on the symmetry
properties of strong interaction, such as the P-invariance, the
C-invariance and the isotopic invariance, without the need to introduce any
additional hypothesis about the reaction mechanism.

On the other hand, the threshold region has some specific problems,
connected mainly to the effects of the final state interaction (FSI) of the
produced particles, which can modify the simple initial picture of the
production mechanism. Corrections to $\phi-$ and $\omega-$production 
ratio due to FSI effects can reach one order of magnitude \cite{Wu94}. In spite of these
difficulties, data on polarization effects in the near-threshold region can
reveal very interesting features, as these effects are usually less
sensitive to FSI.

Experiments aiming at measuring $\phi$-production reactions 
induced by polarized protons have been proposed \cite{Be95} or are being discussed at existing accelerators like the Dubna Accelerator Complex.

We discuss here polarization effects in the reactions: 
$p(n)~+~p \rightarrow   p(n)~+~p~+~V^0$, where $V^0$ is any neutral vector meson ($\omega$, $\phi$ or
$\rho^0$), on the basis of the simplifications which appear naturally in the
threshold region.  In particular we would like to stress that in the
case of $pp-collisions$, the spin structure of the threshold amplitude is so simplified
that it can be compared to the $\overline{p}p-$annihilation (with stopped
antiprotons) through the channel $\overline {p}~+~p ~\rightarrow \phi ~
+~\pi^0$ , which shows such a large yield for the triplet state.  

In principle the threshold  region can be broad: for example, in the reaction $\pi^-+p
\rightarrow n+\omega $ the angular distribution of the produced $\omega$-meson 
 is isotropic up to $p^*_{\omega}=200~ MeV/c$, where $p^*_{\omega}$ is the {\it
CMS} momentum of the $\omega$-meson \cite{Bi73,Ke76}. 

In the final state of the processes $p+p \rightarrow p+p+V^0$, taking into
account the identity of the two produced protons (Pauli principle), the
$pp$-system can be produced only in the singlet state, therefore there is
only one possible configuration for the total angular momentum $\cal J$ and
the parity $ P$, that is ${\cal J^P}=1^- $. In the initial state, 
due to $ P$-parity conservation, only
odd values for the orbital angular momentum $L$ are allowed. As the total
wave function has to be antisymmetric, the two colliding protons have
to be in a triplet state, $S_i=1$.  Therefore only one possible transition
can take place at threshold in the reaction $p+p \rightarrow p+p+V^0$, 
$L=1,~ S_i(pp)=1 ~~\rightarrow~~{\cal J}^{ P}=1^- $ with matrix element:
\begin{equation}
{\cal M}=g_1(\tilde {\chi}_2~\sigma_y \vec{\sigma}\cdot\vec{k} \times\vec
U^*\chi_1)~(\chi^{\dagger}_4 \sigma_y\ \tilde {\chi}^{\dagger}_3 )\label{eq:vm7} ,
\end{equation}
\noindent where  $\vec U$ is the
3-vector polarization of the produced vector meson and $g_1$ is the complex
amplitude corresponding to the triplet interaction of the colliding
particles.  The formula (\ref{eq:vm7}) is universal in the sense that it is valid for
any reaction mechanism which conserves the $P$-parity and does not
contradict the Pauli principle.

The most important consequence that follows from (\ref{eq:vm7}) is that the matrix
element of such a complicated process as $p+p \rightarrow p+p+V^0$ is
defined by a single amplitude $g_1$. All the dynamics of the process is
contained in this amplitude and can be calculated in a framework of a
definite model. But the spin structure of the total amplitude is established
exactly by Eq. (\ref{eq:vm7}) in terms of the 2-component spinors and the vector
polarization $\vec U$. Therefore the polarization effects for any reaction
$p+p \rightarrow p+p+V^0$ can be predicted exactly since they do
not depend on the specific form of the single amplitude $g_1$.  
Of course, $g_1$ depends on the nature of the produced meson and in general
$g_1^{\rho}\neq g_1^{\omega}\neq g_1^{\phi} $, so that the differential cross 
section for the different  
$p+p \rightarrow p+p+V^0$ processes may be different, 
but the polarization observables {\it must be same, independently of the type of vector meson produced}.
 
Let us illustrate this in the calculation of the spin correlation coefficients 
in
the reaction $\vec p +\vec p \rightarrow p+p+V^0$, where both protons in the
entrance channel are polarized:
\begin{equation}
\sigma(\vec P_1, \vec P_2)=\sigma_0(1+\hat{\vec k}\cdot\vec P_1 ~
\hat{\vec k}\cdot\vec P_2)\label{eq:vm8}.
\end{equation}

It is easy to see that  the corresponding
correlation parameter is maximal and equal to $+1$. This correlation
parameter does not contain any information about the dynamics of the
considered processes, because Eq. (\ref{eq:vm8}) is directly derived 
from the $P$-invariance of the strong interaction and from the Pauli principle.  

From (\ref{eq:vm7}), it follows that the 
$V^0-$meson can be polarized even in the collision of
unpolarized protons: $\rho_{xx}=\rho_{yy}=\frac
{1}{2}~~,~~\rho_{zz}=0$, when the $z-$axis is along the initial momentum
direction. Moreover the decay $V^0 \rightarrow \ell^+\ell^-$ (due to the standard
one-photon mechanism) follows the angular distribution:
\begin{equation}
W(\theta) \approx 1+cos^2 \theta \label{eq:vm9},
\end{equation}
\noindent where $\theta$ is the angle between $\vec k$ and the direction of
the momentum of one of the leptons (in the system where the $V^0-$meson is
at rest). 

Here we should emphasize that, at threshold, the distribution (\ref{eq:vm9}) 
is universal and does not depend on assumptions of any
definite mechanism of the process $p+p \rightarrow p+p+V^0$, 
as it was predicted in \cite{El95}, where a similar distribution was obtained 
through the vector current $\overline{s} \gamma_{\mu} s$ acting between
of $s\overline{s}-$pairs in the proton.

Similarly, for the decays $\phi \rightarrow K +\overline{K}$ and $\rho^0
\rightarrow \pi^+ + \pi^-$, the angular distribution of the produced meson
follows a $sin^2 \phi-$dependence, where $\phi$ is the angle between the
3-momentum of the pseudoscalar meson (in the system where the $V^0$ is at
rest) and the direction of the momentum of the colliding particles.

\section{Production of isoscalar vector mesons: $n+p\to n+p+\phi(\omega)$}

The study of polarization effects in  $n+p\rightarrow
n+p+V^0$ is more complicated in comparison with the reaction
$p+p\rightarrow p+p+V^0$. Moreover, for $np$- collisions it is necessary to
treat separately the production of isoscalar ($\omega$ and $\phi$) and
isovector ($\rho^0$) mesons.  This is due to the different isotopic
structure of the amplitudes of the processes $ n+p \rightarrow
n+p+\omega(\phi)$ and $p+p \rightarrow p+p+\omega(\phi)$. 

Due to the isotopic invariance in the strong interactions, the spin structure of 
the amplitudes of the process $ n+p \rightarrow n+p+V^0$ with $I=1$ is described 
by Eq. (\ref{eq:vm7}). For $I=0$, if the final $np$-state is
produced in the $S-$state, then the usual total spin of this system must be
equal to 1 (to satisfy the so-called generalized Pauli principle). This
means that the produced $n+p+V^0$-system can have three values
of $ {\cal J}^{P}$: ${\cal J}^{P}=0^-,~1^-\mbox{~and } 2^-$.

From $P$-invariance, only odd values of the angular momentum $L$ are allowed for
the initial $np-$system: $L=1,~3,....$. One can then conclude that this system 
must be in the singlet state,
$S_i(np)=0$. And, finally, the conservation of the total angular momentum
results in a single possibility, namely: 
$S_i(np)=0,~L=1~\rightarrow~{\cal J^ P}=1^-$. with the following matrix element ${\cal M}_0$:
\begin{equation}
{\cal M}_0=\frac{1}{2}g_0(\tilde {\chi}_2~\sigma_y \chi_1) (\chi^{\dagger}_4
\vec\sigma\times\vec U^*\cdot\vec k\sigma_y \tilde
{\chi}_3^\dagger)\label{eq:vm11} ,
\end{equation}
\noindent where $g_0$ is the amplitude of the process $n+p\rightarrow
n+p+V^0$, which corresponds to $np-$interaction in the initial singlet state.

So, the process $n+p \rightarrow n+p+\omega(\phi)$ is characterized by two
amplitudes, namely $g_0$ and $g_1$. One can see easily that these amplitudes
do not interfere in the differential cross-section of the process $n+p
\rightarrow n+p+V^0$ (with all unpolarized particles in the initial and
final states). Therefore we can obtain the following simple formula for the
ratio of the total cross sections:
\begin{equation}
{\cal R}=\frac{\sigma(p+p \rightarrow p+p+V^0) } {\sigma(n+p \rightarrow
n+p+V^0) } =\frac{4|g_1|^2}{|g_1|^2+|g_0|^2} \label{eq:vm12}.
\end{equation}
In the threshold (or near-threshold) region, this ratio is
limited by:  $0\leq{\cal R}\leq 4 $.

We will see now that the ratio ${\cal R}$ (of unpolarized cross sections)
 contains interesting information on a set of polarization observables for the reaction $n+p\rightarrow
n+p+V^0$.  For example, ${\cal A}_1$ and ${\cal A}_2$ are two independent spin
correlation coefficients, defined only by the moduli square of the
amplitudes $g_0$ and
$g_1$:
\begin{equation}
{\cal A}_1=-\frac{|g_0|^2}{|g_0|^2+|g_1|^2},~~{\cal
A}_2=\frac{|g_1|^2}{|g_0|^2+|g_1|^2}, \label{eq:vm13}
\end{equation}
i.e.
\begin {equation}
\begin{array} {rcl}
 0~(g_0=0) & \leq -{\cal A}_1 & \leq 1~(g_1=0),\\
 0~(g_1=0) & \leq{~\cal A}_2 & \leq 1~(g_0=0).
\end{array}
\end{equation} 
One can easily see that these coefficients are related to the
ratio ${\cal R}$ of the cross sections of $pp-$ and $np-$processes with
unpolarized particles (in the initial and final states) through:
${\cal A}_1=-1+{\cal R}/4,~{\cal A}_2={\cal R}/4.$ 
But the elements of the density matrix of the $V^0$-mesons, produced in
$n+p\rightarrow n+p+V^0$, are independent from the relative values of the
amplitudes $g_0$ and $g_1$ : $\rho_{xx}=\rho_{yy}=\frac
{1}{2}~~,~~\rho_{zz}=0$. 

The interference of the amplitudes $g_0$ and $g_1$
appears only in the polarization transfer from the initial to the final
nucleons:
$$
K_x^{x'}=K_y^{y'}=\frac{-2{\cal R}e g_0 g_1^*}{|g_0|^2+|g_1|^2}.
$$

Returning now to the process $n+p \rightarrow n+p+\phi$ in connection with
the problem of the $s\overline{s}$-component in the nucleon one can mention that 
a measurement of the ratio of cross sections for $p+p\rightarrow p+p+\phi$ 
and $n+p\rightarrow n+p+\phi$,
which are directly related to the relative value of the singlet and triplet
amplitudes would allow to measure the  ratio 
$\displaystyle{\frac{|g_0|^2}{|g_1|^2}}$ and confirm the predicted 
$\phi$-production enhancement from the triplet state
of the $NN$-system. Additional information can be obtained from the measurement 
of spin transfer between the initial and final nucleons.

Starting from a very general analysis, based on the symmetry properties of the 
strong interaction, namely the validity of the Pauli principle,
the $P-$invariance and the isotopic invariance, we can summarize our results as follows.
\begin{itemize}
\item The spin structure of the threshold amplitude of the processes
$p+p\rightarrow p+p+V^0$ is defined by a single spin configuration,
corresponding to the triplet state of the initial protons. This allows very
simple and rigorous predictions for the values of all the polarization observables in these
reactions, independently of the role of a $s\overline{s}-$component
in the nucleon. 

\item The spin structure of the threshold matrix element of the process
$n+p\rightarrow n+p+\omega(\phi)$ is defined by two amplitudes, the triplet
one, $g_1$, (which coincides with the triplet amplitude for the process
$p+p\rightarrow p+p+\omega(\phi)$) and the singlet one, $g_0$, which is not
present in the reaction $p+p\rightarrow p+p+\omega(\phi)$.

\item The vector meson density matrix elements are independent from
the mechanism of the processes $ p+p \rightarrow p+p+V^0$ and $ n+p
\rightarrow n+p+V^0$, so one can obtain for the collision of unpolarized
particles:  $\rho_{xx}=\rho_{yy}=\frac {1}{2}~~,~~\rho_{zz}=0.$ Therefore
the $(1+cos^2\theta)$-distribution of the decay products for the
$V^0\rightarrow \ell^+\ell^-$ is a direct consequence of the P-invariance of
the strong interaction.

\end{itemize}
\begin{figure}
\hspace*{2truecm}\mbox{\epsfxsize=12.cm\leavevmode \epsffile{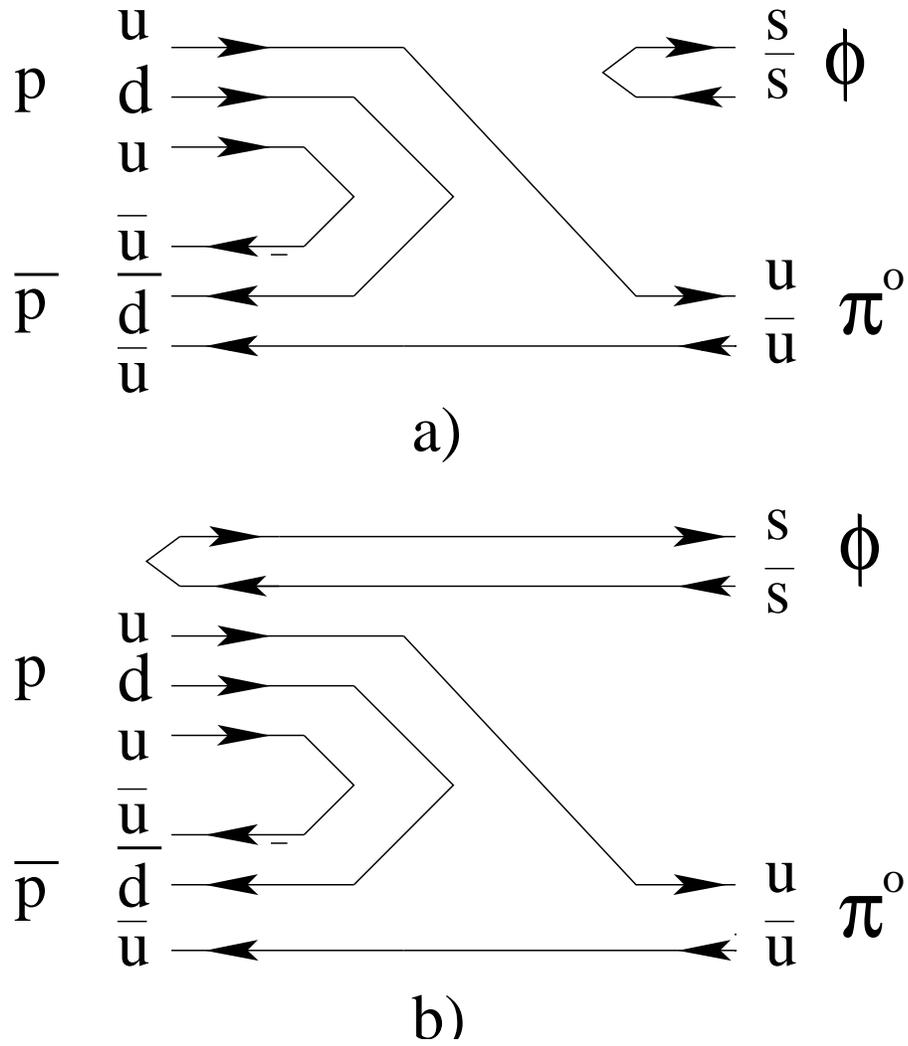}}
\caption{  a) $\phi-$meson production through the disconnected 
diagram. b) $\phi$-meson production by the OZI-allowed process from the 
$|uuds\overline{s}>$ components of the proton wave function.}
\end{figure}

\section{Production of strange particles in NN-collisions}

\subsection{Threshold theorems}
Strange particle production in $NN-$collisions may also bring 
interesting information on the possible presence of a polarized
$\overline {s} s$-component in the nucleon, even at relatively small 
momentum transfer, and on the reaction mechanism. Selection rules applied 
to the $p+p\rightarrow \Lambda(\Sigma)+K+N$ reactions in 
the near-threshold energy region, allow only a spin-triplet
$pp-$interaction. In this sense such reactions are  similar to  $\overline {p} 
p$ annihilation  into a 
 $\phi$ and a $\pi^0$,  where a large yield of  $\phi$-mesons has been observed 
which violates strongly the OZI-rule \cite{Ok63,Zw64,Ii66}.

For hadron interactions as well as for $\gamma N-$ or $eN$-interactions, many 
important low-energy theorems (LET) apply in the threshold region. Let us 
mention some 
of them:
\begin{itemize} 
\item the Thomson limit for the amplitudes of low energy Compton scattering by  
targets  with nonzero electric charges. In these processes electromagnetic 
properties of hadrons such as their electric and magnetic polarizabilities can 
be measured \cite{Lv93}.
\item the Kroll-Ruderman theorem \cite{Kr94} predicts the threshold behavior of pion 
photoproduction
amplitudes. This problem became very actual \cite{Dr92} following the experimental 
results on the $\gamma + p \rightarrow p+\pi^0$ 
cross section near threshold \cite{Ma86,Be90}. 
New data \cite{Ber96,Fu96} with 
tagged photons rise many questions about the limit of validity of the low energy 
theorems for $\pi^0$-photoproduction.
\item predictions of current algebras for the threshold pion production in 
electron and neutrino scattering by nucleons \cite{Am79}: 
$e^-+N\rightarrow  e^-+N+\pi$, $e^-+N\rightarrow e^-+\pi+\Delta,$ 
$\nu_e+N\rightarrow e^-+N+\pi.$ It is interesting to note that 
the electroproduction amplitude contains contributions which 
are proportional to the axial form factors of weak transitions:
$W^*+p\rightarrow n$ and $W^*+p\rightarrow \Delta^0$, where $W^*$ is the virtual 
$W^-$-boson.
\item the value of the $\sigma-$ term which defines the threshold amplitude 
for 
elastic $\pi N$ scattering can  be calculated within LET's. 
Some discrepancies have been found between the theoretical and experimental
values which can be explained by the presence of a
$\overline {s} s$-component in the nucleon.
\item Fundamental characteristics of hadron interactions, such as scattering 
lengths 
may be measured in the threshold region. 
Information \cite{Si94,Vi94} about the low energy $\Sigma N$- and $\Lambda N$- 
interaction
(from 
near-threshold $p+p\rightarrow K+Y+N$ processes) is important for the 
reconstruction of the 
corresponding baryon-baryon potential. 
\end{itemize}

We consider here the polarization effects in threshold production of hyperons
in nucleon-nucleon collisions.  High intensity 
polarized proton beams and the 
detection of the produced hyperons will allow to measure different polarization 
observables, such as the analyzing powers $A$ (for $\vec p+p\rightarrow K+Y+p$), 
the 
polarization $P_Y$ of the hyperons produced in the collision of unpolarized 
protons and the depolarization parameters $D_{ab}$, which give the dependence 
of the $Y$-polarization from the beam polarization \cite{Be95,Ma96}. 
Of course, the numerical values of all these polarization observables can only 
be predicted in the framework of dynamical models
\cite{An83,TdG85,Bo88,Ce91,So92,La91b}, but in the 
threshold region it is possible to find general expressions for the 
polarization observables, independently of the reaction mechanism.

Let us mention the recent experimental data about the
$p+p\to\Lambda(\Sigma^0)+K^++p$, in the threshold region, at COSY 
\cite{Cosy}.

\subsection{Spin structure of threshold amplitudes for  
$p+p\rightarrow Y+K+N$ processes}

The threshold region is again defined as $\ell_1=\ell_2=0$,
where $\ell_1$ is the orbital momentum of the $YN$-system and
$\ell_2$ is the orbital momentum of the K-meson relative to the CMS of the $YN$ system.

Since the P-parity of the kaon is negative (relative to the parity of the 
$N\Lambda$ system), the total angular momentum ${\cal J}$ and the parity P
of the produced $YNK$-system  at threshold are equal to: 
${\cal J}^{P}=0^- ~\mbox{ and }  1^-. $
From parity conservation it follows that the orbital momentum of the colliding 
protons, $L$, must be odd. Then the Pauli principle requires that the 
initial $pp$-system must be in a triplet state, $S_i=1$.

Taking into account the conservation of the total angular momentum, one finds
 that only two transitions are allowed:
$$
 S_i=1,~L =1 \rightarrow {\cal J}^{ P}=0^- , 
$$
$$
 S_i=1,~L =1 \rightarrow {\cal J}^{ P}=1^-.
$$

\begin{figure}
\hspace*{1true cm}
\mbox{\epsfxsize=8.cm\leavevmode \epsffile{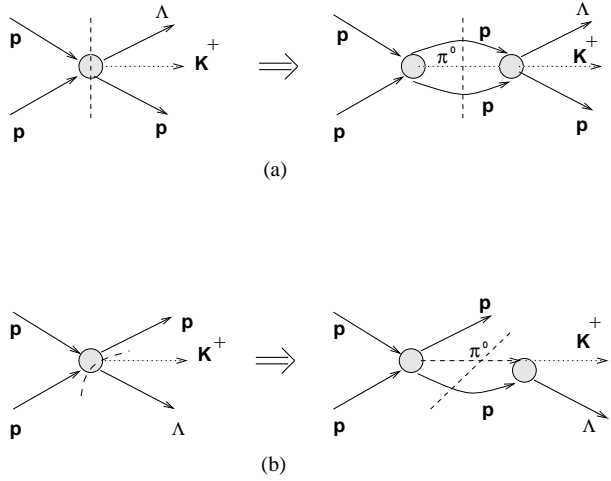}}
\caption{Diagrams contributing to the unitarity conditions for  
$p+p\rightarrow K+N+ p$ : {\it (a)} unitarity condition in s-channel of the 
process  
$p+p\rightarrow K+Y+ N$; {\it (b)} unitarity condition in $\Lambda 
K^+$-channel.}\label{Fig:str}
\end{figure}

Therefore the spin structure of the matrix element for any process 
$p+p\rightarrow Y+K+N$ 
can be written in the following form (in the $CMS$):
\begin{equation}
{\cal M}=~f_0({\chi}_4^{\dagger}~\sigma_y \tilde{\chi}_3^{\dagger})(\tilde 
{\chi}_2~\sigma_y \vec\sigma\cdot\vec k\chi_1)+
if_1({\chi}_4^{\dagger}~\sigma_a \sigma_y \tilde{\chi}_3^{\dagger})
(\tilde {\chi}_2~\sigma_y (\vec\sigma\times\vec k)_a\chi_1),\label{eq:str5} 
\end{equation}
where 
$\chi_1~\mbox {and } \chi_2 $ are the two-component spinors of the colliding 
protons, 
$\chi_3~\mbox {and } \chi_4$  are  the two-component spinors of the final 
nucleon and the
produced hyperon, and 
$f_0 ~(f_1)$ is the  $YN$ production amplitude in the singlet (triplet) 
state. In general, the amplitudes $f_0$ and $f_1$ are complex functions of 
three independent variables: the total energy $\sqrt{s}$ of the colliding 
particles and two energies of the produced particles. Such a complexity 
results from the unitarity conditions in both channels (Fig. \ref{Fig:str}).

The amplitudes $f_0$ and $f_1$ have to be calculated 
using a dynamical model \cite{An83,TdG85}, but polarization effects 
near threshold can be analyzed without knowing these amplitudes.

\subsection{Polarization phenomena in the reactions $p+p\rightarrow 
Y+K+N$}

From the spin structure of the matrix element (\ref{eq:str5}) we can derive rigorous results 
for the 
polarization observables, which are valid for any model of the processes 
$p+p\rightarrow Y+K+N$. 
\begin{itemize}
\item The polarization $\vec P_Y$ of the hyperons produced in collisions of 
unpolarized nucleons is zero
for any value of the amplitudes $f_0$ and $f_1$. This follows from the S-state 
nature of threshold $YKN$-production.

\item Due to the orthogonality of singlet and triplet states of the produced 
$YN$-system, the analyzing powers in $\vec p+p\rightarrow Y+K+N$ and $p+\vec 
p\rightarrow Y+K+N$, must also be zero.

\item The spin correlation coefficients ${\cal A}_1$ and ${\cal A}_1$ for collisions where both particles are
polarized in the initial state are different from zero:
\begin{equation}
{\cal A}_1 =\displaystyle \frac {|f_0|^2}{|f_0|^2+2|f_1|^2},~~
{\cal A}_2 =\displaystyle \frac {2(-|f_0|^2+|f_1|^2)}{|f_0|^2+2|f_1|^2}, 
\label{eq:A1A2}
\end{equation}
from which one deduces:
$$
{\cal A}_1\ge~0,
$$
$$
0(f_0=0)\le ~{\cal A}_1\le 1(f_1=0),
$$
$$
-1(f_1=0)\le ~{\cal A}_2\le 1(f_0=0),
$$
\begin{equation}
3 {\cal A}_1+{\cal A}_2=1,
\end{equation}
One notes that these coefficients are correlated, which results from the pure 
triplet nature of the initial $pp$-system.
Threshold amplitudes do not interfere in collisions of polarized protons.
\end{itemize}
The measurement of the differential cross section $({d\sigma}/{d\omega})_0$ 
(with 
unpolarized particles) and of one spin correlation coefficient only  
$({\cal A}_1$ or ${\cal A}_2)$ allows to determine the  moduli of both 
scalar amplitudes. 

The relative phase of the amplitudes $f_0$ and $f_1$ can be deduced through
a measurement of polarization transfer coefficients from one initial 
proton to the final hyperon. Starting from the P-invariance of the strong 
interaction, one can write the following general formula:
\begin{equation}
\vec P_Y=\rho_1\vec P+ \rho_2\vec k\vec P\cdot\vec k,\label{eq:strp}
\end{equation}
$$
\rho_1=\displaystyle-\frac {2{ \cal R}ef_0 f_1^*}{|f_0|^2+2|f_1|^2},~~~
\rho_2=2\displaystyle\frac {|f_1|^2+ { \cal R}ef_0f_1^*}{|f_0|^2+2|f_1|^2}.
$$
The complete experiment for any $p+p\rightarrow K+Y+N$ reaction at  
threshold must then include the following set of measurements:
\begin{itemize}
\item the differential cross section  ${\displaystyle \left({\frac 
{d\sigma}{d\omega}}\right )_0}$; 
\item a spin correlation coefficient ${\cal A}_1$ or ${\cal A}_2$ in the 
collision of 
polarized protons;
\item a polarization transfer coefficients from the initial proton to the 
produced 
hyperon, $K_y^{y'}$ or $K_x^{x'}$.
\end{itemize}

\subsection{Hyperon production in  $np-$ collisions}

In  the collision of non-identical particles, the isotopic invariance of strong 
interaction 
allows to apply the generalized Pauli principle. Therefore the following 
analysis is 
valid up to electromagnetic corrections and other isotopic 
invariance 
violation effects. The spin structure for $np$-collisions 
corresponding to a total isospin $I=1$,  is the same as for 
$pp-$collisions. 
This part of the amplitude is described in terms of the above mentioned 
amplitudes $f_0$ 
and $f_1$, after introducing appropriate Clebsch-Gordan coefficients. 

For $I(np)=0$, the generalized Pauli principle requires the initial $np$ state 
to be singlet. Therefore a single additional transition is allowed: 
$ S_i=0,~L =1 \rightarrow {\cal J}^{ P}=1^-,$
with the corresponding matrix element:
\begin{equation}
{\cal M}_0=~g_0({\chi}_4^{\dagger}~\vec\sigma\cdot\vec k \sigma_y 
\tilde{\chi}_3^{\dagger})
(\tilde {\chi}_2~\sigma_y \chi_1),\label{eq:mat} 
\end{equation}
where
$g_0$ is the S-wave amplitude.

Let us compare the reactions:
\[
p~+~p\rightarrow p~+~K^+~+~\Lambda,~~
n~+~p\rightarrow  p~+~K^0~+~\Lambda.
\]
From equations (\ref{eq:str5}) and (\ref{eq:mat}) one obtains:
$$
\displaystyle\frac {\sigma(pp\rightarrow pK^+\Lambda)}{\sigma(np\rightarrow 
pK^0\Lambda)}=4\frac {|f_0|^2+2|f_1|^2}{|f_0|^2 +2|f_1|^2+|g_0|^2}=\frac 
{4}{1+r}, 
$$
where $\sigma$ is the total cross section. The ratio 
$r=\displaystyle\frac {|g_0|^2}{|f_0|^2+2|f_1|^2}$
characterizes the relative strength of the $np$-interaction in the singlet and 
triplet states.

The SF's ${\cal A}_1^{(np)}$ and ${\cal A}_2^{(np)}$ for polarized 
$\vec n+ \vec p$ collisions are given by:
$$
\displaystyle {\cal A}_1^{(np)}=\frac {|f_0|^2-|g_0|^2}{|f_0|^2 
+2|f_1|^2+|g_0|^2},~~~{\cal 
A}_2^{(np)}=\frac {2(|f_0|^2+|f_1|^2)}{|f_0|^2 +2|f_1|^2+|g_0|^2}, 
$$
where 
$$3{\cal A}_1^{(np)}+{\cal A}_2^{(np)}-1=-\displaystyle \frac{4r}{1+r},$$
which holds for any value of the amplitudes $f_0$, $f_1$ and $g_0$.
By using two ratios, namely: $\displaystyle\frac {\sigma(pp\rightarrow 
pK^+\Lambda)}{\sigma(np\rightarrow pK^0\Lambda)}$(with unpolarized~ particles), 
and $\displaystyle\frac {|f_0|^2} {|f_1|^2}$ (from $\vec p+\vec p\rightarrow 
K^++\Lambda+p$), it is possible to predict ${\cal A}_1^{(np)}$ and 
${\cal A}_2^{(np)}$, for $\vec n~+~\vec p\rightarrow n~+~K^+~+~\Lambda $:
$${\cal A}_1^{(np)}=({\cal A}_1-r)/(1+r),~~{\cal A}_2^{(np)}={\cal A}_2/(1+r).$$
The general formula for the polarization transfer in $\vec n +p \rightarrow 
\vec Y+K+N$ is similar to equation (\ref{eq:strp}), but with different expressions for the 
SF's $\rho_1$ and $\rho_2$:
$$\rho_1=\displaystyle\frac{2{\cal R}e(f_0 - g_0)f_1^*}{|f_0|^2 
+2|f_1|^2+|g_0|^2}, $$ 
$$\rho_2=-2\frac {\left [|f_1|^2 +{\cal R}e f_0(f_1- g_0)^*- 
{\cal R}eg_0 f_1^* \right ]}{(|f_0|^2 +2|f_1|^2+|g_0|^2)}. $$
So, for S-wave production, each  $p+p\rightarrow Y+K+N$ process is described by 
two independent complex amplitudes which are functions of the energies of the 
colliding and produced particles.To reconstruct experimentally the complete 
spin structure of the amplitude it is necessary to measure at least three 
observables, namely the differential cross section $(d\sigma/d\omega)_0$ 
(with unpolarized particles), the spin correlation coefficients 
(i.e. the asymmetry induced by the collision of two polarized protons, 
$\vec p+ \vec p\rightarrow Y+K+N$), and the polarization transfer from the 
initial proton to the produced hyperon Y. But for a unique determination of 
the relative phase of the two complex amplitudes it is necessary to measure at 
least one T-odd polarization observable. The 
simplest one is a triple polarization correlation of baryons.

P-wave production results generally in non-zero T-odd polarization 
effects. They include one-spin polarization observables such as 
the polarization $\vec P_Y$ of the hyperons produced in the collision of 
unpolarized particles:  
$p+p\rightarrow K+\vec Y+N$ and the analyzing powers for $\vec p+p\rightarrow 
K+Y+N$ and 
$p+\vec p\rightarrow K+Y+N$. A non-zero value of these observables would be an 
evidence for a contribution of P- (or higher order-)waves to the production 
amplitudes.

\section{Processes $n+p\to d + \eta~(\pi^0 )$ and  $n+p\to d+V^0$ and 
test of isotopic invariance of strong interaction}

\subsection{The reactions $n+p\to d+\eta$ and $n+p\to d+\pi^0$.}
The process $n+p\to d+\eta$ is characterized by a relatively large cross section near threshold 
\cite{Pl90}, which favors the experimental study of the polarization 
observables. Following the analysis based on the isotopic invariance (and the 
validity of the generalized Pauli principle), the parity and angular momentum 
conservation, one can show that the $S-$wave $\eta$-production near threshold
for $n+p\rightarrow d+\eta$ is characterized by the single transition: 
$L=1,~ S_i(np)=0 ~~\rightarrow~~{\cal J}^{P}=1^-$, 
with a simple structure of the threshold matrix element:
$$
{\cal M}(np\rightarrow d\eta)=i g_{\eta}~\vec D^*\cdot \vec k
(\tilde {\chi}_2~\sigma_y \chi_1) ,
$$
where $\vec D$ is the spin wave function of the deuteron, $g_{\eta}$ is the
corresponding production amplitude. The initial $np-$system is in the singlet
state.  This might explain the large $n+p\rightarrow d+\eta$ cross-section. Let 
us mention that the equivalent process for
$\pi^0$-production, $n+p\rightarrow d+\pi^0$, is characterized by the
$np-$interaction in the triplet state with another matrix element:
$$
{\cal M}(np\rightarrow d\pi^0)= g_{\pi}~\tilde
{\chi}_2~\sigma_y ~\vec\sigma\times\vec k\cdot\vec D^*\chi_1 .
$$
The differential cross section of $\pi^0$ production near threshold being lower
than the cross section of $\eta-$production, one finds again a correlation
between the singlet or triplet nature of the colliding $np-$particles and
the probabilities of $\pi^0-$ and $\eta-$production. 

Moreover the
polarization observables are different for $\pi^0-$ and $\eta-$production:
\begin{itemize}
\item The dependence of the differential cross section on the polarizations
of the colliding nucleons is described by the following formulas:
\begin{eqnarray}
\displaystyle \frac {d\sigma}{d\omega}(\vec P_1, \vec P_2)&=&\left ({\frac
{d\sigma}{d\omega}}\right )_0(1-\vec P_1\cdot\vec P_2), ~~n+p\rightarrow
d+\eta ,\nonumber\\
\displaystyle \frac {d\sigma}{d\omega}(\vec P_1, \vec P_2)&=&\left ({\frac
{d\sigma}{d\omega}}\right )_0(1+ ~\vec k\cdot\vec P_1~\vec k\cdot\vec P_2),~~
n+p\rightarrow d+\pi^0,\nonumber
\end{eqnarray}
i.e. only the longitudinal components of $\vec P_1$ and $ \vec P_2$ can 
contribute to the polarized cross section in the reaction
$n+p\rightarrow d+\pi^0$.
\item The final deuterons are produced with nonzero tensor polarization even for 
collisions induced by unpolarized nucleons: 
longitudinal polarization ($P_{zz}=1$, where the $z-$axis is along the vector 
$\vec k $) in  
the process $n+p\rightarrow d+\eta$ and transversal polarization in the process  
$n+p\rightarrow d+\pi^0$. In a similar way it is possible to make predictions 
for other polarization observables in the processes $n+p\rightarrow 
d+\pi^0(\eta)$.
\item In principle, the experimental large value of the cross section for the process 
$np\rightarrow d \eta$ near threshold may be directly related to the singlet 
nature of the initial $np$ system.
\item The polarization phenomena for the threshold $\eta$-production 
in $NN$-collisions can be qualitatively predicted in a model independent 
form. Polarization phenomena are important to test the validity of $S+P$ 
approximation and to reconstruct the spin 
structure of the threshold amplitudes.
\end{itemize}
\subsection{The processes  $n+p\to d+V^0$ and $p+p\to d+\rho^+$}

The processes $p+p\rightarrow d+\rho^+$, and $ n+p\rightarrow d+V^0$, with 
$V=\rho,~\omega~or~\phi$, are the simplest two-particle reactions  of vector 
meson production in nucleon-nucleon collisions. Near threshold large momentum 
transfers are associated to these processes, therefore the behavior of the 
deuteron wave function at small distances is important for its description.
As a result the spin structure of the deuteron wave function can be 
investigated, in principle, at high energies through the study of the process 
$p+p\rightarrow d+\rho^+$, where the deuteron is produced at zero degrees. 
Similarly to the backward elastic scattering $d+p\rightarrow p+d$ 
\cite{Pu95,Pu94}, it is possible to suggest polarization experiments with 
polarized proton beam and target and with the measure of the vector and tensor 
polarizations of the outgoing deuterons. It is possible to measure elements of 
the density matrix of the vector mesons, also.

The two-particle nature of the $N+N\rightarrow d+V$ processes simplifies the 
experimental detection and the theoretical interpretation. 

\subsection{Threshold amplitudes for $n+p\rightarrow d+ \omega (\phi)$.}

The S-wave production of the $V^0$-meson induces three possible values of total 
angular momentum $J$ and $P$-parity in the channel $n+p\rightarrow 
d+\omega(\phi)$: ${\cal J}^{P}=0^-,~1^-\mbox{~and } 2^-.$

Due to the  $P$-parity conservation, the orbital angular momentum $L$ of the 
colliding $n$ and $p$ must be odd. According to the generalized Pauli principle 
for the $np$-system (which is correct at the level of the isotopic invariance of 
strong interactions) it is easy to show that the $n+p$-system must be in the 
singlet spin state only, so only one  possible value for $L$ is allowed, namely 
$L=1$. 

So at threshold only one transition is possible: 
$S_i(np)=0,~L=1~\rightarrow~{\cal J}^{ P}=1^-.$ with 
matrix element:
\begin{equation}
{\cal M}= g(\tilde {\chi}_2~\sigma_y \chi_1)
\vec k \times\vec D^*\cdot\vec V^*\label{eq:dv1} ,
\end{equation}
where $\chi_1~(\chi_2)$ is the 2-component spinor of neutron (proton), 
$\vec D (\vec V)$ is the spin wave function of $d~(V^0)$, $\vec D$ is an axial 
vector (as the P-parity of the deuteron is positive), $\vec V$ is a polar 
vector; $\vec k$ is the unit vector along the initial momentum and 
$g$ is a partial amplitude corresponding to the S-production of the $V^0$-meson.

In order to predict the $s$-dependence of $g$, a definite model for the the 
processes $n+p\rightarrow d+V^0$ is necessary, but the analysis of the 
polarization effects can be easily done without any model for $g$. This is a 
consequence of the definite spin structure of the matrix element of this 
reaction, with a single amplitude $g$. This amplitude is different for different 
processes, but the polarization phenomena are universal for any process of 
$V^0$-meson production. Such universality applies also to the processes of 
$\omega$- and $\phi$-radial excitation : $\omega',~\omega'',~\phi'~\phi''..$.
Moreover all polarization observables do not depend on the energy of the 
colliding particles (in the threshold region).

The presence of a single amplitude in (\ref{eq:dv1}) gives very definite predictions for 
numerical values of polarization observables: all non-zero polarization 
observables have their maximum values. Therefore the polarization effects near 
threshold for $n+p\rightarrow d+\omega(\phi)$ do not contain any special 
information on the dynamics of the reaction : the measurement of the 
differential cross section with unpolarized particles represents the complete 
experiment.

\subsection{Polarization effects in  $n+p\rightarrow d+\omega (\phi)$.} 

We discuss here the properties of polarization observables in the processes  
$n+p\rightarrow d+V^0,~V^0=\omega$ or $\phi$ starting from the matrix element 
(\ref{eq:dv1}).
\begin{itemize}
\item The analyzing powers in $\vec n+p\rightarrow d+V^0$ and $n+\vec 
p\rightarrow d+V^0$ vanish.
\item The vector polarization of deuterons produced in unpolarized particle 
collision must be zero, but the tensor polarization is different from zero:
$$T_{20}=1/3 \label{vm2},$$
which is correct for any reaction $n+p\rightarrow d+V^0$ and does not depend on 
the energy of the colliding particles.
\item The dependence of the cross section from the polarizations $\vec P_1$ and 
$\vec P_2$ of the initial nucleons in $\vec n+\vec p\rightarrow d+V^0$ is 
written as:
$$
\frac{d\sigma}{d\Omega} (\vec P_1, \vec 
P_2)={(\frac{d\sigma}{d\Omega})}_0(1-\vec P_1 \cdot\vec P_2).
$$
\item 
$V^0-$mesons, produced in collisions of unpolarized nucleons, are polarized with 
the following nonzero elements of the density matrix (in cartesian coordinates):
$\rho_{xx}=\rho_{yy}=\displaystyle \frac{1}{2}$, if the $z-$axis is along $\vec k$.
\item
the dependence of the $V^0-$meson density matrix from the vector polarization 
$\vec P$ of any initial nucleon can be parametrized at the reaction threshold by 
the following general form:
\begin{equation}
\begin{array}{rc}
& \rho_{ab}=i \epsilon _{abc} P_c \rho_1+i\epsilon _{abc}k_c\vec P\cdot\vec k 
\rho_2 +
( k_a \epsilon _{bcd}  k_c P_d  + k_b\epsilon _{acd} k_c  P_d  )\rho_3,
\end{array}
\end{equation}

where $\rho_i$, i=1,2,3 are the corresponding {\it Structure Functions} (SF) 
,depending only from the energy of the colliding particles.
\end{itemize}

The SF's $\rho_1$ and $\rho_2$ are responsible for the T-even polarization 
characteristics of the $V^0$ mesons, and the SF $\rho_3$ for the T-odd ones. But
the antisymmetric part of $\rho_{ab}$, which characterizes the vector polarization 
of the produced $V^0-$ mesons, cannot be measured through the most probable 
decays of $V^0$mesons (see Section 1).

All the previous statements about polarization phenomena in $n+p\rightarrow 
d+\omega(\phi)$ have a general character and are not related to any hypothesis 
about $s\overline{s}$-component in the nucleon.
It is possible to deduce the following consequences of this hypothesis to the 
threshold $V^0$-meson production:
\begin{itemize}
\item The $\phi$-production is suppressed, due to the fact that the S-wave 
production of the $\phi$-meson is possible here from a singlet state only. This 
means that the $n+p\rightarrow d+\phi$ reaction will not show a large violation 
of the OZI-rule:
$$
{\cal R}=\frac{\sigma(n+p \rightarrow d+\phi) } 
{\sigma(n+p \rightarrow d+\omega) } \simeq\frac{\sigma_p(\overline{p}+p 
\rightarrow \phi+\pi) }   {\sigma_p(\overline{p}+p \rightarrow \omega+\pi) }
\simeq10^{-3}.
$$
where $\sigma_p(\overline{p}+p \rightarrow V^0+\pi^o)$ is the cross section of 
$\overline{p}+p$-annihilation from the P-state, i.e. from the singlet state.

\item The fact that the S-wave production is negligible may enhance the P-wave 
contributions near threshold, which are triplet ones for $n+p \rightarrow 
d+\omega(\phi) $.
Such effect can be experimentally evidenced by different methods: observing the 
angular dependence of the differential cross section, or measuring a nonzero 
vector polarization of deuterons with unpolarized particles, or nonzero values 
of analyzing powers for $\vec n +p\rightarrow d+V^0$ and $n+\vec p\rightarrow 
d+V^0$.
\item Such effects must appear in the process $ n +p\rightarrow d+\phi$ earlier 
than in $ n +p\rightarrow d+\omega$. As the thresholds of these processes are 
different, it is necessary to compare the production at the same value of the 
invariant energy $Q$, $Q=\sqrt{s}-m_d-m_V$.
\end{itemize}
\subsection{The process $N +N\rightarrow d+\rho$.}

The total isotopic spin of the entrance channel is equal to 1. Therefore the 
generalized Pauli principle (for  $ n +p\rightarrow d+\rho^o$) or the usual 
Pauli principle (for  $ p+p\rightarrow d+\rho^+$) allows triplet initial states 
in case of S-wave  $\rho$-meson production. As a result we have the following 
transitions:
$$
\begin{array}{rcl}
 S_i=1,~L =1 &\rightarrow {\cal J}^{ P}=0^- &,~S_f=0, \\
 S_i=1,~L =1 &\rightarrow {\cal J}^{ P}=1^- &,~S_f=1, \\
 S_i=1,~L =1 &\rightarrow {\cal J}^{ P}=2^- &,~S_f=2, 
\end{array}
$$
where $S_i(S_f)$ is the total spin of the initial (final) particles.

The corresponding expressions for the spin structures of these transitions are 
the following:
\begin{eqnarray*}
&f_0~:~\tilde {\chi}_2~\sigma_y \vec \sigma \cdot\vec k \chi_1\vec D^*\cdot\vec 
V^*, \\
&f_1~:~\tilde {\chi}_2~\sigma_y \vec \sigma \times\vec k\cdot \vec D^*\times\vec 
V^*\chi_1, \\
&f_2~:~\tilde {\chi}_2~\sigma_y 
(\sigma_ik_j+\sigma_jk_i-\frac{2}{3}\delta_{ij}\vec\sigma\cdot\vec k)
\chi_1(D_i^*V_j^*+D_j^*V_i^*-\frac{2}{3}\delta_{ij}\vec D^*\cdot\vec 
V^*),
\end{eqnarray*}
where $f_0,~f_1,~f_2~$ are the partial amplitudes corresponding to the $J=0,1$ 
and 2 transitions, respectively.
For the calculation of polarization effects we will use an equivalent but more 
simplified form of the matrix element:
$$
{\cal M}= \tilde{\chi}_2 \sigma_y~[g_0\vec \sigma \cdot\vec k \vec D^*\cdot\vec 
V^*
+ g_1 \vec \sigma \cdot \vec D^* \vec k \cdot\vec V^*
+ g_2 \vec \sigma \cdot \vec V^* \vec k \cdot\vec D^*]\chi_1, 
$$
where $g_i$ are the following combinations of $f_i$:
$$g_0=f_0-\frac{4}{3}f_2,~~g_1=f_1+2f_2,~~g_2=-f_1+2f_2.\label{li4}$$

More complicated spin structure of ${\cal M}$ results in changing polarization 
effects in the process $N+N\rightarrow d+\rho$.
Of course, all one spin T-odd effects in  $N+N\rightarrow d+\rho$ must be zero 
for any values of amplitudes $g_i$. This is an usual property of the S-wave 
production. The dependence of the differential cross section on the 
polarizations $\vec{P_1}$ and $\vec{P_2}$ of the colliding nucleons has the
standard form, Eq. (\ref{eq:poli}), where the coefficients 
${\cal A}_1$ and ${\cal A}_2$ are defined by:
\begin{eqnarray*}
&{\cal A}_1=-\displaystyle{ \frac 
{2|g_0|^2+|g_0+g_1+g_2|^2}{2(|g_0|^2+|g_1|^2+|g_2|^2)+|g_0+g_1+g_2|^2}},\\
&{\cal A}_2=2\displaystyle{ 
\frac{2|g_0|^2-|g_1|^2-|g_2|^2+|g_0+g_1+g_2|^2}{2(|g_0|^2+|g_1|^2+|g_2|^2)+|g_0+
g_1+g_2|^2}}.
\end{eqnarray*}
However the measurement of the spin correlation coefficients:
$$C_{xx}=C_{yy}={\cal A}_1,~~~C_{zz}={\cal A}_1+{\cal A}_2,$$
does not represent a complete experiment for this reaction. Additional 
observables are necessary. One of them is $T_{20}$, the tensor deuteron 
polarization: 
$$
T_{20}=-\frac{F_2}{3F_1+F_2}.
$$
with 
$$
F_1=|g_0|^2+|g_1|^2,~~
F_2=-|g_0|^2-|g_1|^2+2|g_2|^2+|g_0+g_1+g_2|^2.
$$
The unpolarized cross section is related to 
$F_1$ and $F_2$ by $(d\sigma/d\Omega)_0\simeq  3 F_1+F_2$.

The general expression for the density matrix of $V^0$, produced in the 
collision of unpolarized particles, can be written as:
$$
\rho_{ab}=\delta_{ab}q_1+k_ak_bq_2,~~3q_1+q_2=1,
$$
with
$$
q_1=\displaystyle{ \frac 
{|g_0|^2+|g_2|^2}{2(|g_0|^2+|g_1|^2+|g_2|^2)+|g_0+g_1+g_2|^2}}.
$$
The presence of 3 complex amplitudes near threshold of any process 
$N+N\rightarrow d+\rho$ results in T-odd correlations of the $\rho$-meson 
polarization properties with neutron polarization
$\vec n+p \to d+\rho^0$, the nonzero SF $\rho_3$:
$$\rho_3\simeq {\cal I}m(g_0g_1^*+g_0g_2^*+g_1g_2^*).$$
So, the one and two-spin polarization observables near threshold of the 
$N+N\rightarrow d+\rho$ give enough independent combinations of $g_i$ amplitudes
to realize the complete experiment.

\subsection{Test of isotopic invariance through polarization observables in
$n+p\rightarrow d+V$-processes}

The process $n+p\rightarrow d+V$  is one  of those binary reactions, 
where the colliding particles belong to the same isotopic multiplet and only one 
value of the total isotopic spin is allowed in the reaction channel. For such 
reactions the isotopic invariance induces definite properties of symmetry of all 
the polarization observables relative to the exchange $\cos\theta \rightarrow 
-\cos\theta$ \cite{Bi67}, where $\theta$ is the $V^0$-meson production angle in 
the CMS. This means that polarization observables must be odd or even 
functions of $\cos\theta$. 
The theorem of Barshay-Temmer \cite{Ba64} about the symmetry of the differential cross-sections relative to $\theta=90^0$ is the 
simplest example of the above mentioned result. An illustration
in nuclear physics is given by the reaction $^3He(^3H,d)\alpha$ \cite{Ha77}. The isotopic invariance allows to interchange the  $^3He$ and $^3H$, leaving the deuteron and the $\alpha$-particle unaffected. As a result the charge symmetry implies a symmetry about $\theta=90^0$ of the tensor analyzing powers $A_{yy}$ and $A_{xx}$ and the antisymmetry of the tensor $A_{xz}$ and the vector $A_y$ 
analyzing powers. Such behavior of the polarization observables is an 
interesting example of the connection \cite{Ne94} of the polarization effects 
with the properties of internal symmetries which do not affect the magnitude of 
a spin vector of any interacting particle. The isotopic invariance of the strong
interaction belongs to such symmetries.

The experimental study of polarization effects in the reaction $n+p\rightarrow 
d+V$ could be important for the search of violations ot this invariance.
The interpretation of these effects has essentially changed.
If earlier it was assumed that only the electromagnetic corrections are 
responsible for the violation of the isotopic invariance, in the framework of 
QCD the main mechanism is connected with the difference of $u-$ and $d-$ quark 
masses, $\Delta=m_d-m_u\neq 0$. Namely the case of $\Delta\geq 0$ explains the 
signs  of the mass difference of particles for all the known isotopic multiplets of hadrons and nuclei. The scale of such effects of the isotopic invariance 
violation is characterized by the ratios:
$$\displaystyle\frac{m_d-m_u}{\Lambda_{QCD}}\simeq 
\displaystyle\frac{m_d-m_u}{4\pi f_{\pi}}\simeq 
\displaystyle\frac{m_d-m_u}{300 \mbox{~MeV}},$$
where $f_{\pi}$ is a constant of $\pi\rightarrow\mu \nu$ decay. The typical 
scale of $\simeq 300$ MeV can be thought as arising from a constituent quark 
mass, bag model energy or quark condensate. Thus the effects of $\Delta\neq0$ 
are small, compared to the electromagnetic effects. Therefore the charge 
symmetry is not perfect and gives a unique opportunity to find the mass 
difference of $u-$ and $d-$quarks. The violation of the charge independence of 
the strong interaction is connected with the explanation of the masses of the 
fundamental leptons and quarks which is one of the most important problems of 
the Standard Model.

The most evident observation of charge independence breaking effects occurs in 
the $\rho^0 \omega-$ mixing through a nonzero value of the matrix element 
$<\rho^0|H|\omega>$, where H is the QCD Hamiltonian. 
The difference $m_d-m_u\neq 0$ is the main contribution to $<\rho^0|H|\omega>$. The effects of this matrix 
element are observed  in the process $e^++e^-\rightarrow\pi^+ +\pi^-$ 
through the specific behavior of the pion electromagnetic form factor with the 
result:
$$<\rho^0|H|\omega>\simeq-4500~\mbox{MeV}^2.$$
It is natural to use the exchange of a mixed $\rho\omega-$meson as mechanism for the charge symmetry breaking nucleon-nucleon forces. But there is a problem of a significant extrapolation of this matrix element which is determined at 
$q^2=m_{\rho}^2$ ($q$ is the momentum transfer) to the region of NN-forces, 
where the relevant $q^2(\leq 0)$ are space-like. Some models 
\cite{Go92,Pi93,Kr93,Ha94,Mi94} predict a 
strong $q^2$-dependence of this matrix element, the situation is not clear now 
and further experiments are needed.

The study of polarization effects in such processes as $n+p\rightarrow d+V^0$ 
could be important in the search of isotopic invariance violations.
It is necessary to mention also that the large momentum transfers realized at 
the threshold of  $n+p\rightarrow d+V^0$ could be interesting in connection with the possible dependence of the difference $\Delta=m_d-m_u$ on the nuclear 
density and on the momentum transfer.
The $\rho\omega$-mixing can be studied for different regions of momentum 
transfer: for the space-like momentum through the NN-potentials and for 
$q^2=m_{\rho}^2$ through the link between the $n+p\rightarrow d+\omega$ and 
$n+p\rightarrow d+\rho^0$ reactions.

The relation between the polarization effects in $n+p\rightarrow d+V$ 
reactions and the  symmetry properties of the strong interaction as the charge 
independence and in particular with symmetry violations looks as a
 very attractive and unusual application of polarization physics.

Let us summarize the main results obtained in this section:
\begin{itemize}
\item The matrix element of the process $n+p\rightarrow d+\omega(\phi)$ is 
defined at the threshold, by a single amplitude which correspond to the singlet 
interaction of the colliding nucleons.
\item The dependence of the differential cross section for $\vec n+\vec 
p\rightarrow d+\omega(\phi)$ on the polarizations $\vec P_1$ and $\vec P_2$ 
of the initial nucleons has the following form:
$$
\displaystyle\frac{d\sigma}{d\Omega} \left (\vec P_1, \vec 
P_2\right )={\left(\displaystyle\frac{d\sigma}{d\Omega}\right )}_0(1-\vec P_1 ~\vec P_2).
$$
\item The produced particles in $n+p\rightarrow d+\omega(\phi)$ must be 
polarized (even in the collision of unpolarized nucleons): the deuteron must 
have a tensor polarization with $T_{20}=1/3$, the nonzero elements of 
the $V^0$-meson density matrix are $\rho_{xx}=\rho_{yy}=1/2$.
These predictions are universal as they are independent on the type of 
$V^0$-meson and on the energy of the colliding particles in the near-threshold 
region.

\item As the processes $n+p\rightarrow d+V^0$ at the threshold are induced by 
the singlet np-interaction, there is no large violation of the OZI-rule:
$$
\frac{\sigma(n+p \rightarrow d+\phi) } 
{\sigma(n+p \rightarrow d+\omega) } 
\simeq10^{-3}.
$$
\item The matrix element of $\rho$ production, $n+p\rightarrow d+\rho^0$
and $p+p\rightarrow d+\rho^+$, is described by three independent threshold 
amplitudes.
\end{itemize}

\chapter{Application to Nuclear Interaction}

\section{Processes $d+^3\!He\to ^4\!He+p$, $d+d\to ^3\!He+ n$
and thermonuclear fusion}

Nuclear fusion reactions, like  $d +d \to n 
+{^3\!He}$, or $d +{^3\!H} \to n +{^4\!He}$, are characterized by a 
large dependence on the spins of the colliding particles.  
It has been suggested \cite{Ku82} to use this property in magnetic fusion 
reactors with polarized nuclear fuel. A magnetic field of about 1 kG can keep 
the necessary direction of the polarization of the interacting nuclei, during a 
time which is longer in comparison with the reaction time.
Different technical solutions might be used: injection of polarized frozen 
pellets, or polarized targets for inertial fusion.

The strong dependence of the fusion reaction rates on the polarization states
results in an increasing or a decreasing of the cross section (with respect to 
the unpolarized case), depending on the colliding nuclei polarization 
directions. These characteristics can be used to optimize a fusion reactor in 
different ways:
\begin{itemize}
\item
The possible enhancement of the fusion rates for $\vec d+\vec{^3\!He}$ and the 
suppression of $\vec d+\vec d$-collisions would make this fuel competitive with 
$d+{^3\!H}$, as it would, in particular,  result in a {\it clean} reactor.
\item The strong anistropy of the neutron angular dependence in $\vec 
d+\vec{^3\!H}$-collision helps in optimizing the reactor shielding and the 
blanket 
design.
\item $\vec d+\vec {^3\!H}$-collisions can be source of intensive monochromatic 
polarized neutrons, with the choice of the polarization direction.
\end{itemize}

A precise knowledge of the spin structure of the threshold matrix elements for 
the processes induced by: $ d+ {^3\!He}$, $d+{^3\!H}$, ${^3\!He}+ {^3\!He}$, 
${^3\!H}+ ^3\!He$-collisions is required. At energies up to 10 keV, which are 
typical for fusion 
reactors, the S-state interaction of the colliding particles dominates and the   
general analysis of polarization phenomena is essentially simplified.

Our aim is to analyze here in the most general and complete form the reactions 
relevant to magnetic fusion reactors, with polarized fuel. 

The reaction $d +{^3\!He} \to p +{^4\!He}$ was considered in detail in
\cite{Rek98}. Following that methodology, we will give here the general 
parametrization of the threshold amplitudes for $d +{^3\!He}$, and 
$ d+d$-collisions, with special 
attention to the angular distribution of the reaction products for different 
possible polarization states of the colliding particles, without any particular 
assumption about the reaction mechanism.

In a fusion reactor the reaction rates and the 
angular distributions depend on  
the direction of the magnetic field. We use in this analysis a particular set of 
helicity amplitudes, with quantization axis along the direction of the magnetic 
field. We derive the angular dependence of the differential cross 
sections for different polarization states of the colliding particles, and the 
angular dependence of the polarization of the produced neutrons (protons).

\section{The complete experiment for the reaction 
$d +^3\!H (^3\!He) \to n (p) +^4\!He$ }
\subsection{Introductory remarks}

The reaction $d +^3\!H \to n +^4\!He$ in the near threshold region is 
very interesting for the production of thermonuclear energy and plays an 
important role in primordial nucleosynthesis. The  lowest 
$\displaystyle\frac{3}{2}^+$ level of ~$^5\!He$ 
has excitation energy $E_x=16.75$ MeV (only 50 keV above $d +^3\!H$-threshold) 
and has a width of 76 keV. 

The microscopic explanation of the nature and the properties of this resonance 
is very complicated and still under debate in the physics of light nuclei.
The interpretation \cite{Br87} of this resonance as a shadow pole \cite{Ed64} 
introduces a new concept in nuclear physics, after atomic and particle physics. 
The possibility that the corresponding shadow poles for the 
two charge symmetric systems $d +^3\!He$ and $d +{^3\!H}$ (or 
$p +{^4\!He}$ and $n +{^4\!He}$) occupy different Riemann sheets, due to the 
difference in electric charges of the participating particles, can not be 
presently ruled out. Such phenomena can be considered as a new mechanism of 
violation of isotopic invariance of the strong interaction \cite{Cs93}.

Due to the close connection of the three processes $ d +{^3\!He} \to d 
+{^3\!He},~n +{^4\!He} \to n +{^4\!He}$ and $d +{^3\!H} \to n 
+{^4\!He}$ through the unitarity condition, the partial wave analysis 
\cite{Ho66,Je80} can not be performed independently for each reaction. The 
corresponding amplitudes are complex functions of the excitation energy. 
The multilevel ${\cal R}-$matrix approach allows to parametrize this dependence 
in terms of few parameters as shift, penetration factors and hard-sphere phase 
shift 
\cite{La58}. All characteristics of the ${\cal 
J}^P=\displaystyle\frac{3}{2}^+$-resonance, like the position, the width and 
particularly the Rieman sheet, can be found using an $S-$matrix approach
\cite{Br87,Pe89,Cs97,Ba97}.

The polarization phenomena are very important in the near threshold region, even 
for the S-state interaction. In this respect the reaction $d +{^3\!H} 
\to n +{^4\!He}$ plays a 
special role, because the presence of a D-wave in the final state results in 
nonzero one-spin polarization observables, such as, for example, the tensor 
analyzing power. In order to fully 
determine the two possible threshold (complex) amplitudes, two-spin polarization 
observables have to be measured, for example in collisions of polarized deuteron 
with polarized ${^3\!He}$-target. Here we will generalize our previous analysis 
\cite{Rek98}, taking into account the presence of  a magnetic field, which is 
necessary in order to conserve the 
polarization of the fuel constituents in a magnetic fusion reactor \cite{Ku82}.

For thermal colliding energies the 
analysis of polarization 
phenomena for the reaction $d +{^3\!H} \to n +{^4\!He}$ can be carried 
out in a general form.
In the framework of a formalism, based on the polarized structure functions, we 
will point out the observables which have to be measured in order to 
have a full reconstruction of the spin structure of the threshold amplitudes. 
Data on cross section and tensor 
analyzing power exist, at threshold \cite{Dr80} (for a review see \cite{Ti98}).  
Among the two-spin observables, the measurement of a spin correlation 
coefficient, together with the cross section and the tensor analyzing power, 
allows to realize the complete experiment.

\subsection{Spin structure of the matrix element}
Let us first establish 
the spin structure of the matrix element. From the P-invariance of the strong 
interaction and the conservation of the total angular momentum, two partial 
transitions, for  $d +{^3\!He} \to p +{^4\!He}$ (as well as for $d 
+{^3\!H} \to n +{^4\!He}$) are allowed:
\begin{equation}
S_i=\displaystyle\frac{1}{2}~\to {\mathcal 
J}^{P}=\displaystyle\frac{1}{2}^+~\to {\ell}_f=0,~
S_i=\displaystyle\frac{3}{2}~\to {\mathcal  
J}^{P}=\displaystyle\frac{3}{2}^+~\to {\ell}_f=2,
\end{equation}
where $S_i$ is the total spin of the $ d+^3\!He$-system and ${\ell}_f$ is the 
orbital angular momentum of the final proton. The spin structure of the 
threshold matrix element can be parametrized in the form:
\begin{equation}
\begin{array}{ll}
&{\mathcal M}={\chi}_2^{\dagger}{\mathcal F}_{th} {\chi}_1,\nonumber\\
&{\mathcal F}_{th} =g_s\vec \sigma\cdot \vec D+ g_d (3\vec k\cdot\vec D~
\vec \sigma\cdot \vec k-\vec \sigma\cdot \vec D),
\end{array}
\end{equation}
where $\chi_1$ and $\chi_2$ are the two component spinors of the initial 
$^3\!He$ and final $p$, $\vec D$ is the 3-vector of the deuteron polarization 
(more exactly, $\vec D$ is the axial vector due to the positive parity of the 
deuteron), $\vec k$ is the unit vector along the 3-momenta of the proton (in 
the CMS of the considered reaction). The amplitudes of 
the $S-$ and $D-$ production of the final particles, $g_s$ and $g_d$, are complex functions of the excitation energy. Note that, in the general case, the spin structure of the matrix element, for the considered processes, contains six different contributions and the corresponding amplitudes are functions of two variables. 

The general parametrization of the differential cross 
section in terms of the 
polarizations of the colliding particles (in S-state) is given by:
\begin{eqnarray}
&\displaystyle\frac{d\sigma}{d\Omega}(\vec d+\vec {^3\!He} )=
\left (\frac {d\sigma}{d\Omega}\right )_0[ &1+ {\mathcal 
A}_1(Q_{ab}k_ak_b)+
{\mathcal A}_2\vec S\cdot\vec P\nonumber\\
&&+{\mathcal A}_3\vec k\cdot\vec P~\vec k\cdot\vec S+
{\mathcal A}_4\vec k\cdot\vec P\times\vec Q ],
~Q_a=Q_{ab}k_b,
\label{eq:fus3}
\end{eqnarray}
where $(d\sigma/d\Omega)_0$ is the differential cross section with unpolarized 
particles, $\vec P$ is the axial vector of the target ($^3\!He$) polarization, 
$\vec S$ and 
$Q_{ab}$ are the vector and tensor deuteron polarizations. The density matrix of 
the deuteron can be written as:
\begin{equation}
\overline{D_aD_b^*}=\displaystyle\frac{1}{3}(\delta_{ab}-\frac{3}{2}i\epsilon_{a
bc}S_c-Q_{ab}),~~Q_{aa}=0,~~Q_{ab}=Q_{ba}.
\label{eq:fus4}
\end{equation}

After summing over the final proton polarizations one can find the following 
expressions:
\begin{equation}
\begin{array}{rlrl}
{\mathcal A}_1\displaystyle\left (\frac {d\sigma}{d\Omega}\right 
)_0=&-2{\mathcal 
R}e~g_sg_d^*-|g_d|^2,&
{\mathcal A}_2\displaystyle\left (\frac {d\sigma}{d\Omega}\right 
)_0=&-|g_s|^2-{\mathcal 
R}e~g_sg_d^*+2|g_d|^2,\nonumber\\
{\mathcal A}_3\displaystyle\left (\frac {d\sigma}{d\Omega}\right )_0=&3{\mathcal 
R}e~g_sg_d^*-3|g_d|^2,&
{\mathcal A}_4\displaystyle\left (\frac {d\sigma}{d\Omega}\right 
)_0=&-2{\mathcal 
I}m~g_sg_d^*.
\end{array}
\label{eq:fus5}
\end{equation}
The coefficients ${\mathcal A}_i$ are related by the following linear 
relation: ${\mathcal A}_1+{\mathcal A}_2+{\mathcal A}_3=-1$ for any choice of 
amplitudes $g_s$ and $g_d$.
The integration of the differential cross section over the $\vec k$-directions 
gives:
$$\sigma(\vec d+\vec {^3\!H} )=\sigma_0(1+{\mathcal A}\vec S\cdot\vec 
P),~~{\mathcal 
A}={\mathcal 
A}_2+\displaystyle
\frac{1}{3}{\mathcal 
A}_3=\displaystyle\frac{-|g_s|^2+|g_d|^2}{|g_s|^2+2|g_d|^2},$$
and it is independent from the tensor deuteron polarization.

The presence of S-wave contribution (the amplitude $g_s$), decreases the value 
of the integral coefficient ${\mathcal A}$ whereas, in the fusion resonance 
region, 
where the D-wave dominates, the maximum value, 
${\mathcal A}=1/2$, is reached. In the 
complete experiment (which gives $|g_s|^2$, $|g_d|^2$ and 
$Re~g_sg_d^*$), the amplitudes $|g_s|$ and $|g_d|$ can be found in a model 
independent way, with the help of the following formulas: 
\[
9{|g_s|^2}=\left (5+2{\mathcal A}_1-4{\mathcal A}_2\right )\left 
(\displaystyle\frac 
{d\sigma}{d\Omega}\right )_0,
\]
\begin{equation}
9{|g_d|^2}=\left (2-{\mathcal A}_1+2{\mathcal A}_2\right )\left 
(\displaystyle\frac 
{d\sigma}{d\Omega}\right )_0,
\end{equation}
\[
-9{\mathcal R}e{g_s}{g_d^*}=\left (1+4{\mathcal A}_1+{\mathcal A}_2\right 
)\displaystyle\left (\frac {d\sigma}{d\Omega}\right )_0.
\]
One can see that the {\it 
integral} 
coefficient ${\mathcal A}$ can be determined from  polarized 
nuclei collisions by measuring:
\begin{itemize}
\item the tensor analyzing power ${\mathcal A}_1$ in  $\vec d +{^3\!He} 
\to p +{^4\!He}$,
\item the spin correlation coefficient $C_{xx}= C_{yy}={\mathcal A}_2$ (if the 
$z-$axis is 
along $\vec k$-direction.)
\end{itemize}

Let us study now the polarization properties of the outgoing nucleons.
We will show that it can be predicted only from the tensor analyzing power, 
${\mathcal A}_1$. The polarization 
$\vec P_f$ of the produced nucleon depends on the polarization $\vec P$ of the 
initial ${^3\!He}$ (or ${^3\!H}$) as follows:
$\vec P_f=p_1\vec P+p_2\vec k~\vec k\cdot\vec P$, where the real coefficients 
$p_i,~i=1,2$, characterize the spin transfer coefficients (from the initial 
${^3\!He}$ or ${^3\!H}$ to the final nucleon): $K_x^{x'}=p_1+p_2\cos^2\theta 
,~K_x^{z'}=p_2~\sin\theta ~\cos\theta, $ where 
$\theta$ is the angle between $\vec k$ and $\vec P$.
Averaging over the polarizations of the initial deuteron, we can find:
$$
p_1\left (\frac {d\sigma}{d\Omega}\right )_0=-\displaystyle\frac{1}{3}\left (
|g_s|^2+4 Re~ g_sg_d^*+4|g_d|^2\right ),$$
$$
p_2\displaystyle\frac{d\sigma}{d\Omega}_0=4 Re~ g_sg_d^*+2|g_d|^2,
$$
$$3p_1=-1+2{\mathcal A}_1,~p_2=-2{\mathcal A}_1,~~3p_1+p_2=-1.$$
This analysis  holds in the presence of S-state only, in the entrance 
channel. The validity of this assumption can be experimentally verified with the 
measurement of T-odd one-spin
polarization observables, as the analyzing powers in $\vec d +{^3\!He} 
\to p +{^4\!He}$ induced by vector deuteron polarization or $d +\vec 
{^3\!He} \to p +{^4\!He}$. These observables are very sensitive to the 
presence of even a small P-wave contribution, due to its interference with the 
main S-wave amplitude. 

\subsection{Helicity amplitudes}
We calculate here the helicity amplitudes $ 
F_{\lambda_1\lambda_2,\lambda_3}$, with ${\lambda_1}={\lambda_d}$, 
${\lambda_2}={\lambda_{^3He}}$, ${\lambda_3}={\lambda_p}$ (or ${\lambda_n}$), in 
terms of the partial amplitudes $g_s$ and $g_d$. This formalism is very well 
adapted for the analysis of angular distributions of the reaction products, in 
conditions of fusion reactors (with polarized fuel) and to the description of 
polarization phenomena. The direction of magnetic field $\vec B$ can be chosen 
as 
the most preferable quantization axis ($z-$axis). The formalism of the helicity 
amplitudes allows to study the angular dependence of the polarization 
observables, relative to $\vec B$. For example, the polarization properties of 
the neutron in 
$\vec d +\vec {^3\!H} \to n +{^4\!He}$ can be easily calculated in terms 
of these 
amplitudes.

The  peculiar strong angular dependence of all observables 
is due to the presence (in conditions of fusion polarized reactor) of two 
independent physical directions, $\vec k$ and $\vec B$. So even for the S-state 
interaction, a non trivial angular dependence of the reaction products appears, 
i.e. some angular anisotropy, related to the initial polarizations. As all the 
polarizations of both colliding particles depend on the same magnetic field 
$\vec B$, the results for these observables depend only on the angle
$\theta$, between $\vec k$ and $\vec B$. The case of the collision of polarized 
beam with polarized target, where the beam and the target may have different 
directions of polarization is more complicated, but it can also be treated in 
the framework of the helicity formalism.

The deuteron polarization vector $\vec D^{(\lambda)}$ 
(with a definite helicity 
$\lambda$), can be chosen as:  $\vec D^{(0)}=(0,0,1)$ and $\vec 
D^{(\pm)}={1}/{\sqrt{2}}(\pm 1,i,0)$. So the following expressions for the 
six possible helicity amplitudes can be found:
\begin{equation}
\begin{array}{ll}
F_{0+,+}=g_s-(1-3\cos^2\theta)g_d,&F_{++,-}=\displaystyle\frac{3}{
\sqrt{2}}\sin^2\theta g_d,\nonumber\\
F_{0+,-}=\displaystyle\frac{3}{2}\sin~2\theta g_d,&
F_{-+,+}=\displaystyle\frac{3}{2\sqrt{2}}\sin~2\theta g_d,\nonumber\\
F_{++,+}=-\displaystyle\frac{3}{2\sqrt{2}}\sin~2\theta g_d,&
F_{-+,-}=-\frac{1}{\sqrt{2}}\left [ 2 g_s+ (1-3\cos^2\theta)g_d\right ],
\label{eq:helic}
\end{array}
\end{equation}
where $\theta$ is the nucleon production angle, relative to  the $\vec B$ 
direction. Other possible helicity amplitudes, with reversed helicities of all 
particles, can be obtained from (\ref{eq:helic}), by parity reversion.

\subsection {Collision of polarized particles}
The angular dependence of the reaction 
products in $d +{^3\!H} \to n +{^4\!He}$ for different polarization 
states of the colliding particles can be  derived from (\ref{eq:helic}).
\begin{itemize}
\item Collisions of longitudinally polarized deuterons ($\lambda_d=0$), with 
polarized $^3\!H$ or $^3\!He$: 
\end{itemize}
\begin{equation}
\sigma_{0+}(\theta)=|F_{0+,+}|^2+|F_{0+,-}|^2=|g_s|^2+ 
2Re~g_sg_d^*(-1+3\cos^2\theta)+|g_d|^2(1+3\cos^2\theta).
\label{eq:fus8}
\end{equation}
\begin{itemize}
\item $\vec d+\vec{^3\!He}$ collisions with parallel polarizations (relative to 
$\vec B$):
\end{itemize}
\begin{equation}
\sigma_{++}(\theta)=|F_{++,+}|^2+|F_{++,-}|^2=\frac{9}{2}|g_d|^2\sin^2\theta.
\label{eq:fus9}
\end{equation} 
\begin{itemize}
\item $\vec d+\vec{^3\!He}$ collisions with antiparallel polarizations:
\end{itemize}
\begin{equation}
\sigma_{+-}(\theta)=|F_{+-,+}|^2+|F_{+-,-}|^2=2|g_s|^2+2Re~g_sg_d^*
(1-3\cos^2\theta)+ \frac{1}{2}(1+3\cos^2\theta)|g_d|^2.
\label{eq:fus10}
\end{equation}
The sum of all these polarized cross sections is independent from polar angle 
$\theta$: the unpolarized 
cross section is isotropic, as expected for $S-$state interaction.
 
For the pure fusion resonance (with $g_s=0$), the angular distribution of the 
reaction products depends specifically on the direction of the polarizations of 
the colliding 
particles: the $\sin^2\theta$-dependence for parallel (++) collisions, becomes a 
dependence in $(1+3\cos^2\theta)$ for (+-) and (0+) collisions, to be compared  
with the isotropic behavior of the  
unpolarized collisions. Such definite and strong anisotropy can play a very 
important role in the design of the neutron shield of a reactor and of the 
blanket, where energetic 
neutrons  
(from $d +{^3\!H} \to n +{^4\!He}$ ) can produce ${^3\!H}$ through the 
reaction $n +{^6\!Li}\to {^3\!H} +{^4\!He}$. Once a $d +{^3\!H} 
$-reactor is beginning to operate,   ${^3\!H}$-fuel can be produced  in 
$^6\!Li$-blanket. In principle, this blanket can contain polarized $\vec 
{^6\!Li}$, 
for a more efficient ${^3\!H}$-production in $\vec n+\vec {^6\!Li}$-collisions. 

\begin{figure}
\begin{center}
\mbox{\epsfxsize=14.cm\leavevmode \epsffile{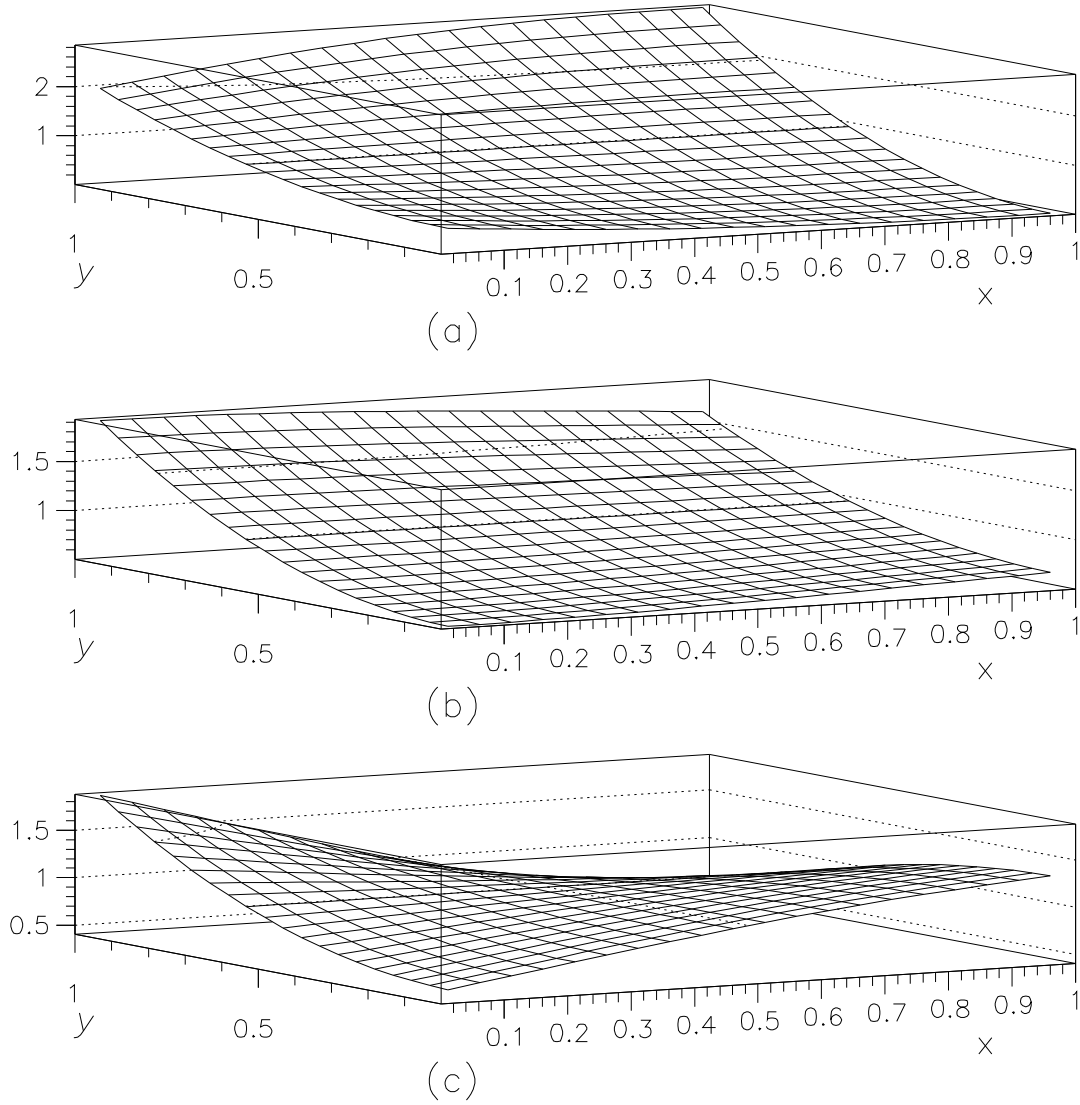}}
\end{center}
\caption{Ratio
$\sigma_{0+}/\sigma_{00}$, as a function of  $x=|g_s|/|g_d|$ and $y=\cos\theta$
for the reaction
$\vec d+^3\!He\rightarrow\vec n+^4\!He$,for different values of the phase
$\delta$: (a)
$\delta=0$, (b) $\delta=\frac{\pi}{2}$ and (c) $\delta={\pi}$, from Eq. (8).
}
\label{fig:fus1}
\end{figure}

\begin{figure}
\begin{center}
\mbox{\epsfxsize=14.cm\leavevmode \epsffile{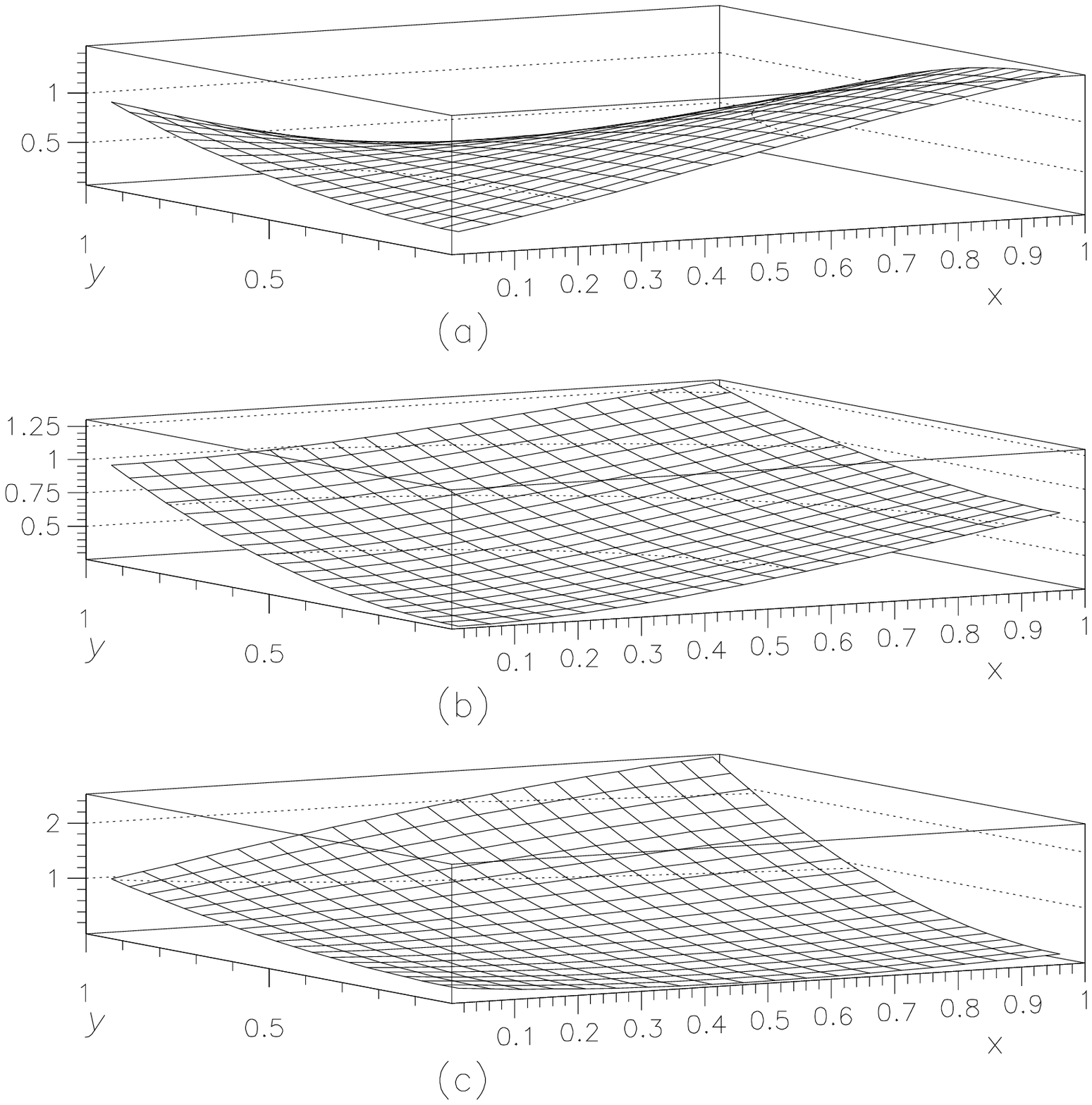}}
\end{center}
\caption{ Ratio $\sigma_{+-}/\sigma_{00}$ as a function of
$x=|g_s|/|g_d|$ and $y=\cos\theta$ for the reaction
$\vec d+^3\!He\rightarrow\vec n+^4\!He$, for different values of the phase
$\delta$: (a)
$\delta=0$, (b) $\delta=\frac{\pi}{2}$ and (c) $\delta={\pi}$, from 
Eq. (\ref{eq:fus10}).}
\label{fig:fus2}
\end{figure}

From Figs. \ref{fig:fus1} and \ref{fig:fus2}, one can see that the angular dependence of the cross 
sections for polarized collisions, is essentially influenced by the presence of 
the S-wave amplitude and its relative phase.

\newpage

Let us calculate now the following ratios:
$$R_{\lambda_1\lambda_2}=\displaystyle\frac{\int_{-1}^{+1}\sigma_{\lambda_1
\lambda_2}(\theta )d\cos\theta}
{\int_{-1}^{+1}d\cos \theta (d\sigma/d\Omega)_0,}$$
from Eqs. (\ref{eq:fus8},\ref{eq:fus9},\ref{eq:fus10}) 
for $\sigma_{\lambda_1\lambda_2}(\theta)$:
\begin{equation}
R_{0+}=1,~~R_{++}=\frac{3}{2}f,~~~~R_{+-}=\displaystyle\frac{1}{2}(4-3f).
\end{equation}
So we can write the following limits:
$$0\le R_{++}\le 3/2~{(g_s=0)},~~1/2\le R_{+-}\le 2~(g_d=0).$$
In the fusion resonance region, ($f$=1)\footnote{In particular the ratio of 
amplitudes 
$f=2|g_d|^2/(|g_s|^2+2|g_d|^2)$ was 
firstly defined in \protect\cite{Ku82}.}, the 
(++)-collisions increase the reaction yield (in comparison with collisions of 
unpolarized particles) with a maximum coefficient $\le$ 3/2, for pure D-wave 
fusion 
resonance. 
Using the notations of \cite{Ku82} one can obtain the following 
general formula for the differential cross section of  $\vec d +\vec {^3\!H}$ 
(or $\vec d +\vec {^3\!He}$)-collisions:
$$\displaystyle\frac{d\sigma}{d\Omega}
(\vec d+\vec {^3\!H})=6|g_d|^2\left \{ \frac{3}{4}a \sin^2\theta+\frac{b}{6} 
\left [\frac{2}{f}-(1-3\cos^2\theta)\left (1+\displaystyle\frac{2Re~ 
g_sg_d^*}{|g_d|^2}\right)\right]+\right .
$$
\begin{equation}
\left .+\frac{c}{12}\left [ \frac{8}{f}
-6-(1-3\cos^2\theta )\left( 1-\displaystyle\frac
{4Re~g_sg_d^*}{|g_d|^2}\right )\right ]\right \}.
\label{eq:fus12}
\end{equation}
Here $a=d_+t_++d_-t_-$, $b=d_0$, $c= d_+t_++d_-t_+$ and  $d_+$, $d_0$, $d_-$ are
the fractions of deuterons with polarization respectively parallel, 
transverse, antiparallel to $\vec B$, while  $t_+$ and $t_-$ are the 
corresponding fractions for ${^3\!H}$. The relations $d_++d_0+d_-$=1 and 
$t_++t_-=1$ hold. The 
case $a=b=c=1/3$ corresponds to unpolarized collisions.

Note that the predicted angular dependence for $b$ and $c$ contributions, Eq. 
(\ref{eq:fus12}), differs essentially from the corresponding expression 
of \cite{Ku82}. It coincides only for the special case $f=1$, $g_s=0$. 
The denominator for Eq. (2) in ref. \cite{Ku82} must be also different.

From (\ref{eq:fus12}) one can find the following expression for the differential cross 
section of collisions of polarized deuterons with unpolarized ${^3\!H}$:
$$\displaystyle\frac{d\sigma}{d\Omega}(\vec 
d+{^3\!H})=2|g_d|^2\left[\frac{1}{f}+P_{zz}
\displaystyle\frac{1-3\cos^2\theta}{4}
\left(1+\displaystyle\frac{2Re~g_sg_d^*}{|g_d|^2}\right)\right],
$$
i.e. it depends on the tensor deuteron polarization only. We used above the 
standard definition: 
$P_{zz}=d_+-2d_0+d_-$. Due to the $ (1-3\cos^2\theta)$ dependence, 
after integration 
over $\theta$, the cross section, again, does not depend on the deuteron 
polarization. 
\subsection{Polarization of neutrons in 
$\vec d + \vec {^3\!H}$ collisions}

Using the helicity amplitudes (\ref{eq:helic}) it is possible to predict also the angular 
dependence of the neutron polarization in $\vec d+\vec {^3\!H}\to n 
+{^4\!He}$, in the general case of polarized particle collisions:
$$
(n_+-n_-)\displaystyle\frac{d\sigma}{d\Omega}(\vec d+\vec{^3\!H})= 
\frac{9}{2}(d_-t_--d_+t_+)\sin^2\theta(1-2\cos^2\theta)|g_d|^2+$$
$$+d_0(t_+-t_-)\left[|g_s|^2-2(1-3\cos^2\theta)Re~g_sg_d^*+(1-15\cos^2\theta+18
\cos
^4\theta)|g_d|^2\right ]+$$
$$+ 
(d_+t_--d_-t_+)\displaystyle\frac{1}{2} \left 
[4|g_s|^2+4(1-3\cos^2\theta)Re~g_sg_d^*+(1-15\cos^2\theta+18\cos
^4\theta)|g_d|^2\right ],$$
where $n_{\pm}$ is the fraction of neutrons, polarized parallel and 
antiparallel to the direction of the magnetic field.

\begin{figure}
\begin{center}
\mbox{\epsfxsize=14.cm\leavevmode \epsffile{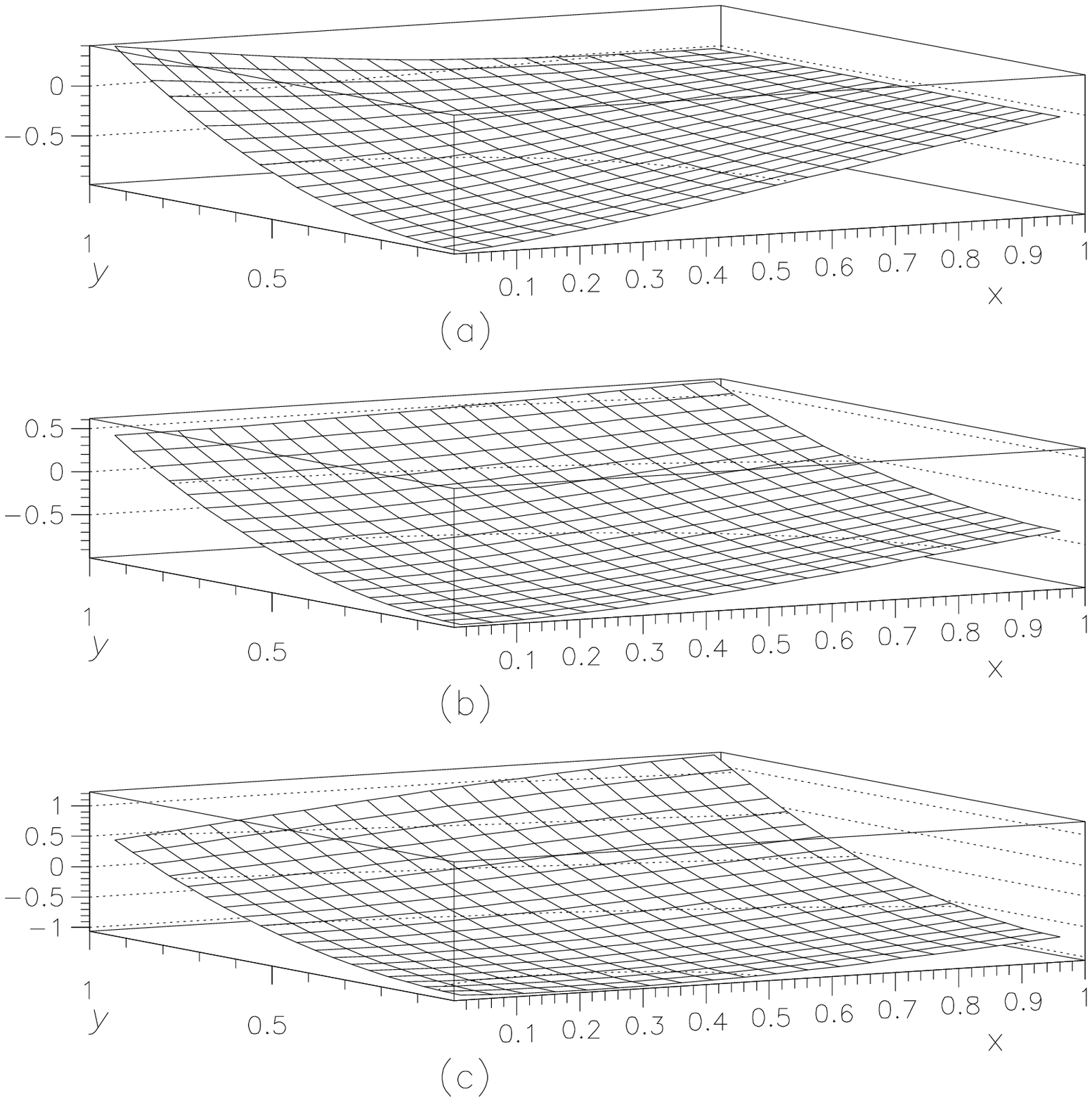}}
\end{center}
\caption{
Neutron polarization $P_n=\frac
{\left[-2+3y^2+xcos\delta(1-3y^2) +x^2\right]}{(2+x^2)},$ 
as a function of $x=|g_s|/|g_d|$ and $y=\cos\theta$ in
$\vec d+^3\!He$-collisions, for different values of the phase
$\delta$: (a) $\delta=0$, (b) $\delta=\frac{\pi}{2}$ and (c) $\delta={\pi}$.
}
\label{fig:fus3}
\end{figure}

\begin{figure}
\begin{center}
\mbox{\epsfxsize=14.cm\leavevmode \epsffile{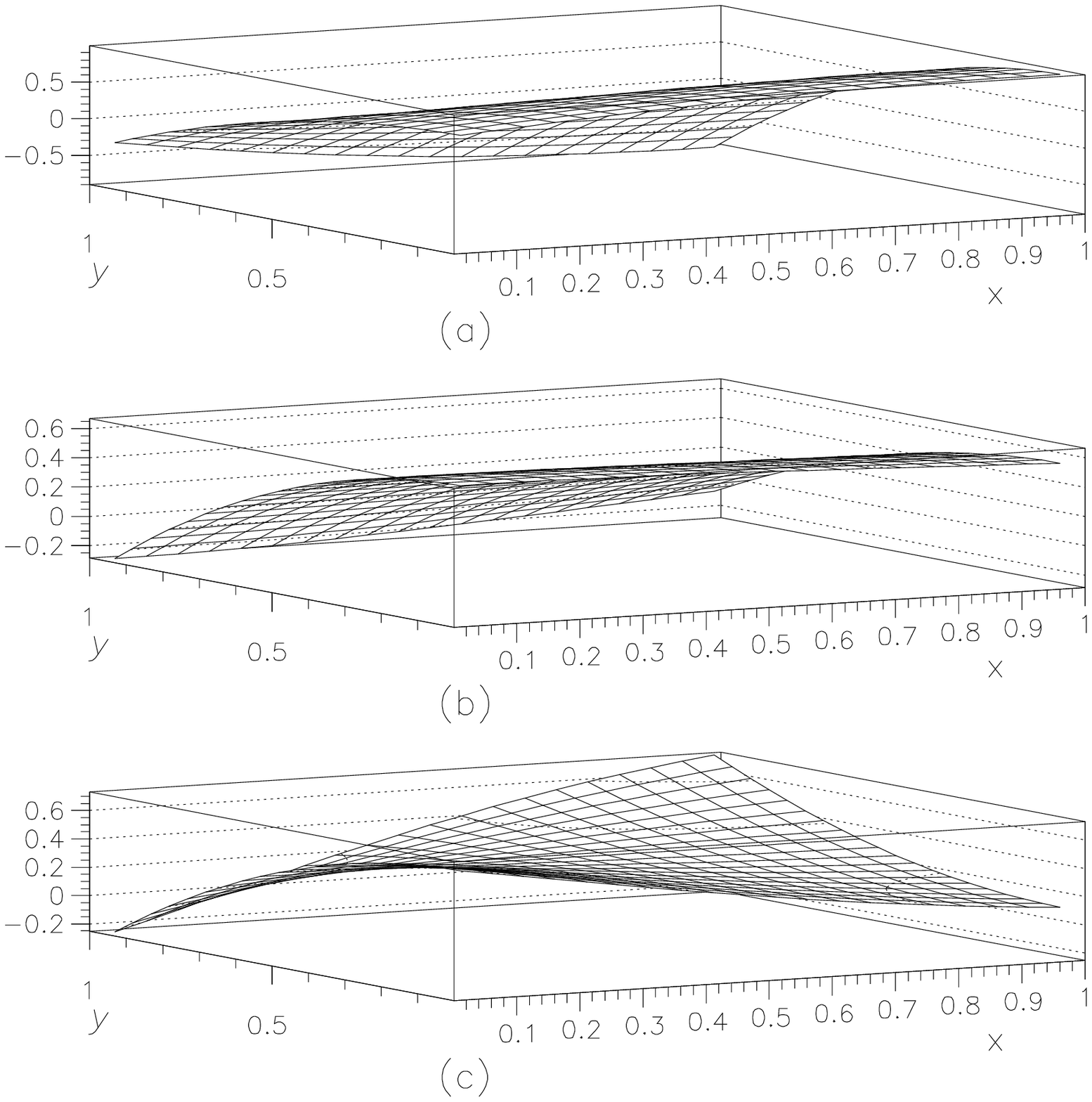}}
\end{center}
\caption{Neutron polarization $P_n=\frac{1}{3}\frac{
\left[2(2-3y^2)+4xcos\delta(1-3y^2)+x^2\right ]}{(2+x^2)}$, as a function of
$x=|g_s|/|g_d|$ and $y=\cos\theta$ in
unpolarized $d+^3\!He$-collisions, for different values of the phase
$\delta$: (a) $\delta=0$, (b) $\delta=\frac{\pi}{2}$ and (c) $\delta={\pi}$.}
\label{fig:fus4}
\end{figure}

Let us write some limiting cases of this general formula:
\begin{itemize}
\item[(a)] Collisions of polarized deuterons with unpolarized ${^3\!H}$-nuclei:
$$
(n_+-n_-)\displaystyle\frac{d\sigma}{d\Omega}(\vec 
d+{^3\!H})=(d_+-t_-)\left[ |g_s|^2 
+(1-3\cos^2\theta)Re~g_sg_d^*-\right .
$$\begin{equation}
\left .
 (2-3\cos^2\theta)|g_d|^2\right ].
\end{equation}
\item[(b)] Collisions of unpolarized deuterons with polarized ${^3\!H}$-nuclei:
\end{itemize}
$$(n_+-n_-)\displaystyle\frac{d\sigma}{d\Omega}(d+\vec{^3\!H})=\displaystyle
\frac{t_--t_+}{3}\left
[|g_s|^2 +4(1-3\cos^2\theta)Re~g_sg_d^*\right .
$$
\begin{equation}
\left . +2(2-3\cos^2\theta)|g_d|^2\right ].
\end{equation}
In the case of fusion resonance ($g_s=0$), these formulas reduce to:
$$
(n_+-n_-)\displaystyle\frac{d\sigma}{d\Omega}(\vec d+\vec 
{^3\!H})=\frac{9}{4}\sin^2\theta
(1-2\cos^2\theta)( d_+t_--d_+t_+)+ 
$$
\begin{equation}
\frac{1}{2}\left [d_0(t_+-t_-)+\frac{1}{2}( d_+t_--d_-t_+) 
(1-15\cos^2\theta+18\cos^4\theta)\right ].
\end{equation}
Averaging over the polarizations of $d$ (or $^3\!H$) one can find
particular expressions:
$$
(n_+-n_-)\displaystyle\frac{d\sigma}{d\Omega}(d+\vec{ ^3\!H})=\frac{1}{2}
(t_--t_+)(2-3 \cos^2\theta)
$$
and
$$
(n_+-n_-)\displaystyle\frac{d\sigma}{d\Omega}(\vec d+^3\!H)=-\frac{1}{3}( 
d_--d_+)(2-3\cos^2\theta).
$$
The angular dependence of most of these polarization observables is sensitive to 
the relative value of the $g_s$ and 
$g_d$ amplitudes, due to the $g_sg_d^*$-interference contributions. Of course, 
in the region of the fusion resonance the $g_d$ amplitude is dominant. However 
the 
temperature conditions, typical for a fusion reactor, correspond to collision 
energies lower than the energy of the fusion resonance. Even a small $g_s/g_d$ 
ratio can change the angular behavior of the polarization observables. In Figs. 
\ref{fig:fus3} and \ref{fig:fus4} we show, in a 3-dimensional plot, the dependence of the neutron polarization 
on the ratio $x=|g_s|/g_d|$ and on the production angle $\theta$ for three 
values of the relative phase 
$\delta$, $\delta=0,~\pi/2,~\pi$, for $\vec d+^3\!H$ and $ 
d+\vec{^3\!H}$-collisions.

\newpage

The exact determination of the parameters $x$ and $\delta$, is crucial for 
thermonuclear 
processes. This is a reason to perform a complete experiment for this reaction 
as discussed earlier \cite{Rek98}. The important point is 
that  even at very low energies, where the spin structure is simplified, a 
complete experiment must include the scattering of polarized beam on polarized 
target. The full reconstruction of the threshold matrix elements requires this 
type of experiment.
\section{Processes $d+d\to n+^3\!He$ 
and $d+d\to p+^3\!H$}

The $d+d\to n+^3\!He$ and $d+d\to p+^3\!H$ reactions at low 
energy have a very wide spectrum of fundamental and practical applications, from
the discovery of tritium and helium isotopes \cite{Ol33}, to the 
important role for primordial nucleosynthesis in the early Universe and fusion 
energy production with 
polarized and unpolarized fuel \cite{Ku82,Ku86}. These processes are of large 
interest in nuclear theory: for example, in a four nucleon system, contrary to 
three nucleon system, broad resonant states can be excited \cite{Ti92}.
The angular dependence of the differential cross sections \cite{Br90,Kr87} and 
the polarization observables \cite{Cs97,Ba97,Dr80,Ti98} for these charge symmetric reactions 
constitutes a good test of the isotopic invariance for the low energy nuclear 
interaction.
The $dd-$ interaction is also connected to muon catalyzed processes ($\mu d 
d)\to \mu+p+^3\!H$ or $(\mu dd)\to \mu +n+^3\!He$ \cite{Br89}, 
where only the P-state of the $dd$-system is present, at low energy. 

In the general case the spin structure of the matrix element for  
$d+d\to n+^3\!He~(p+^3\!H)$ is quite complicated, with 
18 independent spin combinations, and therefore with 18 complex scalar 
amplitudes, which are functions of the excitation energy and the scattering 
angle. However, at thermal collision energies, where the 
$S-$state deuteron interaction has to dominate, this structure is largely 
simplified. The identity of the colliding deuterons, which are bosons, 
is an important guide for the partial amplitude analysis in order to determine  
the spin structure of the reaction amplitude. The determination of the 
polarization observables is indispensable, for this purpose. The four possible 
analyzing powers for $\vec d+d$-collisions, $A_y$, $A_{zz}$, $A_{xz}$ and 
$A_{xx}-A_{yy}$ were measured at $E_d\le 100$ keV, as well as the angular 
dependence of the differential cross section \cite{Br90,Kr87,Fl94}.

The knowledge of the relative role of different orbital angular momenta (and 
 the corresponding partial amplitudes) is essential for the solution of 
different 
fundamental problems concerning these processes, like the possibility to build a 
thermonuclear "clean" reactor with polarized $d+^3\!He$-fuel. The main reaction 
$d+^3\!He\to p+^4\!He$ does not produce radioactive nuclei, and the 
possibility to decrease the cross section of $\vec d+\vec d$-collisions (which 
produces $n+^3\!He$ or $p+^3\!H$) with 
parallel polarizations, will decrease the production of neutrons and the 
tritium.
Direct experimental data about $\vec d+\vec d$- low energy collisions are 
absent, so the dependence of the cross section on the polarization states of the 
colliding particles can be  calculated  only from theoretical predictions or 
from 
different multipole analysis.

The theoretical predictions and the results of multipole analysis seem very 
controversial now, even at very low 
energy.
In the first partial wave analysis \cite{Ad69,Ad81} it was found that the 
S-state $dd-$interaction in the quintet state (i.e. with total spin $S_i=2$) is 
smaller in comparison with the $S_i=0$ interaction. This was consistent with the 
conclusion of ref. \cite{Ku82}, that in a polarized reactor it is possible to 
suppress $\vec d+\vec d$-collisions. Later \cite{Ho84}, it was pointed out that 
strong central forces with $D-$state in $^3\!He$ can induce a large $dd-$ 
interaction in the quintet state and resonating-group calculations \cite{Ho84} 
found that polarized collisions are not suppressed. On the other hand, DWBA 
calculations give a large suppression for the ratio of polarized on unpolarized 
cross section, $\sigma_{++}/\sigma_0\simeq 0.08$ in the range $E_d=20-150$ keV, 
even after inclusion of the $^3\!He$ D-state. A more recent analysis 
\cite{Le90,Pa92} based on $R-$matrix approach, concludes that this ratio does 
not decrease with energy. Note that in principle, it can be energy dependent 
\cite{Zh95}.

Again, a direct measurement of polarized $dd$-collisions would greatly help in 
solving 
these problems and the complete experiment will allow to reconstruct the spin 
structure of the reaction amplitude.  Therefore, the considerations based on 
$S-$wave only, have to 
be considered as the first necessary step which can illustrate the possible 
strategy of the complete experiment for this case.

\subsection{Partial amplitudes}
We establish here the spin structure of the threshold matrix element for the 
$d+d\to n+^3\!He ~(p+^3\!H)$ process. For S-state $dd-$interaction the 
following 
partial transitions are allowed:
\begin{eqnarray*}
&S_i=0~\to {\mathcal 
J}^{P}=0^+~\to S_f=0,~{\ell}_f=2,\\
&S_i=2~\to {\mathcal 
J}^{P}=2^+~\to S_f=0,~{\ell}_f=2,\\
&S_i=2~\to {\mathcal 
J}^{P}=2^+~\to S_f=1,~{\ell}_f=2,
\end{eqnarray*}
where $S_i$ is the total spin of the colliding deuterons, ${\ell}_f$ is the 
orbital angular momentum of the final nucleon. Note that the Bose statistics for 
identical deuterons allows only even values of initial spin, that is 
$S_i=0$ and $S_i=2$ for the S-state. The resulting  spin structure of the 
threshold matrix element can be written as:
\begin{equation}
\begin{array}{ll}
{\mathcal M}=i({\chi}_3^{\dagger}\sigma_2 \widetilde{{\chi}_1^{\dagger}})&\left[ 
g_1\vec 
D_1\cdot \vec D_2+g_2( 3 \vec k\cdot\vec D_1~\vec k\cdot \vec D_2-\vec D_1\cdot 
\vec D_2)\right .\nonumber\\
&\left .+g_3(\vec \sigma\cdot \vec k\times\vec D_1~\vec k\cdot \vec D_2+ 
\vec \sigma\cdot \vec k\times\vec D_2~\vec k\cdot \vec D_1) \right ],
\end{array}
\label{eq:fus16}
\end{equation}
where $\chi_1$ and $\chi_3$ are the 2-component spinors of the produced 
nucleon and 
$^3\!He$ (or $^3\!H$), $\vec D_1$ and $\vec D_2$ are  the 3-vectors of the 
deuteron polarization, $\vec k$ is the unit vector along the 3-momenta of the 
nucleon (in 
the CMS of the considered reaction). The amplitudes $g_1$ and $g_2$ describe 
the production of the singlet $n+^3\!He$-state, and the amplitude $g_3$- the 
triplet state. The complete experiment in $S-$state $dd$-interaction implies the 
measurement of 5 different 
observables, to determine 3 moduli and two relative phases of partial 
amplitudes.

The validity of the S-state approximation in the near threshold region can be 
checked by measuring any T-odd polarization observable, the simplest of which 
are the one-spin observables as the vector analyzing power in the reaction 
$ \vec d+d \to n+^3\!He$ \cite{Fl94}. Note that Eq. (\ref{eq:fus16}) is correct also 
for the threshold matrix elements of the inverse process: 
$n+^3\!He\to d+d $ (or $p+^3\!H\to  d+d $). 
\subsection{Helicity amplitudes}
In order to establish the angular dependence of the reaction products, for 
collisions of polarized particles, in the presence of magnetic field, let us 
derive 
the  helicity amplitudes.  The spin structure of the $d+d$ reactions is more 
complex 
in 
comparison to $d+^3\!He$. The analysis of polarization phenomena is also more 
complicated. It was mentioned in \cite{Ku82}, that an enhancement factor, equal 
to 2 
can be obtained in a polarized plasma \footnote{Note that this holds only for 
the partial wave analysis \cite{Ad69,Ad81}.}, for the reaction $d+d\to 
n+^3\!He$, if 
the deuterons are polarized transversally to the direction of the magnetic 
field, i.e. 
for (00)-collisions, in an ordinary thermal ion distribution. Alternatively, if 
colliding beams or beam and target methods are used (inertial fusion), the two 
ions 
should 
be polarized in opposite direction, relatively to the field. In case of 
collisions of 
deuterons with parallel polarizations i.e (++) or ($--$), a large suppression of 
the 
reaction rate is expected.

It is then interesting to analyze all possible configurations of the 
polarization of 
the colliding deuterons. We can classify the helicity amplitudes according to 
the 
following scheme:
\begin{description}
\item [ I)]\underline{00~collisions}: the polarization is transverse to the 
magnetic 
field 
$\rightarrow$ 2 independent amplitudes;
\item [ II)]\underline{++~collisions}: the polarization parallel to the magnetic 
field 
$\rightarrow$ 4 independent amplitudes;
\item [III)]\underline{+$-$~collisions}: collisions with deuterons with 
antiparallel 
polarization, in the same direction as the magnetic field $\to $ 4 
independent 
amplitudes;
\item [IV)] \underline{0+~collisions}: collisions of one  deuteron with  
polarization 
transverse to the magnetic field with the other deuteron polarized along  the 
magnetic 
field $ \to $ 4 independent amplitudes;
\end{description}

The corresponding helicity amplitudes ${\mathcal 
F}_{\lambda_1\lambda_2,\lambda_3\lambda_4}$, (with ${\lambda_1}\equiv 
\lambda_{d_1}$, 
${\lambda_2}\equiv\lambda_{d_2}$, ${\lambda_3}\equiv \lambda_{^3\!He}$, 
${\lambda_4}\equiv \lambda_N$) are 
given in 
terms of partial amplitudes:
\vspace{.2truecm}
\begin{equation}
\begin{array}{ll}
(I)&{\mathcal F}_{00,++}=-\sin~2\theta g_3,~~
{\mathcal F}_{00,+-}=g_1-(1-3\cos^2\theta)g_2,\nonumber \\
(II)&{\mathcal F}_{++,++}=\sin~2\theta g_3,
~~{\mathcal F}_{++,+-}=\sin^2\theta
(\displaystyle\frac{3}{2}g_2+g_3),\nonumber \\
&{\mathcal F}_{++,--}=0,~~{\mathcal F}_{++,-+}=\sin^2\theta
(-\displaystyle\frac{3}{2}g_2+g_3),\nonumber \\
(III)&{\mathcal F}_{+-,++}=
{\mathcal F}_{+-,--}=-\displaystyle\frac{1}{2}\sin~2\theta 
g_3,\label{eq:fus17}, \\
&{\mathcal F}_{+-,+-}=-{\mathcal F}_{+-,-+}=-g_1-  
\displaystyle\frac{1}{2}(1-3\cos^2\theta)g_2,\nonumber \\
(IV)&
{\mathcal F}_{0+,++}=\displaystyle\frac{1}{\sqrt{2}}(-1+3\cos^2\theta) 
g_3,~~
{\mathcal F}_{0+,+-}=\displaystyle\frac{1}{2\sqrt{2}}\sin~2\theta 
(3g_2+g_3),\nonumber \\
&{\mathcal F}_{0+,--}=-\displaystyle\frac{1}{\sqrt{2}}\sin^2\theta g_3,
~~{\mathcal F}_{0+,-+}=\displaystyle\frac{1}{2\sqrt{2}}\sin~2\theta 
(-3g_2+g_3).\nonumber
\end{array}
\end{equation}
where $\theta$ is the nucleon production angle relative to $\vec B$ 
direction.
\subsection{Angular dependence for collisions of polarized deuterons}
After summing over the polarization states of the produced particles,  the cross 
section of the process $\vec d+\vec d\to n+^3\!He$, for definite 
deuteron polarizations, can be written as:
$$\sigma_{00}(\theta)=2\left (|{\mathcal F}_{00,++}|^2+|{\mathcal 
F}_{00,+-}|^2\right 
)=2|g_1-g_2(1-3\cos^2\theta)|^2+8 \sin^2\theta \cos^2\theta |g_3|^2,$$
$$\sigma_{++}(\theta)=\sum_{\lambda_3,\lambda_4}|{\mathcal 
F}_{++,\lambda_3\lambda_4}|^2=\sin^2\theta \left [\displaystyle\frac{9}{2} 
\sin^2\theta |g_2|^2+2(1+\cos^2\theta) |g_3|^2\right ],$$
\begin{eqnarray}
\sigma_{+-}(\theta)=&\sum_{\lambda_3,\lambda_4}|{\mathcal 
F}_{+-,\lambda_3\lambda_4}|^2= 
2|g_1|^2+ 2 Re ~g_1g_2^*(1-3\cos^2 \theta +  \label{eq:fus18}\\
&\displaystyle\frac{1}{2} (1-3\cos^2\theta)^2 |g_2|^2+ 2
\sin^2\theta \cos^2\theta |g_3|^2,\nonumber
\end{eqnarray}
$$\sigma_{0+}(\theta)=\sum_{\lambda_3,\lambda_4}|{\mathcal 
F}_{0+,\lambda_3\lambda_4}|^2= 
9\sin^2\theta \cos^2\theta |g_2|^2
+(1-3\cos^2 \theta +4\cos^4\theta)|g_3|^2.$$
With the help of these formulas we can estimate the corresponding integral 
ratios:
$$R_{\lambda_1\lambda_2}=\displaystyle\frac{\int_{-1}^{+1}\sigma_{\lambda_1
\lambda_2}(\theta )d\cos\theta}
{\int_{-1}^{+1}d\cos \theta (d\sigma/d\cos\theta)_0},$$
which characterize the relative role of polarized collisions with respect to 
unpolarized ones:
\begin{equation}
R_{00}=\displaystyle\frac{3}{5}\displaystyle\frac{15+4r}{3+2r},
~R_{++}=\frac{36}{5}\frac{r}{3+2r},~R_{+-}=\frac{12}{5}\frac{15+r}{3+2r},
~R_{0+}=\displaystyle\frac{9}{5}\displaystyle\frac{r}{3+2r},
\end{equation}
where $r=(3 |g_2|^2+2 |g_3|^2)/|g_1|^2)$. It is interesting that all these 
ratios 
depend on a single contribution of the moduli of the partial amplitudes, the 
ratio 
$r\ge 0$. The ratios $R_{\lambda_1\lambda_2}$ are limited by:
$$1.2\le R_{00}\le 3,~~0\le R_{++}\le 3.6,~~1.2\le R_{+-}\le 12,~~0\le R_{0+}\le 
0.9,$$
where the upper limits correspond to $g_2=g_3=0$, (when only the $g_1$ amplitude 
is 
present), and the lower limits correspond to $g_1=0$ (for any amplitudes $g_2$ 
and 
$g_3$). But the exact values of $R_{\lambda_1\lambda_2}$ depend on the relative 
value 
of the partial amplitudes, through one parameter, $r$.

The general dependence of the differential cross section for $\vec d+\vec 
d$-collisions, can be written in terms of partial cross sections 
$\sigma_{\lambda_1\lambda_2}$ as follows:
\begin{equation}
\displaystyle\frac {d\sigma}{d\Omega}(\vec d+\vec 
d)=(d_+^2+d_-^2)\sigma_{++}(\theta)+d_0^2\sigma_{00}(\theta)+2d_+d_-
\sigma_{+-}(\theta)+2d_0(d_++d_-)\sigma_{0+}(\theta),
\label{eq:fus20}
\end{equation}
where we used the evident relations between $\sigma_{\lambda_1\lambda_2}$: $
\sigma_{++}(\theta)=\sigma_{--}(\theta)$, 
$\sigma_{0+}(\theta)=\sigma_{0-}(\theta)$, 
$\sigma_{+-}(\theta)=\sigma_{-+}(\theta)$, due to the P-invariance of 
the 
strong interaction, and the standard notation: $d_+$, $d_0$ and  $d_-$ for 
different 
deuteron fractions in polarized plasma.

Using Eq. (\ref{eq:fus20}) one can find some interesting limiting cases. Setting for 
example, 
$d_+=d_-$ (deuterons with tensor polarization only: $P_{zz}=1-3d_0$, $P_z=0$), 
one can 
obtain the following dependence of the differential cross section on $P_{zz}$:
\begin{equation}
\displaystyle\frac{d\sigma}{d\Omega}(\vec d+\vec 
d)=a_0(\theta)+2P_{zz}a_1(\theta)+\displaystyle\frac{1}{2}P_{zz}^2a_2(\theta),
\end{equation}
where the coefficients $a_i(\theta),i=0-2$, are linear combinations of the  
helicity 
cross sections $\sigma_{\lambda_1\lambda_2}$:
$$
9a_0(\theta)=2\left[ \sigma_{++}(\theta)+\sigma_{+-}(\theta)\right 
]+\sigma_{00}(\theta)+4\sigma_{+0}(\theta),$$
\begin{equation}
9a_1(\theta)=\sigma_{++}(\theta)+\sigma_{+-}(\theta)-\sigma_{00}(\theta)-\sigma_
{+0}
(\theta),
\end{equation}
$$
9a_2(\theta)=\sigma_{++}(\theta)+\sigma_{+-}(\theta)+2\sigma_{00}(\theta)-
4\sigma_{+0}(\theta).$$
So, measuring the $P_{zz}$-dependence of the cross section for $\vec d+\vec d$ 
collisions, one can determine all 3 coefficients $a_i(\theta)$ (at each angle 
$\theta$). This allows to determine the individual helicity partial cross 
sections  
$\sigma_{\lambda_1\lambda_2}(\theta)$:
$$\sigma_{00}(\theta)=a_0(\theta)-4a_1(\theta)+2a_2(\theta),$$
\begin{equation}
\sigma_{0+}(\theta)=a_0(\theta)-a_1(\theta)-a_2(\theta),
\end{equation}
$$\sigma_{++}(\theta)+\sigma_{+-}(\theta)=2a_0(\theta)+4a_1(\theta)+a_2(\theta).
$$
In order to disentangle the  $\sigma_{++}(\theta)$ and $\sigma_{+-}(\theta)$ 
contributions, an additional polarization observable has to be measured, from 
the 
collisions of vector polarized deuterons ($d_{\pm}=\displaystyle\frac 
{1}{3}\pm\displaystyle\frac {1}{2}P_z,~d_0=\displaystyle\frac{1}{3}$):
\begin{equation}
\displaystyle\frac{d\sigma}{d\Omega}(\vec d+\vec 
d)=a_0(\theta)+\displaystyle\frac{P_z^2}{2}(\sigma_{++}(\theta)-\sigma_{+-}
(\theta)).
\end{equation}

The linear $P_z$ 
contribution is forbidden by the P-invariance of the strong interaction. Only
the measurement of the $P_z^2$ contribution allows to separate the cross 
sections 
$\sigma_{++}(\theta)$ and $\sigma_{+-}(\theta)$.

This analysis is equivalent to the discussion  of the complete experiment (in 
terms of 
helicity cross sections $\sigma_{\lambda_1\lambda_2}(\theta)$).

Finally let us derive the polarization properties of the neutrons in the process 
 $\vec d+\vec d\to n+^3\!He$. Using Eqs. (\ref{eq:fus17}) 
for the helicity amplitudes, one 
can find for the $\theta$ dependence of the neutron polarization (for the 
different 
spin configurations of the colliding deuterons):
$$ (n_+-n_-)\sigma_{++}(\theta)=2\sin^2\theta d_+^2\left [ 3 Re 
g_2g_3^*+2\cos^2\theta 
|g_3|^2\right ],$$
\begin{equation}
(n_+-n_-)\sigma_{0+}(\theta)=2d_0d_+ \cos^2\theta\left [ 
-(1-2\cos^2\theta)|g_3|^2+3\sin^2\theta Re~ g_2g_3^*\right ],
\end{equation}
$$(n_+-n_-)\sigma_{00}(\theta)=(n_+-n_-)\sigma_{+-}(\theta)=0,$$
where $n_+$ and $n_-$ are the fractions of polarized neutrons with spin 
parallel and antiparallel relative to the $\vec B$ direction.

The production of unpolarized neutrons for 00-collisions of deuterons results 
from 
P-invariance, and for $-+$ collisions  results from  the identity of colliding 
deuterons and from the P-invariance.
\subsection{Complete experiment for $d+d\to n+^3\!He$}

Due to three complex partial amplitudes for the S-wave $dd-$interaction for  the 
process
$d+d\to n+^3\!He$,
 the measurement of a large number of observables is necessary, in order to 
perform  the 
complete 
experiment. This study  will be based on the formalism of the 
polarized structure functions, previously used in \cite{Rek98} for the process 
$d+^3\!H\to n+^4\!He$.

Let us consider the collisions of polarized deuterons 
$\vec d+\vec d\to n+^3\!He$. The differential cross section can be 
parametrized in the following general form:
\begin{equation}
\begin{array}{ll}
\displaystyle\frac{d\sigma}{d\Omega}=&\left 
(\displaystyle\frac{d\sigma}{d\Omega}\right 
)_0\left [ 1+{\mathcal A}_1(\vec k\cdot\vec Q_1+\vec k\cdot \vec Q_2)+{\mathcal 
A}_2 \vec S_1\cdot \vec S_2+{\mathcal A}_3\vec k\cdot\vec S_1~\vec k\cdot\vec 
S_2\right 
.\\
& +{\mathcal A}_4\vec k\cdot\vec Q_1~\vec k\cdot \vec Q_2+{\mathcal A}_5\vec 
Q_1\cdot \vec Q_2+{\mathcal A}_6 Q_{1ab}Q_{2ab}\\
&\left .
+{\mathcal A}_7(\vec k\cdot\vec S_1\times\vec Q_2+\vec k\cdot\vec S_2\times\vec 
Q_1)\right 
], ~Q_{1a}=Q_{1ab}k_b,~Q_{2a}=Q_{2ab}k_b,
\end{array}
\end{equation}
where $\vec S_1$ and $\vec S_2$ ($Q_{1ab}$ and $Q_{2ab}$) are the vector 
(tensor) 
polarizations of the colliding deuterons. The real coefficient ${\mathcal A}_1$ 
describes 
the tensor analyzing power in $\vec d+ d\to n+^3\!He$, ${\mathcal 
A}_2-{\mathcal A}_7$ are the spin 
correlation 
coefficients in $\vec d+\vec d\to n+^3\!He$. The coefficients ${\mathcal 
A}_1-{\mathcal 
A}_6$ are T-even polarization observables and ${\mathcal A}_7$ is the T-odd one 
(due to 
the specific correlation of the vector polarization of one deuteron and the 
tensor 
polarization of the other deuteron). Note that these coefficients ${\mathcal 
A}_i$ 
can not 
fix the relative phases of the singlet amplitudes $g_1$ and $g_2$ (from 
one side) 
and the triplet amplitude $g_3$ (from the other side). The complete experiment 
has to 
be more complex than the determination of the polarization observables 
${\mathcal 
A}_i$. 
The polarization transfer coefficients from the initial deuteron to the produced 
fermion ($n$ or $^3H$) have to be measured, too.

After summing over the polarizations of the produced particles in $\vec d+\vec 
d\to n+^3\!He$, the following expressions can be found, for the 
coefficients 
${\mathcal A}_i,~i=1-7$, in terms of the partial amplitudes $g_k,~k=1-3$:
$$
-\displaystyle\frac{9}{2}{\mathcal A}_1\left (\displaystyle\frac 
{d\sigma}{d\Omega}\right)_0=3|g_2|^2+|g_3|^2+6 Re~ g_1g_2^*,$$
$$
 {\mathcal A}_2\left (\displaystyle\frac{d\sigma}{d\Omega}\right)_0= 
-|g_1|^2+2|g_2|^2+|g_3|^2- Re~ g_1g_2^*,$$
 $$
{\mathcal A}_3\left(\displaystyle\frac{d\sigma}{d\Omega}\right)_0=
-3|g_2|^2-|g_3|^2+3Re~ g_1g_2^*,$$
\begin{equation}
 \displaystyle\frac{9}{4}{\mathcal A}_4
\left 
(\displaystyle\frac{d\sigma}{d\Omega}\right)_0= 9|g_2|^2-4|g_3|^2
\end{equation}
$$
\displaystyle\frac{9}{2}{\mathcal A}_5\left (\displaystyle\frac 
{d\sigma}{d\Omega}\right)_0= -6|g_2|^2+ 6Re~ g_1g_2^*+2|g_3|^2,
$$
$$
 \displaystyle\frac{9}{2}{\mathcal A}_6\left (\displaystyle\frac 
{d\sigma}{d\Omega}\right)_0= |g_1|^2+|g_2|^2-2Re~ g_1g_2^*,$$
$$
{\mathcal A}_7 \left 
(\displaystyle\frac{d\sigma}{d\Omega}\right)_0=-2~Im ~g_1g_2^*,
$$
where $({d\sigma}/{d\Omega})_0$ is the differential cross section with 
unpolarized 
particles:
$$\left (\displaystyle\frac {d\sigma}{d\Omega}\right )_0=\frac{2}{9}\left [ 3 
|g_1|^2+6|g_2|^2+4|g_3|^2\right ]=\displaystyle\frac{2}{9}|g_1|^2(3+2r).$$
Using these expressions, the following relations can be found between the 
coefficients 
${\mathcal A}_i$:
\begin{itemize}
\item [(a)] linear: between T-even polarization observables,
$${\mathcal A}_2+{\mathcal A}_3+\displaystyle\frac{9}{2}{\mathcal A}_6
={\mathcal A}_1+{\mathcal A}_4-\displaystyle\frac{1}{3}
{\mathcal A}_3+\displaystyle\frac{7}{4}{\mathcal A}_5=0$$
\item[(b)] quadratic, relating the T-odd asymmetry ${\mathcal A}_7$ with the 
T-even 
coefficients ${\mathcal A}_i,~i=1-6$;
\end{itemize}
$$ \displaystyle\frac{9}{4}(1+{\mathcal A}_1^2-{\mathcal A}_7^2)
={\mathcal A}_2^2+({\mathcal A}_2+{\mathcal A}_3)^2
+6({\mathcal A}_1{\mathcal A}_2+{\mathcal A}_1{\mathcal A}_3+{\mathcal 
A}_2{\mathcal A}_3)$$

Therefore, the measurements of $({d\sigma}/{d\Omega})_0$ and 3 coefficients 
${\mathcal 
A}_i,~i=1-3$, allow to find the moduli of all S-wave partial amplitudes 
$g_k,~k=1-3$, and the relative phase of the singlet amplitudes $g_1$ and $g_2$:
$$18|g_1|^2=(9-12{\mathcal A}_2-4{\mathcal A}_3)\left (\displaystyle\frac 
{d\sigma}{d\Omega}\right )_0,$$
$$-18|g_2|^2=(9+18{\mathcal A}_1+ 6{\mathcal A}_2+10{\mathcal A}_3)\left 
(\displaystyle\frac 
{d\sigma}{d\Omega}\right )_0,$$
$$2|g_3|^2=(3+3{\mathcal A}_1+ 2{\mathcal A}_2+2{\mathcal A}_3)\left 
(\displaystyle\frac 
{d\sigma}{d\Omega}\right )_0,$$
$$18 Re~g_1g_2^*=(-9{\mathcal A}_1+2{\mathcal A}_3)\left (\displaystyle\frac 
{d\sigma}{d\Omega}\right )_0.$$
So these measurements can be considered as the first step of the complete 
experiment 
for the process 
$ d+d\to n+^3\!He$ in the  near threshold conditions.

Using these expressions, one can find the following expression for the ratio 
$r$:
$$ r=3\displaystyle\frac{1+a}{1-2a},~a=
\displaystyle\frac{2}{9}(3{\mathcal A}_2+{\mathcal 
A}_3).$$

The results obtained here on the angular dependence and 
the reaction rate dependence on the nuclei polarizations, can be used as a guideline in 
the conception of magnetic fusion reactors. The polarization of the produced 
particles is also important, as it can help the fusion process in a working 
reactor. 
For example, in a a reactor based on $d+^3\!H$-fuel, the intensive flux of 14 
MeV 
neutrons can be used in the $Li-$blanket, not only for its heating, with 
consequent production of electric power, but also to produce extra $^3\!H$- 
fuel, through the processes: 
$n+^6\!Li\to ^3\!H+^4\!He \mbox{~and ~}n+^7\!Li\to 
n+^3\!H+^4\!He.$
Due to the definite polarization properties of these reactions, 
one can increase, in principle, the yield of $^3\!H$.

We showed that the polarization and the angular distribution of the neutrons, 
produced in the process $d+^3\!H\to n+^4\!He$ depends strongly on the 
relative value of the two possible partial amplitudes. The presence of a 
contribution (even relatively small) of the  ${\mathcal J}^P =1/2^+$ amplitude 
is 
very important for polarization phenomena.

For the reaction $d+d\to n+^3\!He$ (with three independent threshold 
partial amplitudes) the situation is more complicated. The $d+d$-reactions 
produce energetic  neutrons and tritium, and should be suppressed in a 
$d+^3\!He$ reactor. 

The detailed information about partial amplitudes of different reactions can be 
obtained, in a model independent way, through the realization of the complete 
experiment. Even at low energy, where the spin structure of all matrix 
elements is highly simplified, the complete experiment includes the scattering 
of a polarized beam on a polarized target. These experiments, which are absent
up to now,  allow the full reconstruction of the spin structure of the threshold 
matrix elements.

The main results derived above can be summarized as follows:
\begin{itemize}
\item We give a model independent parametrization of the spin structure of the 
threshold matrix elements for the reactions: 
$d+d \to n+^3\!He$ and $d+^3\!H\to n+^4\!He$.
\item The angular distributions of the reaction products for $\vec d+\vec d$ and 
$\vec d+\vec {^3\!He}$-collision shows a strong dependence on the polarizations 
of the colliding particles, 
and 
it can be very important to optimize the blanket and the shielding of a reactor.
\item The polarization properties of neutrons, produced in the processes 
$d+^3\!H$ $\to $$n+^4\!He$ and $d+d $$\rightarrow$$ n+^3\!He$  are 
derived for 
collisions of polarized particles.
\end{itemize}

\chapter{Polarization phenomena in astrophysical processes}
\section{Low energy polarized collisions and astrophysics}
The experimental and theoretical study of the collisions of light nuclei ($p$, 
$d$, $^3\!H$, $^3\!He$, $^4\!He$..)  at very low energies has always been 
motivated by questions in fundamental physics and by interesting possible 
applications in particular in the astrophysics domain \cite{La96}. Many works in 
the past have been devoted to the experimental study of these reactions, in 
particular cross section measurements, but only recently some polarization 
observables such as, for example, analyzing powers have become available 
\cite{Ba94}. Polarization 
phenomena are  important also in order to understand the mechanisms of the 
electromagnetic processes such as $n+p \to n+\gamma,~p+d \to 
^3\!He+\gamma,~n+d \to ^3\!H+\gamma, d+d \to ^4\!He+\gamma,~ 
d+^3\!He\to ^5\!Li+\gamma$ etc.,\cite{Ba94} which are at the basis of  
models of primordial nucleosynthesis in early Universe.

The strong magnetic field \mbox{($B\simeq 10^{12-14}~G$)} on the surface of 
neutron stars  \cite{Ta85} must induce large degree of polarization for heavy
particles like protons, neutrons, deuterons, etc. The reaction rates for all 
the above mentioned reactions at low energy depend strongly on the polarization 
properties of the colliding nuclei.  The astrophysical S-factors, determining the threshold behavior of cross sections, are very important parameters in models of big-bang nuclear synthesis, stellar hydrogen burning, solution of the Sun-neutrino puzzle etc. Moreover possible large magnetic fields in the early 
Universe \cite{Ho83,Tu88,Qu89,Va91,Che94}, B$\simeq 10^{20}$G, may have influenced the process of nucleosynthesis of light elements. The most evident effect of a strong magnetic field concerning the "deformation" of electrons in atoms and in specfic Landau quantization of the electron behavior has been extensively studied for different electromagnetic conversion processes such as magnetic bremsstrahlung (synchrotron radiation), magnetic $e^+e^-$-pair production, magnetic Cherenkov radiation, photon splitting, etc. This deformation is especially important in 
calculations of reaction rates for the weak processes involving an electron:
\begin{equation}
e^-+p\rightleftharpoons n+\nu_e,~\overline{\nu_e}+p\rightleftharpoons 
n+e^+,~n\to p+e^-+ \overline{\nu_e},
\label{eq:as1}
\end{equation}
which play an essential role in the big-bang nucleosynthesis and neutron star 
cooling \cite{Ch93,Gr95}.

Of course, magnetic deformation is not so important for heavier particles: 
protons, neutrons, deuterons..., due to the small value of magnetic moment in 
comparison with electrons, but a strong magnetic field {\it can polarize these 
particles}. So, due to the strong dependence of 
the corresponding reaction rates on 
the polarization states of colliding particles, this effect must have important 
consequences on nucleosynthesis. For example, it would change the standard 
predictions of 
abundances of light elements in Universe, such as $d$, $^{3}\!He$, $^{4}\!He$, 
$^7\!Li$ and $^9\!Be$, because all these elements are produced 
in processes with an essential spin dependence of the corresponding matrix elements. Note, that the 
relative abundances of these elements provide now a reliable method to determine such important characteristics of the Universe as the baryon mass density 
parameter $\Omega_B$ \cite{ka97}. This parameter is very important to 
discriminate between different models. The precise determination of  $d$ and 
$^3\!He$ abundances will definitely constrain the upper bound of 
$\Omega_B$-values, 
whereas  $^7\!Li$ and $^9\!Be$ abundances lead to a constraint on the lower 
bound of the  $\Omega_B$-values .

Reaction rates for processes as $n+p\to d+\gamma$, $n+d\to 
^3\!H+\gamma$, \mbox{$p+d\rightarrow$} $^3\!He+\gamma..$, which are the most 
important in nucleosynthesis in early Universe and in hydrogen burning in stars 
and in the cooling process of neutron stars, essentially are changed in case of 
collision of polarized particles. As far as we know, spin degrees of freedom 
have not been taken into account in the analysis of possible nuclear processes 
in nucleosynthesis and neutron stars. Polarization phenomena in collisions of 
light nuclei may represent an additional effect in any estimation of the 
abundances of the light elements in the Universe. Discussions about the relative role of various effects as anisotropy and baryon inhomogeneities, about the 
special 
neutrino properties (oscillations, degeneracy, electromagnetic characteristics, 
massive neutrinos), cosmic strings etc. \cite{ka97}) must consider 
polarization phenomena, also.

Our main goal here is to predict the dependence of the cross sections on the 
polarization of colliding particles in some cases, where this problem can be 
treated in model independent way: this can be done at threshold, in the 
framework of our well adapted formalism. 

\section{The reaction $p+p\to d+e^++\nu_e$}
Let us consider the process $p+p\to d+e^++\nu_e$. At the $keV$ 
energy scale for the colliding protons, we can consider that  the S-wave 
approximation for colliding protons is correct. This allows to establish the spin structure of the 
corresponding matrix element in a model independent way. 
Due to Pauli principle, we have in this case only one initial state, 
${\cal J}^{P}=0^+$, that produces the following transitions (taking into 
account the strong violation of P-invariance in this weak process) 
$0^+\to M1,E1_t,~E1_{\ell},~E1_s,$
where we describe the intermediate $W^*$-boson in the process $p+p\to 
d+W^*$ as a virtual photon, using a formalism of 
multipole decomposition; the indexes $\ell$ and $t$ correspond to transversal 
and longitudinal components of the $W-$boson polarizations, the index $s$ 
corresponds to the $4-th$ component of the axial hadron current (due to the 
non-conservation of this current). The dynamics of the considered process is 
contained in the $k^2-$ dependence of the corresponding form factors, which 
describe the above mentioned multipole transitions, making the matrix element
quite complicated, even at threshold.

However some of the polarization observables can be calculated without knowing this 
dynamics. For example, the dependence of the cross section on the polarizations 
$\vec{P_1}$ and $\vec{P_2}$which
of the colliding protons can be predicted exactly, using only the singlet nature of the initial $pp-$state for collisions in S-state:
\begin{equation}
\sigma(\vec{P_1},\vec{P_2})=\sigma_0(1-\vec{P_1}\cdot\vec{P_2}).
\label{eq:as2}
\end{equation}
This simple dependence has a model independent nature and shows the strong 
effects of colliding particle polarizations. It results in a decrease of cross 
section in collision of particles with parallel polarizations. This condition 
can typically be realized on the surface of neutron stars, where protons can be 
polarized by the strong magnetic field. Note that the deuterons, produced in 
such $pp-$collisions, must be polarized, with respect to the direction of the 
magnetic 
field, so this polarization must be taken into account in the following 
reactions which take place in the proton-proton chain of hydrogen burning:
 $d+p \to ^3\!He+\gamma$, $^3\!He+^3\!He\to ^4\!He+2p$ etc., 
which are also characterized by a strong spin dependence of the corresponding 
matrix element.

\section{The big-bang and the reaction $n+p\to d+\gamma$.}
In the standard big-bang model the ratio of the proton to neutron numbers was 
determined through the reactions (\ref{eq:as1}) mediated by the weak interaction. 
Deuterons are produced through the process $n+p\to d+\gamma$, but at large 
temperature the probability that they undergo photo-dissociation is very high, 
due to the large number of energetic photons 
(the ratio of photons to baryons is $\simeq 10^9$). So a significant concentration of deuterons can not be found 
until the temperature has dropped below the deuteron binding energy. At 
\mbox{$T\simeq 10^9~K$} deuterons begin to be formed and then the following 
reactions of primordial nucleosynthesis proceed rapidly: 
$p+d\to ^3\!He+\gamma $, , $n+d\to ^3\!H+\gamma $, $d+d\to ^3\!H+p$, 
$d+d\to ^3\!He+n $, $^3\!H+d\to ^4\!He+n$, $^3\!He+d\to 
^4\!He+p$ and $^3\!He+^3\!He\to ^4\!He+2p$. With the production of 
$^4\!He$, the primordial nucleosynthesis comes to the end. It is the reaction 
$n+p\to d+\gamma$ which essentially starts this chain. 

In presence of strong magnetic field in the early Universe, 
polarization effects for all these processes may become important. The capture of thermal neutrons in $n+p\to d+\gamma$
is characterized by the following transitions:
$$S_i=0\to M1,~~ S_i=1\to M1 ~\mbox{and}~ E2,$$
where $S_i$ is the total spin of the system $n+p$. From the classical deuteron 
electrodynamics, one can find that singlet $M1$ transition dominates at low 
energy, with the following structure of the corresponding matrix element 
(Fermi radiation)\cite{Fe35}:
\begin{equation}
{\cal M}(M1)=g_M~\left (\vec e^*\times\vec k \cdot D^*\right )~\left 
(\tilde{\chi_2}\sigma_y\chi_1\right ),
\label{eq:as3}
\end{equation}
where $\vec e$ is the 3-vector polarization of the produced photon and $\vec k$ 
is the unit vector along the 3-momentum of the photons. Due to the singlet nature of the considered transition, the cross section has a definite dependence on the 
polarizations $\vec{P_1}$ and $\vec {P_2}$ of the colliding nucleons (after 
summing over the polarizations of the produced $\gamma$ and $d$):
\begin{equation}
\displaystyle\frac{d\sigma}{d\Omega}(\vec n \vec p\to d\gamma)=
\left (\frac {d\sigma}{d\Omega}\right )_0\left(  1- \vec{P_1}\cdot\vec 
P_2\right).
\label{eq:as4}
\end{equation}
 
The presence of a single dynamical constant $g_M$ in the matrix element
(\ref{eq:as3}) 
allows to predict numerical values for all polarization observables of the 
considered process in a model independent form. For example, the polarization 
properties of the produced 
deuterons are: $\vec S=0$ and $Q_{zz}=1/2$ (see Eq.(\ref{eq:eqrho}), for any polarization of colliding nucleons. Such specific  polarization properties strongly affect the production rates of 
the deuteron induced reactions in the primordial nucleosynthesis.

In the considered reaction, $n+p\to d+\gamma$, the range of the 
$M1-$radiation is limited by the thermal energy of neutrons, and at higher 
temperatures the $E1-$ radiation must be more important. In the general case the 
$E1-$ radiation is characterized by a large number of independent multipole 
transitions, 
corresponding to different values of ${\cal J}$ and $S_i$. But the situation is 
essentially simplified, due to the Bethe-Peierls \cite{Be35} Ansatz on the spin 
independence 
of the $E1-$radiation. The resulting matrix element can be written as:
$${\cal M}(E1)=g_E~\left (\vec e\cdot\vec q\right ) ~\left 
(\tilde{\chi_2}\sigma_y\vec\sigma~\cdot D^*\chi_1\right ),$$
where  $\vec q$ is the unit vector along the 3-momentum of the colliding 
nucleons and $g_E$ is the amplitude of the $E1$-transition. After summing over 
the polarization states of the produced $\gamma$ and $d$ one can find the 
following dependence of the cross section on the polarization of the colliding 
particles:
\begin{equation}
\sigma_E(\vec n\vec p\to d\gamma)=\sigma_0\left( 1+ \vec{P_1}\cdot\vec 
P_2\right).
\label{eq:as5}
\end{equation}
Neglecting the possible interference of $M1-$ and $E1-$ transitions, one can 
predict the resulting dependence:
\begin{equation}
\sigma(\vec n \vec p\to d\gamma)=\sigma_0\left( 1+ {\cal 
A}\vec{P_1}\cdot\vec P_2\right),~~{\cal A}=(|g_E|^2-|g_M|^2)/(|g_E|^2+|g_M|^2),
\label{eq:as6}
\end{equation}
where  the asymmetry coefficient ${\cal A}$ strongly depends on the temperature.

\section{Radiative capture of nucleons by deuterons}
In the $keV$ energy region, the processes $n+d\to ^3\!H+\gamma$ and 
$p+d\to ^3\!He+\gamma$ play an important role in nuclear astrophysics 
and in nuclear physics. In the standard big-bang nucleosynthesis theory the 
corresponding reaction rates are necessary to estimate the $^3\!He$-yield as 
well as the abundances of other light elements. In nuclear physics these 
reactions are also very interesting since one can expect large contributions of 
meson exchange currents.

The spin structure of the matrix elements for $N+d$ radiative capture is 
complicate also for the low energy interaction, as we have here 3 independent 
multipole transitions (allowed by the $P-$parity and the total angular momentum 
conservation): ${\cal J}^{P}=\displaystyle\frac{1}{2}^+\to 
M1$ and 
$ {\cal J}^{P}=\displaystyle\frac{3}{2}^+\to M1$ and $E2$, with the following parametrization of the corresponding contributions to the 
matrix element:
$$i(\chi_3^\dagger\chi_1)(\vec D\cdot\vec{e^*}\times\vec k),$$
\begin{equation}
(\chi_3^\dagger\sigma_a\chi_1)(\vec D\times [\vec{ e^*}\times\vec k])_a,
\label{eq:as7}
\end{equation}
$$\chi_3^\dagger(\vec\sigma\cdot\vec{e^*}~\vec D\cdot\vec k+\vec\sigma\cdot
\vec k~ \vec D\cdot\vec e^*)\chi_1,$$
where $\chi_1$ and $\chi_3$ are the 2-component spinors of initial nucleon and 
final $^3\!He$ (or $^3\!H$).

The first two structures in (\ref{eq:as7}) correspond to the M1 radiation. To obtain the 
spin structure, which corresponds to a definite value of ${\cal J}$ for the 
entrance channel, it is necessary to build special linear combinations of 
products
$\vec D\chi_1$ and $\vec \sigma\times\vec D\chi_1$, with ${\cal 
J}^{P}=\displaystyle\frac{1}{2}^+$ or ${\cal 
J}^{P}=\displaystyle\frac{3}{2}^+$:
$$\vec {\phi}_{1/2}=(i\vec D+\vec\sigma\times\vec D)\chi_1~\mbox{and}~(2i\vec 
D-\vec\sigma\times\vec D)\chi_1.$$
For both possible magnetic dipole transitions with 
 ${\cal J}^{P}=\displaystyle\frac{1}{2}^+$ (amplitude $g_1$) and ${\cal 
J}^{P}=\displaystyle\frac{3}{2}^+$ (amplitude $g_3$) we can write:
$$g_1:~~\chi_3^\dagger(i\vec D\cdot\vec{e^*}\times\vec k+\vec\sigma\times\vec 
D\cdot\vec{e^*}\times\vec k)\chi_1,$$
\begin{equation}
g_3:~~\chi_3^\dagger(i\vec D\cdot\vec{e^*}\times\vec k+\vec \sigma\times\vec 
D\cdot\vec{e^*}\times\vec k)\chi_1.
\label{eq:as8}
\end{equation}
The general dependence for the cross section of S-wave particle collisions 
(with spins 1/2 and 1) can be parametrized by the following formula:
\begin{eqnarray}
\displaystyle\frac{d\sigma}{d\Omega}(\vec N \vec d\to ^3\!He\gamma)=
&\left (\displaystyle\frac {d\sigma}{d\Omega}\right )_0 [ 
1+a_1(Q_{ab}k_ak_b)+a_2 \vec{S}\cdot\vec P+a_3\vec k \cdot \vec P~\vec 
k\cdot\vec S \nonumber \\
& +a_4~\vec k\cdot\vec Q\times\vec P ],~Q_a\equiv Q_{ab}k_b,
\label{eq:as9}
\end{eqnarray}
where $\vec P$ is the pseudovector of proton polarization, $\vec S$ and $Q_{ab}$ 
are the vector and tensor deuteron polarizations, defined above. The real 
coefficients $a_2-a_4$ in (\ref{eq:as9}) characterize the spin correlation coefficients and 
the coefficient $a_1$ is the tensor analyzing power for the collisions of 
unpolarized protons with polarized deuterons. 

After integration in Eq. (\ref{eq:as9}) 
over the $\vec k$-direction, one can find for the total cross section:
$$\sigma(\vec N\vec d) =\sigma_0(1+a\vec P\cdot\vec S),~a=a_2+a_3/3,$$
i.e. the dependence from the tensor deuteron polarization disappears.

Using expressions (\ref{eq:as8}), 
one can obtain the following formulas for the 
corresponding differential cross-sections of radiative capture of polarized 
nucleons by polarized deuterons:
$$\displaystyle\frac{d\sigma}{d\Omega}(\vec N \vec d)=
\left (\frac {d\sigma}{d\Omega}\right )_0\left(1-\vec{S}\cdot\vec P\right 
),~\mbox{if}~g_1\ne 0,$$
\begin{equation}
\displaystyle{\frac{d\sigma}{d\Omega}(\vec N \vec d)=
\left (\frac {d\sigma}{d\Omega}\right )_0\left(1-\frac{1}{4}\vec{S}\cdot\vec P
+\frac{3}{4}\vec{k}\cdot\vec P~\vec k\cdot\vec S+Q_{ab}k_ak_b 
\right ),~\mbox{if}~g_3\ne 0.}
\label{eq:as10}
\end{equation}
So, after $\vec k$-integration one can find for the total cross sections:
$${\sigma}(\vec N \vec d)=
 {\sigma}_0\left(1-\vec{S}\cdot\vec P
\right ),~\mbox{if}~g_1\ne 0,$$
$${\sigma}(\vec N \vec d)=
 {\sigma}_0,~\mbox{if~only}~g_3\ne 0,$$
i.e. the amplitude $g_3$ cannot produce polarization dependence in the total 
cross section of $N+d$-interactions.

\section{Magnetic field and Polarization}

\begin{figure}
\mbox{\epsfxsize=14.cm\leavevmode \epsffile{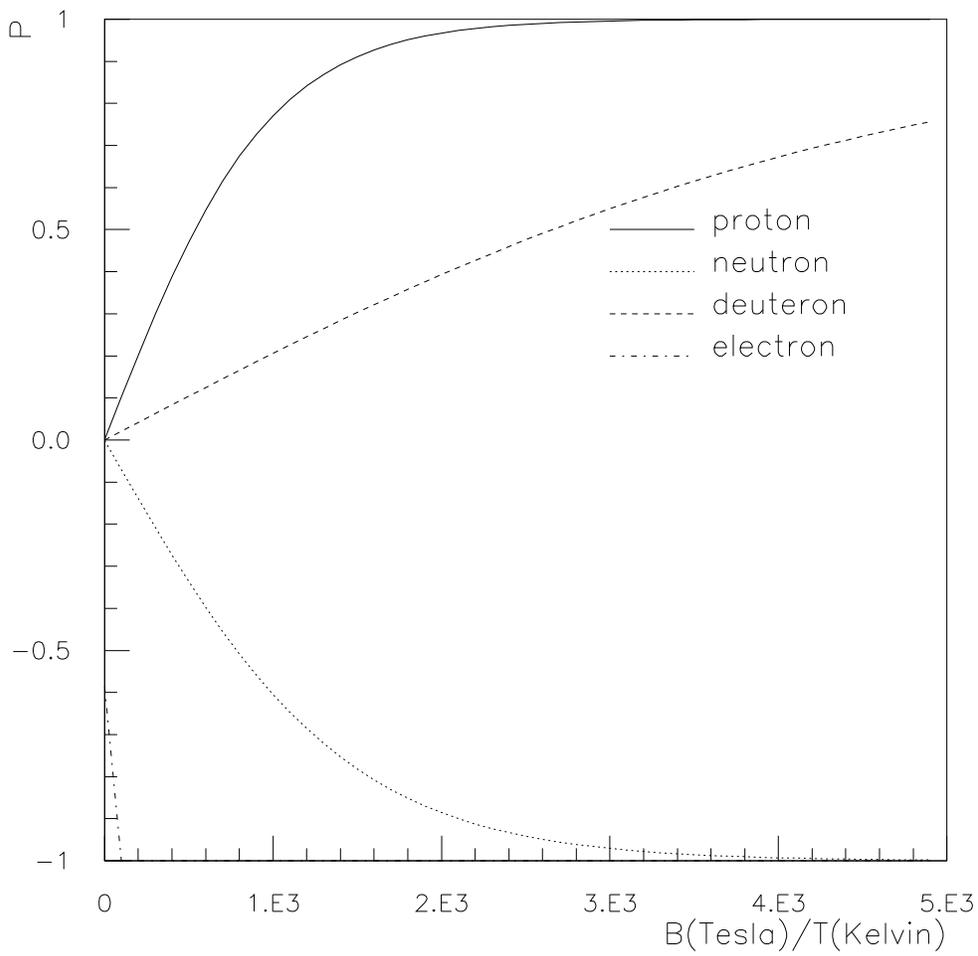}}
\caption{Dependence of polarization on the ratio of the 
magnetic field [Tesla] over the temperature [K], 
for different particles:   proton (full line),
neutron (dotted line),  deuteron (dashed line) and  electron (dashed-dotted 
line)}.
\label{fig:as1}
\end{figure}
For the reactions discussed above, namely  $p+p\to d+e^+ +\nu_e$ 
and 
$n+p\to 
d+\gamma$, we have derived a simple  dependence of the cross section on the 
polarizations of the colliding particles, which does not depend on the model 
chosen to describe the dynamics of the reaction and the structure of the 
particles involved. For these cases, it is then possible to find a simple 
relation between the ratio of polarized/unpolarized cross section and the ratio 
of magnetic field/temperature. 

\begin{figure}
\mbox{\epsfxsize=14.cm\leavevmode \epsffile{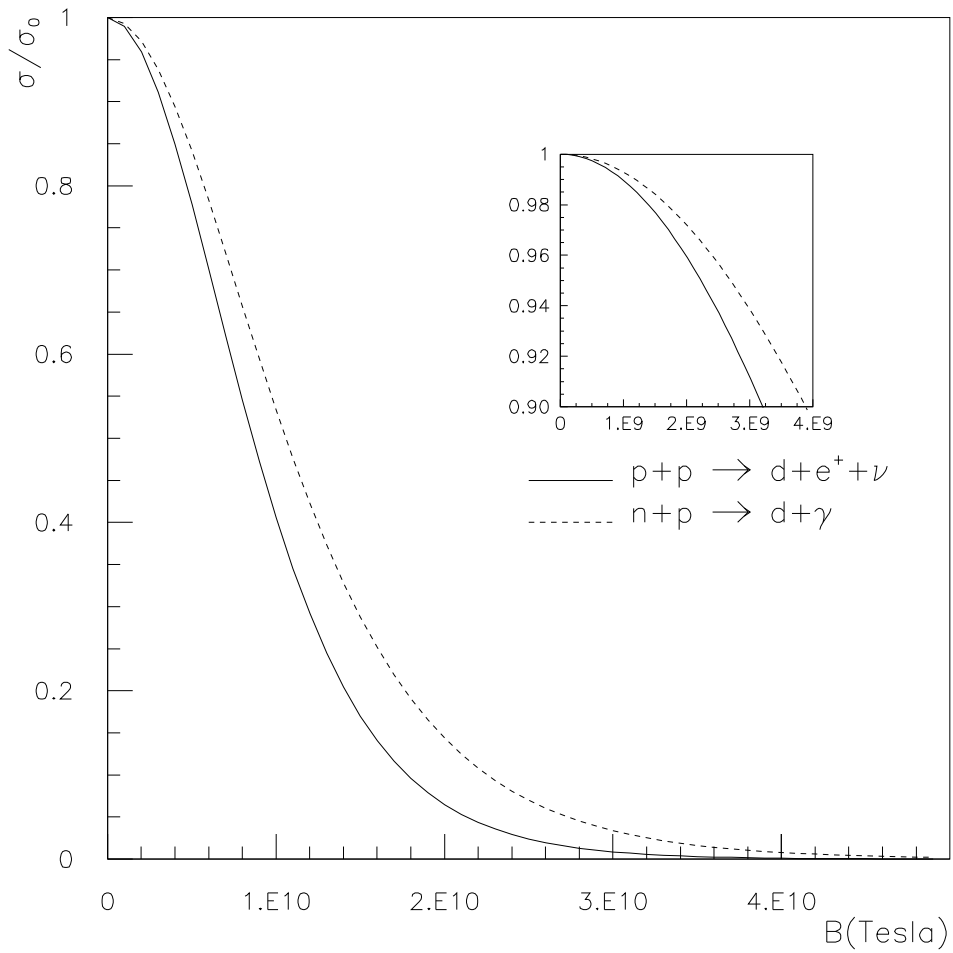}}
\caption{Dependence of  the polarized on unpolarized cross section ratio on 
the magnetic field expressed in Tesla  for a temperature of  $10^7~$K, for two 
reactions: $p+p\to d+e^++\nu_e$ (full line) and $n+p\to 
d+\gamma$ (dashed line).}
\label{fig:as2}
\end{figure}

The polarization $P$, of $I$-spin particles, induced by a magnetic field B,  at 
thermal equilibrium with temperature $T$, is given by the Brillouin function:
$$P_I(x)=\displaystyle\frac{2I+1}{2I}\cot\left( 
\displaystyle\frac{2I+1}{2I}x\right ) 
-\displaystyle\frac{1}{2I}\cot \left( \displaystyle\frac{1}{2I}x\right ),$$
with $x=\hbar\gamma I B/kT$, $\gamma$ the gyromagnetic ratio and $k$ the 
Boltzmann 
constant. For spin $I=1/2$ we find at thermal equilibrium: 
$P=th\displaystyle\frac{\gamma\hbar B}{2kT}.$ 
For a given kind of particles, the polarization depends only on the ratio 
$B/T$ (Fig. \ref{fig:as1}).
As an example, it is possible to apply  the previous formalism, to evaluate the 
changing of  the cross 
section due to the magnetic field for the reactions $p+p\to 
d+e^++\nu_e$ and 
for the $E1$-radiation in $n+p\to 
d+\gamma$. Typically for these cases, we have shown that  a model independent 
expression of  the cross section as a function of the polarizations of the 
colliding particles can be found: Eqs. (\ref{eq:as2}) and (\ref{eq:as5}). We illustrate this 
dependence in Fig. \ref{fig:as2}, for $T=10^7 ~$K, which is a 
typical value for the temperature in the center of the Sun. In these two cases 
the 
polarized cross section (for collisions of particles with parallel 
polarizations) is lower than the unpolarized one. Recently a limit
on possible deviations of the cross section for the reaction $p+p\to 
d+e^++\nu_e$ from SSM based on heliosysmology 
constraints, has been given: $0.94\le S/S_{SSM}\le 1.18$ \cite{SDI98}. Assuming 
the 
existence of a magnetic field in the Sun (which has the effect to polarize 
protons), its upper limit, allowed by this 
constraint, would be $2.5\times10^9$ T. Such small sensitivity is a result of a 
quadratic dependence of the cross section on proton polarization.

The  limit given here is some order of magnitude larger than the current 
estimations.  However, here,  it is not necessary to asssume that the magnetic 
field in the Sun is the uniform and constant in time. Our estimate 
is correct also in case of a magnetic field, resulting from some local 
fluctuations in plasma, with eventually different directions in  different 
regions. Therefore the possibility \cite{Pa74}, that the  strong
magnetic field in the core of the Sun,  present at its creation, would have been
raised to surface during a short time, would not affect the 
considered analysis. It must also be noticed that in the SSM 
predictions no magnetic field is included.

\begin{figure}
\mbox{\epsfxsize=14.cm\leavevmode \epsffile{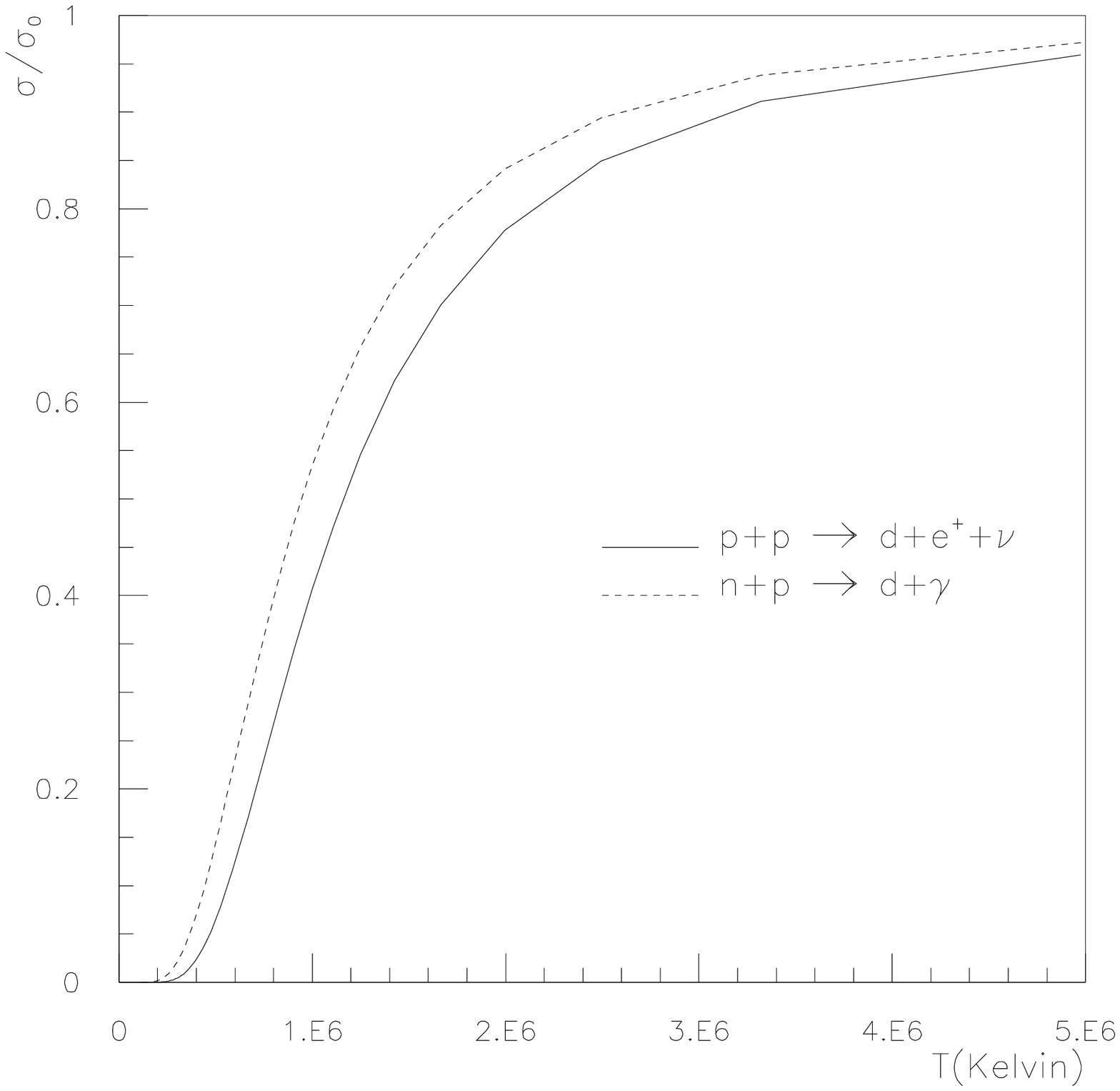}}
\caption{Dependence of  the  polarized on unpolarized cross 
section ratio
on the  temperature expressed in Kelvin, for a magnetic field of $10^9$ T, 
typical of 
neutron 
stars, for two reactions: $p+p\to d+e^++\nu_e$ (full line) nd 
$n+p\to 
d+\gamma$ (dashed line)}
\label{fig:as3}
\end{figure}

In Fig. \ref{fig:as3} we report, for the same reactions as in 
Fig. \ref{fig:as2}, the dependence of the 
ratio of polarized on unpolarized cross section on the temperature, for a 
magnetic 
field of $B=10^{9}$ ~T which is the value usually quoted for the magnetic 
field at  
the surface of  neutron stars. This ratio varies rapidly  from 0 to 1 for about 
one order of magnitude of variation in the temperature (in the range 
$2\times10^9~$K to $5\times10^{10}~$K).

In this chapter, we showed the importance of polarization phenomena for collisions of light 
nuclei at thermonuclear energies. The results of our analysis can be summarized in the 
following way:
\begin{itemize}

\item In general case the polarization effects are large, in absolute value, for 
all 
reactions, which are responsible for the primordial nucleosynthesis in the 
Universe, 
and for the nuclear processes in usual stars, like Sun, and  in neutron stars.

\item A strong magnetic field, which is present in neutron stars and in the early Universe, can polarize protons, neutrons, deuterons.. This polarization 
changes the 
reaction rates for fundamental  processes, participating in primordial 
nucleosynthesis.

\item The ejectiles of the considered reactions are also polarized, if at least 
one of the colliding particles is polarized. This induced polarization has to 
change the relative role of different reactions in the chain of primordial 
nucleosynthesis.
\item The model-independent result on the dependence of the cross section for 
the process   $p+p\to d+e^++\nu_e$
on the polarizations of the colliding protons, namely,  
$\sigma(\vec{P_1},\vec{P_2})=\sigma_0(1-\vec{P_1}\cdot\vec{P_2}),$ has to be 
taken into account for the analysis of processes in hydrogen burning stars, 
like Sun.
\item In the presence of magnetic field, the cross section of radiative capture 
of neutrons by protons, $n+p\to d+\gamma$, has to show a large 
dependence on temperature, as a result of the contributions of magnetic and 
electric dipole radiations.
\item The polarization observables for $n+^3\!He\to ^4\!He+\gamma$ and
$p+^3\!H\to ^4\!He+\gamma$ can be predicted exactly. On the other hand, 
for the calculation of polarization effects in the processes  
 $n+d\to ^3\!H+\gamma$ and
\mbox{$p+d\to ^3\!He+\gamma$}  it is necessary to have dynamical  
information relative to multipole amplitudes.

\item The limitations on the deviations of the cross section for the 
reaction $p+p\to d+e^++\nu_e$ from the SSM value, given by 
helioseysmology constraints,
can give a {\it model independent} estimation of the  maximum possible value of 
the magnetic field in Sun core, $B< 2\cdot10^9$ T.

\end{itemize}


\chapter{Conclusions}

We showed above that the threshold region for different hadronic and nuclear
reactions (induced by all known interactions, weak, electromagnetic and strong)
has some universal properties, concerning the analysis of the spin structure of
the corresponding matrix elements and the polarization phenomena. It is
important that this universal behavior, being essentially model independent, is
dictated by the most general symmetry of fundamental interactions, such as the
Pauli principle, the conservation of total angular momentum, the C- and P-
invariance and the isotopic invariance.

The above developed polarization formalism allows to express definite and
transparent statements about polarization phenomena in different reactions. Even
complicated processes, such as $p+p\to p+p+V^0$, of vector meson production,
where all five particles are with non zero spin, can be exactly 
described at threshold in terms of a single amplitude.

Note that such symmetry analysis of the spin structure of different threshold
matrix elements must be considered as the first necessary step, which allows to
separate strong kinematical predictions from dynamical, model-dependent
assumptions. Note also, that, as a rule, the threshold polarization phenomena in hadronic and
nuclear collisions, which are non-zero, due to symmetry properties, take their
maximal vaue.

We proved above that in some cases the polarization phenomena are nonuseful for
testing the dynamics of the considered reaction, because such polarizations are
often model independent.

We showed the connection of polarization phenomena in different reactions of
$NN-$collisions with isotopic invariance of the strong interaction, therefore
namely polarization observables in $np-$collisions can be used as an independent
and original method of testing the isotopic invariance.

We stressed the importance of polarization phenomena in some non standard
applications to the thermonuclear fusion reaction with polarized fuel. In this
way it is possible to solve such principal problems, as the essential decreasing
of production of radioactive $^3He$ and intensive neutron beams as well as the
effective arrangement of the reaction shielding and blanket.
Nontrivial application of polarization phenomena can be found in astrophysics,
where the strong magnetic field can change essentially the reaction rates, due
to non-zero polarization of heavy particles, such as protons, neutrons,
deuterons, etc..

So, finally, we can conclude that the analysis of polarization
phenomena in threshold region, for hadron and nuclear interactions, can be done
in a general form, with many interesting applications.

\addcontentsline{toc}{chapter}{\protect\numberline{5} {Bibliography}}

\end{document}